\documentclass[%
readprint,
twocolumn,
superscriptaddress,
showpacs,preprintnumbers,
nofootinbib,
aps,
prd
]{revtex4-2}

\usepackage{subfigure}
\usepackage{amsmath}
\usepackage{amssymb}
\usepackage{mathtools}
\usepackage{hyperref}
\usepackage{placeins}
\usepackage{graphicx}
\usepackage{dcolumn}

\usepackage{color}

\usepackage{ulem}

\usepackage{comment}

\def\be{\beta}

\def\Om{\Omega}

\newcommand{\ben}{\begin{equation}}
\newcommand{\een}{\end{equation}}
\newcommand{\bea}{\begin{eqnarray}}
\newcommand{\eea}{\end{eqnarray}}
\newcommand{\ba}{\begin{array}}
\newcommand{\ea}{\end{array}}
\newcommand{\bit}{\begin{itemize}}
\newcommand{\eit}{\end{itemize}}

\newcommand{\mpl}{m_{\text{P}}}

\newcommand{\half}{\frac12}

\newcommand{\bx}{\textbf{x}}

\newcommand{\Hc}{H_*} % Hubble rate at transition (vaguely defined)
\newcommand{\Nb}{N_\text{b}} % Bubble number
\newcommand{\Rc}{R_\text{c}} % Critical droplet radius
\newcommand{\Rstar }{R_*} % Mean droplet radius at collision
\newcommand{\vw}{v_\text{w}} % Wall velocity
\newcommand{\Mb}{M_\text{b}} % Mass of scalar field in broken phase
\newcommand{\lambar}{\overline{\lambda}} % Parameter showing how close we are to thin wall.

\newcommand{\rGW}{\rho_\text{gw}} % Energy density GW
\newcommand{\OmGW}{\Omega_\text{gw}} % Fractional energy density GW

\newcommand{\gmStar}{\gamma_*} % gamma_*=R_*/2R_c
\newcommand{\gmStarOneD}{\gamma_{*\text{,1D}}} % lattice version of \gmStar
\newcommand{\gmStarThreeD}{\gamma_{*\text{,3D}}} % lattice version of \gmStar

\newcommand{\phiAtMin}{\phi_\text{b}} % Phi at minimum of potential
\newcommand{\phiPeak}{\phi_\text{max}}

\newcommand{\VPeak}{V_\text{max}} % V(\phi_{max}) - V(0)
\newcommand{\rVac}{\rho_\text{vac}} %Vacuum energy density
\newcommand{\OmVac}{\Omega_\text{vac}} % Fractional vacuum energy density
\newcommand{\siTW}{\sigma^\text{tw}}
\newcommand{\phiCrit}{\phi_\text{c}}

\newcommand{\rIn}{r_\text{in}} % Inner radius of wall
\newcommand{\rOut}{r_\text{out}} % Outer radius of wall
\newcommand{\rMid}{r_\text{mid}} % Bubble radius
\newcommand{\lw}{l_\text{w}}

\newcommand{\rhoIn}{\rho_\text{in}} % Inner Euclidean radius of wall
\newcommand{\rhoOut}{\rho_\text{out}} % Outer Euclidean radius of wall

\newcommand{\nb}{n_\text{b}} % Number of bubbles

\newcommand\Tstrut{\rule{0pt}{2.4ex}}       % top strut
\newcommand\Bstrut{\rule[-1.3ex]{0pt}{0pt}} % bottom strut

\definecolor{newgreen}{RGB}{10,100,20}
\definecolor{purple}{rgb}{0.5,0,0.5}
\definecolor{BLUE}{rgb}{0,0,1}

\def\lsi{\raise0.3ex\hbox{$<$\kern-0.75em\raise-1.1ex\hbox{$\sim$}}}

\graphicspath{
  {./}
  {./figures/}
}

\begin{document}

\newcommand{\Sussex}{\affiliation{
Department of Physics and Astronomy,
University of Sussex, Falmer, Brighton BN1 9QH,
U.K.}}

\newcommand{\HIPetc}{\affiliation{
Department of Physics and Helsinki Institute of Physics,
PL 64, % (Gustaf H\"{a}llstr\"{o}min katu 2),
FI-00014 University of Helsinki,
Finland
}}

\newcommand{\Nottingham}{\affiliation{
    School of Physics and Astronomy,
    University of Nottingham,
    Nottingham NG7 2RD,
    U.K.
}}

\title{Gravitational waves from vacuum first-order phase transitions II: \\
  from thin to thick walls}

\author{Daniel Cutting}
\email{d.cutting@sussex.ac.uk}
\Sussex
\author{Elba Granados Escartin}
\email{elva.granadosescartin@gmail.com}
\HIPetc
\author{Mark Hindmarsh}
\email{m.b.hindmarsh@sussex.ac.uk}
\Sussex
\HIPetc
\author{David J. Weir}
\email{david.weir@helsinki.fi}
\HIPetc
\Nottingham

\preprint{HIP-2020-13/TH}

% Style to use for fixed dates:
% \date{April 20, 2017}

\date{\today}

\begin{abstract}
  In a vacuum first-order phase transition, gravitational waves are
  generated from collision of bubbles of the true vacuum. The spectrum
  from such collisions takes the form of a broken power law. We
  consider a toy model for such a phase transition, where the dynamics
  of the scalar field depends on a single parameter $\lambar$, which
  controls how thin the bubble wall is at nucleation and how close to
  degenerate the vacua are relative to the barrier. We extend on our
  previous work by performing a series of simulations with a range of
  $\lambar$. The peak of the gravitational-wave power spectrum varies
  by up to a factor of $1.3$, which is probably an unobservable
  effect. We find that the ultraviolet (UV) power law in the
  gravitational-wave spectrum becomes steeper as
  $\lambar \rightarrow 0$, varying between $k^{-1.4}$ and $k^{-2.2}$
  for the $\lambar$ considered. This provides some evidence that the
  form of the underlying effective potential of a vacuum first-order
  phase transition could be determined from the gravitational-wave
  spectrum it produces. \end{abstract}
\maketitle

\section{Introduction}

Upcoming space-based gravitational-wave detectors like the Laser
Interferometer Space Antenna~\cite{Audley:2017drz} (LISA) are
anticipated to dramatically increase our capability to probe early
universe cosmology through gravitational waves~\cite{Caprini:2018mtu}.
In particular, LISA will be sensitive to first-order cosmological
phase transitions at the electroweak
scale~\cite{Caprini:2015zlo,Caprini:2019egz}.

In the Standard Model, the electroweak transition is a
crossover~\cite{Kajantie:1996mn,Kajantie:1996qd}, and as such there
are no first-order phase transitions at the electroweak scale.
However, there is a multitude of well-motivated extensions to the
Standard Model that produce first-order phase transitions, ranging
from singlet
extensions~\cite{Profumo:2007wc,Espinosa:2011ax,Cline:2012hg,Profumo:2014opa,Beniwal:2018hyi},
two-Higgs doublet
models~\cite{Kakizaki:2015wua,Dorsch:2016nrg,Basler:2016obg}, models
in which a conformal symmetry is spontaneously
broken~\cite{Randall:2006py,PhysRevD.82.083513,Konstandin:2011dr,vonHarling:2017yew,Dillon:2017ctw,Megias:2018sxv,Bruggisser:2018mrt},
to models with a phase transition in a hidden
sector~\cite{Schwaller:2015tja,Addazi:2017gpt,Aoki:2017aws,Croon:2018erz,Breitbach:2018ddu,Okada:2018xdh,Hasegawa:2019amx,Hall:2019rld,Hall:2019ank}.
The gravitational wave signal generated by phase transitions in these
Beyond the Standard Model extensions will enable LISA to detect or
constrain their existence.

In a first-order cosmological phase transition, some effective scalar
field is trapped in a metastable state (symmetric phase), separated by
a potential barrier from the true vacuum state (broken
phase)~\cite{Coleman:1977py,Linde:1981zj,Steinhardt:1981ct}. When the
transition proceeds, bubbles of the true vacuum nucleate, expand and
eventually collide, sourcing transverse-traceless modes of shear
stress, which in turn source gravitational
waves~\cite{Witten:1984rs,1986MNRAS.218..629H}. The dynamics of the
resulting phase transition can be split qualitatively according to
whether the bubble wall reaches a terminal velocity before colliding,
or whether the bubble wall continues to accelerate until collision. We
denote these different transition types respectively as `thermal' and
`vacuum-like'.

In a thermal phase transition, bubbles nucleate in the presence of a
hot relativistic plasma made up of the early universe particle
content. The friction felt between the plasma and the expanding bubble
wall is sufficient to eventually result in the wall approaching a
terminal wall velocity. Shells of hotter plasma develop around the
expanding bubbles, and after collision continue to propagate as
long-lasting sound waves. Eventually, the sound waves are expected to
decay, and the flow may become
turbulent~\cite{Witten:1984rs,KurkiSuonio:1984ba}.

In a vacuum-like transition, the vacuum pressure driving the phase
transition overcomes the resulting friction from the plasma and the
bubble wall continues to accelerate before collision. An early study
predicted that in most electroweak scale phase transitions the bubble
wall would undergo `run away' acceleration, provided sufficient
supercooling~\cite{Bodeker:2009qy}. More recently, it has been shown
that if the scalar field couples to heavy gauge bosons,
next-to-leading-order effects cause the friction to grow
proportionally to the Lorentz factor $\gamma$ of the bubble
wall~\cite{Bodeker:2017cim}. In this case, the runaway condition in
Ref.~\cite{Bodeker:2009qy} is no longer fulfilled.

Several scenarios have been proposed that can still result in
`vacuum-like' behaviour. The friction term proportional to $\gamma$ is
generated by transition radiation of gauge bosons acquiring a mass as
they cross the bubble wall. If the phase transition occurs in the
absence of gauge fields, such as during the spontaneous breaking of an
approximate global symmetry, then the dominant transition radiation
process may grow as $\sim\log \gamma$~\cite{Caprini:2019egz}.

Alternatively, if there is an extreme level of supercooling, then
there could be a sufficient dilution of the early universe plasma and
resulting plasma friction to allow for the bubble walls to accelerate
until collision. The levels of supercooling required for this to occur
is large but can be achieved in transitions with a classically
scale-invariant potential~\cite{Ellis:2019oqb}. In other models,
sufficient supercooling is difficult to achieve; at large supercooling
the universe can start to inflate, meaning that the bubbles cannot
percolate and the transition does not complete~\cite{Ellis:2018mja}.

Finally, in a dark sector that is decoupled at the time of transition,
runaway-type transitions are achieved more easily than in the visible
electroweak sector~\cite{Breitbach:2018ddu}.

The first attempts to model the gravitational-wave power spectrum from
first-order phase transitions employed a seminumerical simulation
method termed the `envelope approximation'. In this calculation, the
stress-energy is assumed to be located in an infinitesimally thin
shell at the bubble wall which disappears upon
collision~\cite{Kosowsky:1992vn}. This technique was first applied to
vacuum transitions, and then to thermal
transitions~\cite{Kamionkowski:1993fg,Huber:2008,Konstandin:2017sat}.
When the bubble wall velocity is ultra-relativistic the resulting
gravitational-wave spectrum was shown to be a broken power law rising
in the infrared (IR) and falling in the ultraviolet (UV) as
approximately $k^{2.9}$ and $k^{-0.9}$, respectively. Analytic studies
which build upon the envelope approximation have confirmed the broken
power laws found from numerical simulations~\cite{Jinno:2016vai}.

Extensions to the envelope approximation in which the shell of
shear-stress continues to propagate after collision have also been
studied~\cite{Jinno:2017fby,Konstandin:2017sat}. We will follow
Ref.~\cite{Konstandin:2017sat} in referring to this as the bulk flow
model. Simulations of the bulk flow model with ultra-relativistic wall
velocities have found power laws of approximately $k^{0.9}$ and
$k^{-2.1}$.

Many developments have been made in the study of thermal phase
transitions. 3D hydrodynamical simulations of weak and intermediate
strength
transitions~\cite{Hindmarsh:2013xza,Giblin:2014qia,Hindmarsh:2015qta,Hindmarsh:2017gnf}
demonstrated that sound waves form the dominant contribution to the
gravitational-wave signal and that while the contribution from bubble
collisions is subdominant, for thermal phase transitions it is well
represented by the envelope approximation~\cite{Weir:2016tov}.
Modelling has shown weak and intermediate strength transitions are
well represented shortly after the transition by a linear
superposition of propagating sound
waves~\cite{Hindmarsh:2016lnk,Hindmarsh:2019phv}.

Simulations of stronger first-order phase transitions indicated that
for walls moving slower than the speed of sound, the formation of hot
droplets of the symmetric phase in the later stages of the collisions
can significantly reduce the gravitational-wave
signal~\cite{Cutting:2019zws}. The gravitational wave production in
extremely strong phase transitions has also been studied using a
combination of 1D simulations and modelling~\cite{Jinno:2019jhi}. It
has been shown that for almost all observable transitions, the
timescale on which nonlinearities in the fluid are expected to play a
role (given by the ratio of the bubble radius to the root-mean-square
fluid velocity) is shorter than a Hubble time~\cite{Ellis:2020awk}. On
longer timescales, the flow may become turbulent. The gravitational
wave spectrum from freely decaying turbulence has been
modelled~\cite{Kamionkowski:1993fg,Caprini:2007xq,Gogoberidze:2007an,Caprini:2009yp,Caprini:2009fx,Niksa:2018ofa},
and recently numerically simulated~\cite{Pol:2019yex}.

Full 3D lattice simulations have also been employed to test the
envelope approximation within vacuum transitions~\cite{Child:2012qg,
Cutting:2018tjt}. In Ref.~\cite{Child:2012qg}, it was seen that after
percolation, the gravitational-wave signal in a vacuum transition was
amplified by more than an order of magnitude during a what was termed
a period of coalescence. A more recent study by the present authors
identified this growth of gravitational waves with oscillations of the
scalar field around the true vacuum, producing gravitational waves
peaked at the broken-phase mass scale~\cite{Cutting:2018tjt}. With a
realistic separation of scales between the mass scale and the mean
bubble separation, the signal generated by these oscillations would be
negligible in comparison to that from bubble collisions, and peak at
too high a frequency to be observable. It also found that while the
peak frequency and amplitude of the spectrum were roughly predicted by
the envelope approximation, the UV power law was slightly steeper at
around $k^{-1.5}$.

Early studies of two colliding bubbles in the thin wall limit
demonstrated that the scalar field in the overlap region rebounds to
the false vacuum, and can become temporarily
trapped~\cite{Hawking:1982ga,Kosowsky:1991ua,Watkins:1991zt}. Further
simulations have shown that far away from the thin wall limit, the
trapping is reduced~\cite{Braden:2014cra}.

The question of trapping has recently been revisited in light of the
recent interest in the dynamics of cosmological first-order phase
transitions. In Ref.~\cite{Jinno:2019bxw} the collision of two
ultra-relativistic planar bubble walls was studied for a variety of
potential shapes. Depending on the shape of the potential, it was seen
that the scalar field could become trapped temporarily in the false
vacuum in the collision region. The authors proposed that if trapping
indeed occurred, then the gravitational-wave power spectrum should be
given by the envelope approximation, but if it did not then the bulk
flow model should apply. This has been investigated by colliding two
vacuum bubbles and measuring the resulting gravitational-wave power
spectrum~\cite{Lewicki:2019gmv}, where small changes were observed
when varying the potential shape.

In this paper, we conduct a series of 3D lattice simulations of
colliding vacuum bubbles, with the intention of exploring how
modifying the shape of the effective potential changes the
gravitational-wave spectrum. We consider a vacuum phase transition in
a toy model with a quartic effective potential with a cubic term. We
show that for this model, the effect of the potential on the scalar
field dynamics can be shown to depend on a single parameter $\lambar$.
As $\lambar\rightarrow 1$ the potential approaches thin-wall limit,
whereas for $\lambar\rightarrow 0$ the bubble wall becomes thick in
comparison to the critical radius. We see that as $\lambar\rightarrow
0$, less trapping occurs in the collision region. Our simulations span
a range of $\lambar$ and we analyse the effect changing $\lambar$ has
on the gravitational-wave power spectrum. The peak of the
gravitational-wave power spectrum varies by a factor of up to $1.3$
between the $\lambar$ values we consider. We find that UV power law in
the gravitational-wave spectrum becomes steeper as $\lambar
\rightarrow 0$, varying between $k^{-1.4}$ for $\lambar=0.84$ and
$k^{-2.2}$ for $\lambar =0.07$.

We also find some evidence that the power law in the IR continues to
evolve after the bubbles finish colliding. While at early times it is
proportional to $k^3$ (as one would expect from
causality~\cite{Caprini:2009fx}), it becomes shallower after the
bubbles have finished colliding. By visualising our simulations, we
can see that there are outward-propagating shells of
transverse-traceless shear-stress that continue to propagate outward.
This provides some evidence for the bulk flow model which assumes the
continued outward propagation of shear-stress after bubbles have
collided.

In Section~\ref{sec:scalardynamics} we discuss the dynamics of the
scalar field, and how it varies according to the potential shape in
our toy model. We illustrate how the variation of $\lambar$ modifies
the critical bubble profile, the evolution of an isolated bubble, and
the dynamics of the scalar field in the overlap region of two
colliding bubbles. We detail the linearised gravity approach we employ
within our simulations in Section~\ref{sec:intrograv}, and outline the
current understanding of the gravitational-wave power spectrum
produced in a vacuum transition. We describe the numerical methods we
employ in our simulations in Section~\ref{sec:Methods}. The results we
obtain are split into two sections, in Section~\ref{sec:ResultsScalar}
we show the behaviour of the scalar field and transverse traceless
shear-stress, whereas in Section.~\ref{sec:ResultsGrav} we analyse the
resulting gravitational-wave power spectrum and provide a fit for the
power spectrum over time for each $\lambar$. Finally, in
Section~\ref{sec:Conclusions} we list our conclusions.

\section{Scalar field dynamics}
\label{sec:scalardynamics}

In a vacuum first-order phase transition, the universe transitions
from a metastable false vacuum state into a true vacuum state. In a
first-order phase transition, a potential barrier will separate these
two states. Local patches of the universe will transition into the
true vacuum state via quantum tunneling. These patches of the true
vacuum state will form bubbles, with the interface between the true
and false vacuum forming the bubble wall. After nucleating, these
bubbles will expand, eventually reaching cosmological sizes and
ultra-relativistic speeds before they collide.

We can describe the transition by using a scalar field order parameter
$\phi$ which corresponds to the vacuum expectation value of the field
transitioning. The equation of motion for this scalar field in our
simulations is:
\begin{equation}
\Box \phi - V'(\phi)=0\text,
\end{equation}
where we choose the scalar field potential $V(\phi)$ to be given by
\begin{equation}
  V(\phi)= \frac{1}{2} M^2 \phi^2 +\frac{1}{3} \delta \phi^3 +\frac{1}{4}\lambda \phi^4\text.
\end{equation}
Note here that we are neglecting the expansion of the universe in the
dynamics of the scalar field. This is equivalent to making the
assumption that the duration of the phase transition is much shorter
than a Hubble time at the time of the transition $\Hc^{-1}$.

This potential has a degenerate second ground state when the mass $M$ is equal to the critical mass value,
\begin{equation}
  M^2_\mathrm{c} = \dfrac{2\delta^2}{9 \lambda}.
\end{equation}
It is useful to introduce the parameter $\lambar= M^2
/M^2_\mathrm{c}$.  When $\lambar<1$ this potential has two ground
states, one of which is metastable. The metastable state (or symmetric
phase) is at $\phi=0$ and the true vacuum state (or broken phase) is
at
\begin{equation}
  \phiAtMin=\frac{3 M_\mathrm{c}}{2\sqrt{2\lambda}}\left[ 1 + \sqrt{1-8\lambar /9} \right]\text,
\end{equation}
The symmetric phase at $\phi=0$ is separated
from the broken phase at $\phiAtMin$ by a potential barrier, which peaks
at
\begin{equation}
  \phiPeak= \frac{3 M_\mathrm{c}}{2\sqrt{2\lambda}}\left[ 1 - \sqrt{1-8\lambar /9} \right]\text.
\end{equation}

The broken phase mass is given by
\begin{equation}
\Mb=\frac{3 M_\mathrm{c}}{2} \sqrt{1- \frac{8 \lambar }{9} + \sqrt{1- \frac{8 \lambar }{9}}}\text.
\end{equation}
Furthermore the potential difference between the two minima is given by
\begin{equation}
\rVac = \frac{1}{12 \lambda} \left(\Mb^4- M^4\right)\text,
\end{equation}
and the height of the potential barrier $\VPeak = V(\phiPeak) - V(0)$,
\begin{equation}
\VPeak = \frac{M^6\left(M^2 + 2 \Mb^2\right)}{\left(M^2 - \Mb^2\right)\left(M^2 + \Mb^2\right)^3}\text.
\end{equation}

The total energy density in the scalar field $\rho_\phi$ can be split into three components,
\begin{equation}
\rho_\phi=\rho_\text{K}+\rho_{V}+\rho_\text{D}\text,
\end{equation}
with the kinetic energy density,
\begin{equation}
\rho_\text{K}=\frac{1}{2}\dot{\phi}^2\text,
\end{equation}
the gradient energy density,
\begin{equation}
\rho_\text{D}=\frac{1}{2} (\nabla \phi)^2\text,
\end{equation}
and the potential energy density,
\begin{equation}
\rho_{V}=V(\phi)-V(\phiAtMin)\text.
\end{equation}

We are free to rescale the potential by some constant value, and
likewise the field, i.e $V \rightarrow V'= cV$ or
$\phi \rightarrow \phi' = k\phi $. After accounting for this there are
only two interesting dimensionless quantities that describe our
potential which can affect the dynamics.

The first of these quantities is the ratio of $\VPeak$ and $\rVac$,
\begin{equation}
  \dfrac{\VPeak}{\rVac} = \frac{\left(\sqrt{9-8 \lambar}-3\right)^2 \left(4 \lambar+\sqrt{9-8 \lambar}-3\right)}{\left(\sqrt{9-8 \lambar}+3\right)^2 \left(-4 \lambar+\sqrt{9-8 \lambar}+3\right)}\text.
 \end{equation}

 The second is the ratio of $\Mb^2$ and $M^2$. This is given by
 \begin{align}
   \dfrac{\Mb^2}{M^2}= \frac {4 \lambar}{9-8 \lambar+3 \sqrt{9-8 \lambar}}\text.
 \end{align}
 Both of these ratios depend solely on $\lambar$ rather than any other
 combination of the potential parameters.

 Furthermore, we can reparameterize the scalar field as
 $\psi=\phi/\phiAtMin$ and rewrite the coordinates
 $x'^{\mu}=x^{\mu} M$ in order to obtain the following equation
 of motion,
  \begin{widetext}
 \begin{equation}
0=M^2 \phiAtMin \left( \Box' \psi - \left(\psi-\frac{3 \left(\sqrt{9-8 \lambar}+3\right)}{4 \lambar}\psi^2 + \frac{9 -4 \lambar+3 \sqrt{9-8 \lambar}}{4 \lambar} \psi^3\right)\right)\text,
\end{equation}
\end{widetext}
where
$\Box'=\dfrac{\partial}{\partial x'^{\mu}}\dfrac{\partial}{\partial
  x'_{\mu}}$. Clearly the dynamics here depend only on the value of
$\lambar$.

\begin{figure}
\centering
\includegraphics[width=0.48\textwidth]{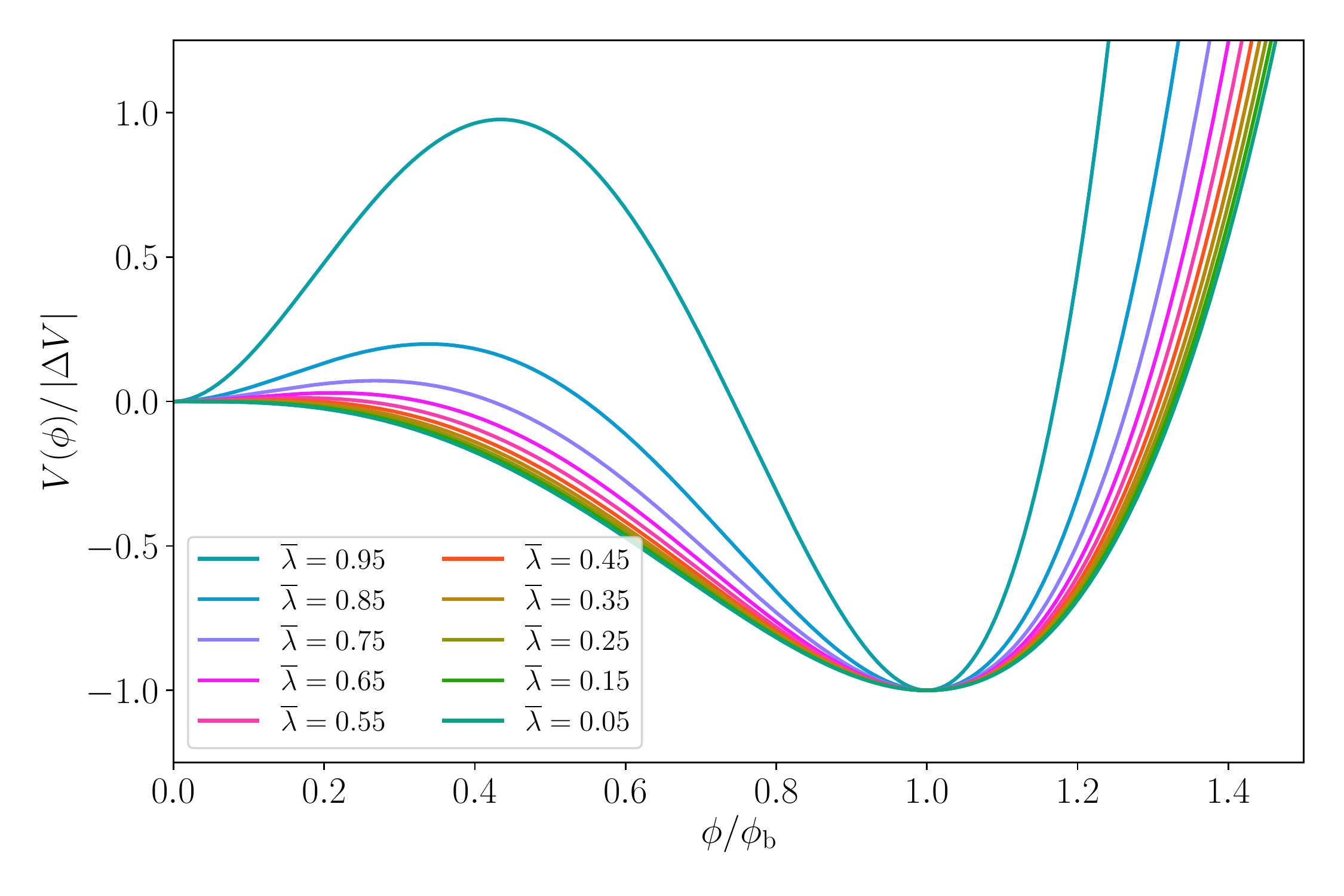}
\caption{ The effect on the potential due to the variation of $\lambar$.}
\label{fig:Potential}
\end{figure}

In the limit of $\lambar\rightarrow 1$ the minima of the potential
become degenerate, and this corresponds to the thin wall limit of our
potential. In the limit $\lambar\rightarrow 0$ the potential barrier
becomes infinitesimally small in comparison to the potential energy
difference. We call this the thick wall limit. To see how varying
$\lambar$ affects the scalar field potential, see Fig.~\ref{fig:Potential}.

By varying $\lambar$ between one and zero we are able to fully explore
the physically meaningful parameter space of our potential.

\subsection{Nucleation}

The probability of nucleating a bubble per unit volume per unit time $p(t)$ is given by \cite{Linde:1981zj}
\begin{equation}
p(t)=p_\text{n}\exp(-S_4)\text,
\end{equation}
where $S_4$ is the Euclidean action,
\begin{equation}
S_4(\phi) = \int d^4x \left[ \frac{1}{2}\left(\frac{d\phi}{dt}\right)^2 + \frac{1}{2}\left(\nabla \phi\right)^2 + V(\phi)\right]\text.
\end{equation}

In the previous work Ref.~\cite{Cutting:2018tjt}, several nucleation
scenarios were investigated. These were denoted exponential
nucleation, simultaneous nucleation, and constant nucleation.

An exponential nucleation rate can occur if there is a change in
temperature or background field. Then the Euclidean action
decreases slowly in time resulting in the following nucleation
probability
\begin{equation}\label{eq:exp_nuc}
p(t)=p_f\exp[\beta(t-t_f)]\text,
\end{equation}
where $\be = - \left. d \ln p(t) /d t \right|_{t_f} $ and $t_f$ is the
time at which the fraction of the universe in the symmetric phase is
$h(t_f)=1/e$~\cite{Enqvist:1991xw}.

For an exponential nucleation rate, the number density of bubble
nucleation sites at the end of the transition can be shown to be
\begin{equation}
\nb = \frac{1}{8\pi} \frac{\be^3}{\vw^3},
\end{equation}
where for a vacuum transition the wall velocity $\vw$ can be
approximated to unity.

Simultaneous nucleation can occur if  there is a minimum in $S_4(t)$ which is reached at
time $t_0$ before a transition completes. Then the probability of nucleating a bubble
per unit volume evolves as
\begin{equation} \label{eq:sim_nuc}
p(t)=p_0\exp[{- \textstyle\half}\beta^2_2 (t-t_0)^2 ]\text,
\end{equation}
where $\beta_2= \sqrt{S''(t_0)}$. Nucleation is then concentrated
around time $t_0$ \cite{Jinno:2017ixd}.  In the limit of
$\beta_2 \rightarrow \infty$, the number density of nucleation sites
tends towards
\begin{equation}
  \nb =  \sqrt{2\pi}\frac{p_0}{\be_2}.
\end{equation}

A constant nucleation rate can occur if $S_4(t)$ tends to a constant,
see e.g Ref.~\cite{GarciaGarcia:2016xgv}. The nucleation probability
in this scenario is then simply
\begin{equation} p(t)=p_\text{c}.
\end{equation}
with the nucleation site number density given by
\begin{equation}
  \nb = \frac{1}{4} \left(\frac{3}{\pi }\right)^{1/4} \Gamma\left(\frac{1}{4}\right) \left(\frac{p_\text{c}}{v_\text{w}}\right)^{3/4} .
\end{equation}

An important parameter for the gravitational wave power spectrum is
the mean separation between bubble centres at the end of the
transition, $\Rstar$. This is simply given by
\begin{equation}
  \Rstar = \frac{1}{\nb^{1/3}}\text.
\end{equation}

\subsection{Critical profile}
\label{ss:CriPro}

During a vacuum phase transition, the critical profile corresponds to
the most likely field configuration for a nucleated bubble. The
profile of a vacuum bubble is invariant under four-dimensional
Euclidean rotations \cite{Coleman:1977py}, i.e it obeys an
$\mathrm{O}(4)$ symmetry. We can therefore express the field profile
$\phi(\rho)$ as a function of a single variable
$\rho = \sqrt{\tau^2 + r^2} $ with $r$ the spatial radius from the
bubble centre and $\tau$ the Euclidean time.

In the thin wall limit the scalar field profile of the critical bubble is given by
\begin{equation}
  \phiCrit(r)=\frac{\phiAtMin}{2}\left[1-\text{tanh}\left(\frac{r-\Rc^\text{tw}
      }{\lw^\text{tw}}\right)\right]\text,
  \label{eq:tw_profile}
\end{equation}
where $\lw^\text{tw}$ is thickness of the critical bubble wall in the thin wall
limit,
\begin{equation}
  \lw^\text{tw}=\frac{2}{\Mb}\text,
\end{equation}
and $\Rc ^\mathrm{tw}$ is the radius of the critical bubble,
\begin{align}
  \Rc ^\mathrm{tw}&=\frac{3 \sigma^\text{tw}}{\rVac}\text, \\
  &= \frac{12}{M\left(\dfrac{\Mb^4}{M^4} - 1\right)}
\end{align}
Here
\begin{equation}
\siTW = \frac{M^3}{3\lambda}\text,
\end{equation}
is interpreted as the surface tension of the bubble. Note that both
the combination $\lw^{\text{tw}} M$ and $\Rc^{\mathrm{tw}} M$ are
constructed from quantities that depend only on $\lambar$.

Taking inspiration from the thin wall approximation, we can define the
``wall'' of the bubble to correspond to the section of the field
profile between $\rIn(t)$ and $\rOut(t)$ where
$\phi(t,\rIn) = \phi_0 (1 - \text{tanh}\left( - 1/2 \right)) /2$ and
$\phi(t,\rOut) = \phi_0 (1 - \text{tanh}\left( 1/2 \right)) /2$. Here
$\phi_0$ is the value of the scalar field at the centre of the
critical bubble, $\phi_0 = \phi(0)$. We then say that the radius of
the bubble $\rMid(t)$ is defined by $\phi(t,\rMid) = \phi_0 / 2$.

\begin{figure}
\centering
\includegraphics[width=0.485\textwidth,clip]{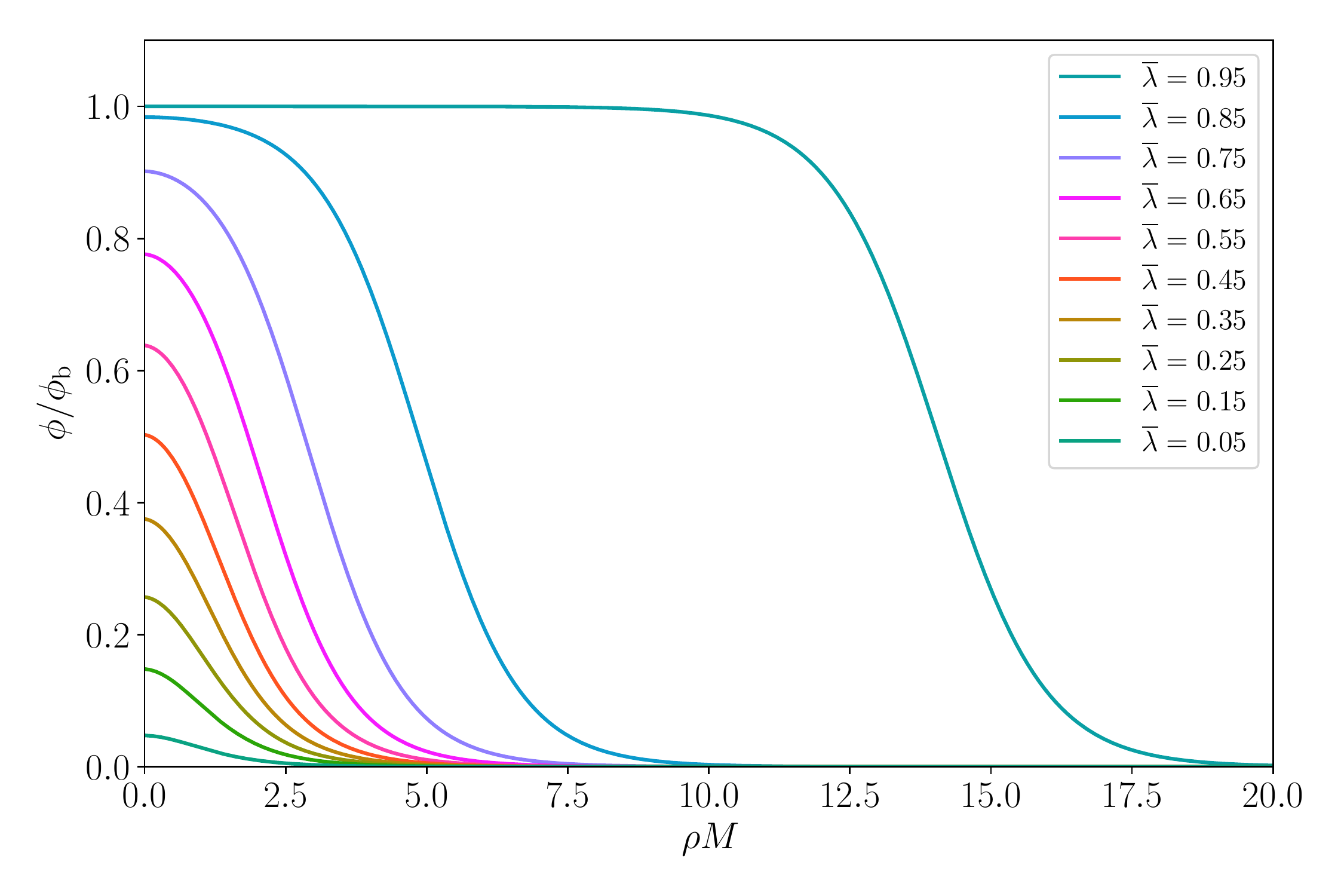}
\caption{The critical profile for a series of
  potentials with different values of $\lambar$. }
\label{fig:Profiles}
\end{figure}

For potentials with $\lambar$ close to 1, we find that the profile of
the critical bubble is close to a hyperbolic tangent, as expected from
Eq.~\ref{eq:tw_profile}. At the centre of the critical bubble the
field sits very close to $\phiAtMin$. As $\lambar$ is reduced, we see a
deviation of the critical bubble radius, $\Rc$ and initial wall width,
$\lw$, from those predicted in the thin wall limit. The lower the value
of $\lambar$, the smaller the critical radius of the bubble becomes in
comparison to the thickness of the wall. For small values of $\lambar$, the
field profile can be approximated by a Gaussian, and the value of $\phi_0$
decreases such that as $\lambar \rightarrow 0$, we find that
$\phi_0 / \phiAtMin \rightarrow 0$. We plot the critical bubble
profile for a series of $\lambar$ in Fig.~\ref{fig:Profiles}. Note
that with $\lambar$ fixed, the profile $\phi/\phiAtMin$ as a function
of $\rho M$ is invariant under changes of the potential parameters.

\subsection{Expansion}
\label{sec:scalarExp}

The energetically favourable state inside the bubble exerts
an outward pressure on the bubble wall. Bubbles with the critical
profile will begin to expand due to the pressure difference between
the false and true vacuum states.

\begin{figure*}
\subfigure[\,$\lambar = 0.84$]{\includegraphics[width=0.485\textwidth,clip]{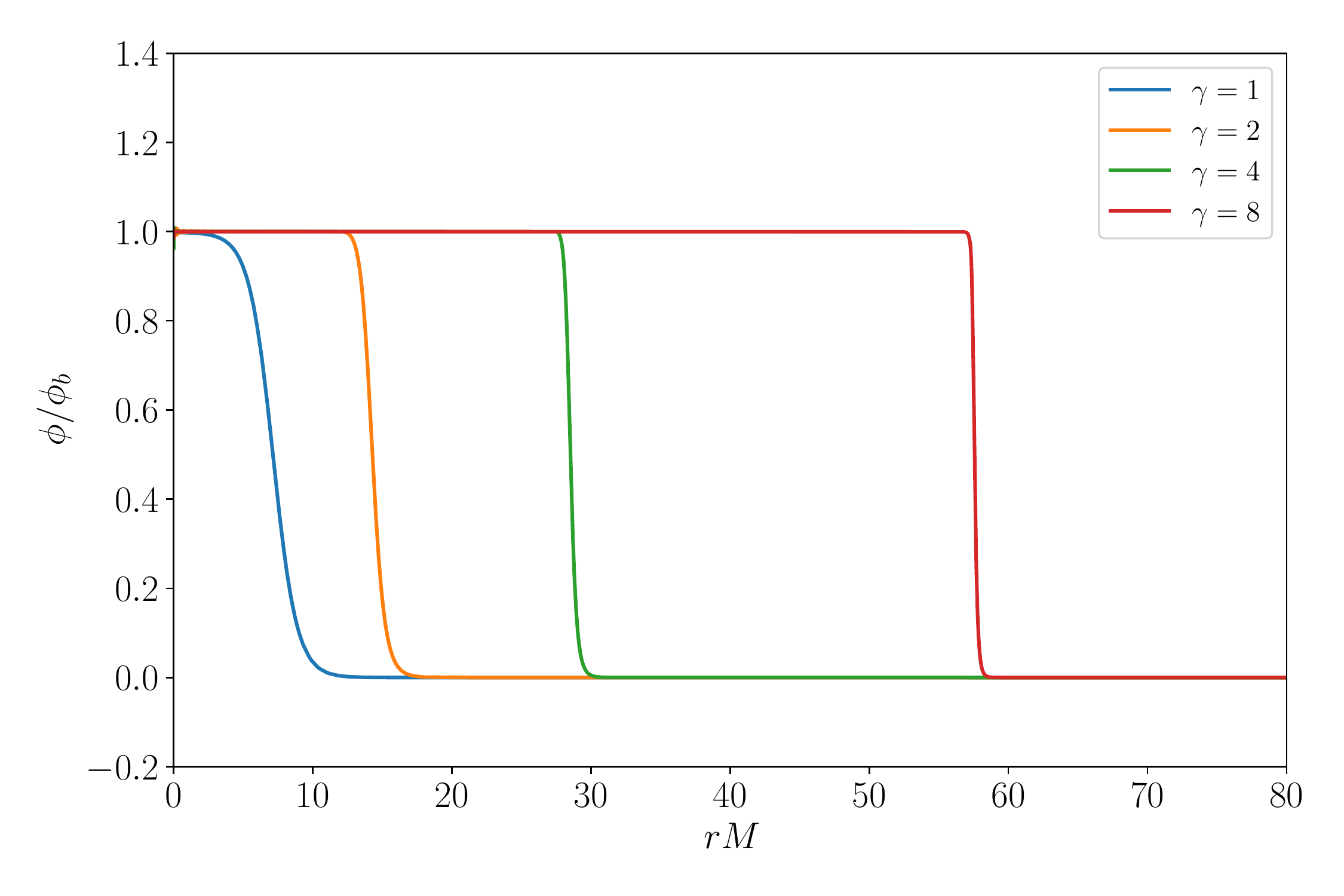}}
\hfill
\subfigure[\,$\lambar = 0.07$]{\includegraphics[width=0.485\textwidth,clip]{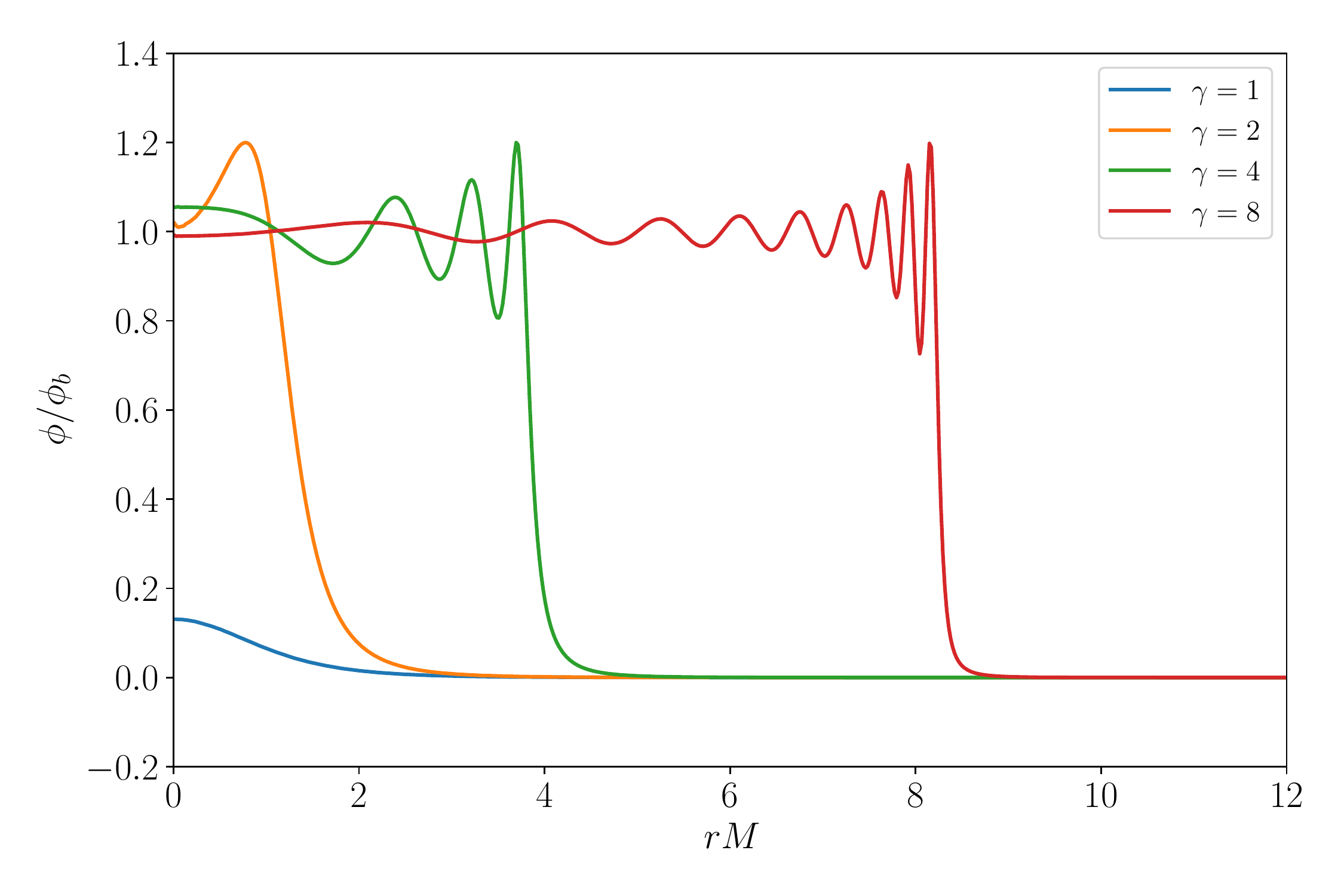}}
\caption{ Field profiles of bubbles when the bubble walls have
  accelerated up to various $\gamma$ factors. Note that $\gamma=1$
  corresponds to the critical bubble profile. }
\label{fig:Expansion}
\end{figure*}

As bubbles with high $\lambar$ expand, the field profile inside the
bubble remains close to $\phiAtMin$. In the frame in which the center of the
bubble is at rest, the bubble wall will become thinner due to Lorentz
contraction. Thick wall bubbles have substantially different dynamics. For these the field
at the centre of the bubble will move towards $\phiAtMin$ from its
initial value of $\phi_0$ as the bubble starts to expand. It will then proceed
to oscillate around $\phiAtMin$, resulting in outgoing waves of the scalar field
following the bubble wall. We depict this behaviour for a thin wall
bubble and a thick wall bubble in Fig.~\ref{fig:Expansion}.

\begin{figure}
  \includegraphics[width=0.485\textwidth,clip]{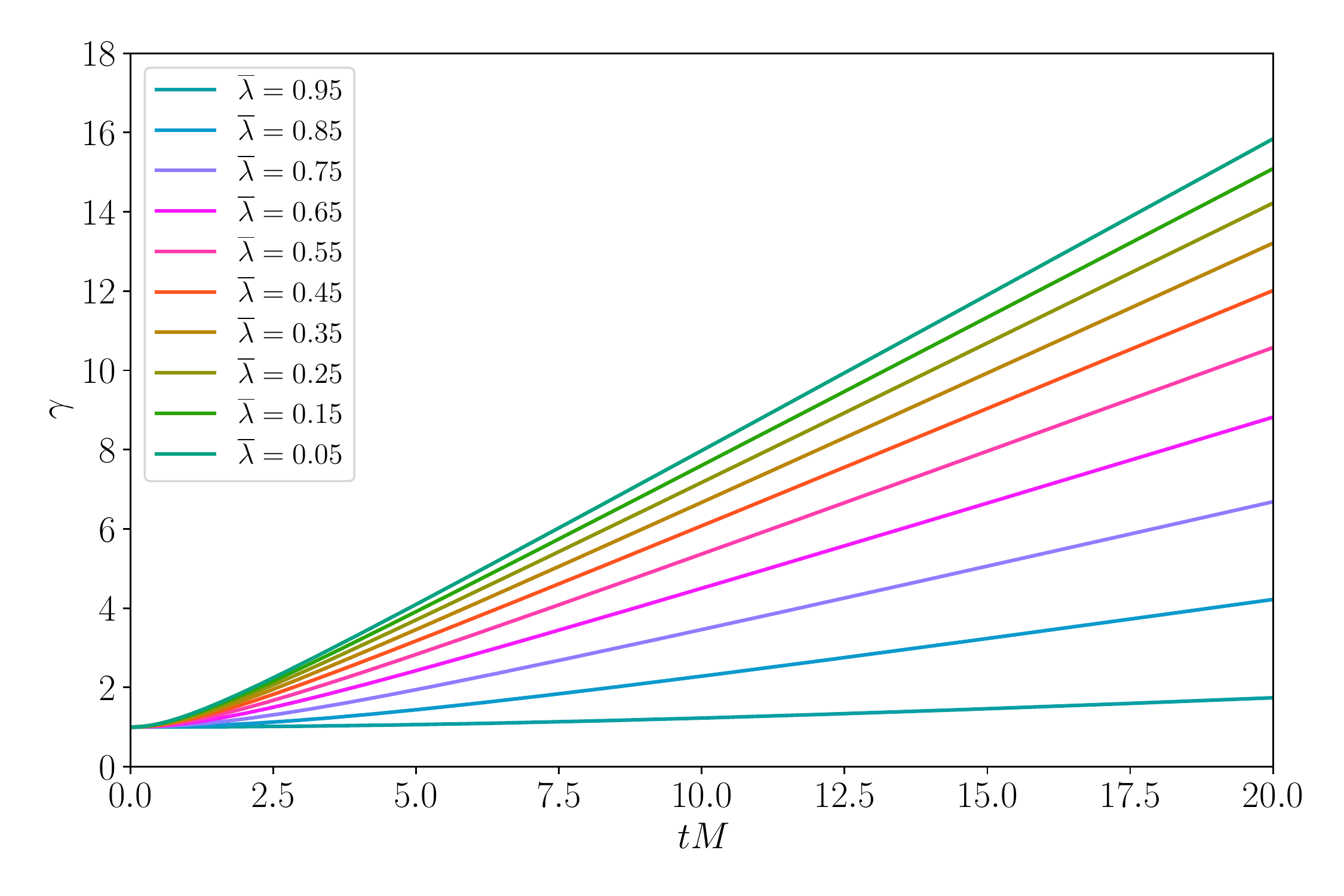}
\caption{ Evolution of $\gamma$ as defined in Eq.~\ref{eq:gammath} for
  a series of values of $\lambar$.}
\label{fig:gammath}
\end{figure}

For times $t > 0$ and for $r \geq t$, the profile of the bubble should
be given by $ \phi(t , r) = \phi(\sqrt{r^2 - t^2}) $. Therefore
$\rOut(t) = \sqrt{\rhoOut^2 + t^2}$ and $\rIn(t) = \sqrt{\rhoIn^2 +
  t^2}$. We define the Lorentz factor of the bubble wall by measuring
how much the wall contracts, $\gamma(t) = \lw(0)/\lw(t)$. This can be
expressed as
\begin{equation}
\gamma(t) = \dfrac{\rhoOut - \rhoIn}{ \sqrt{\rhoOut^2 +
    t^2} - \sqrt{\rhoIn^2 + t^2}}.
\label{eq:gammath}
\end{equation}
We show how $\gamma$ increases for a series of $\lambar$ at early
times in Fig.~\ref{fig:gammath}. It can be clearly seen that as
$\lambar \rightarrow 0$, where $\rhoIn$ and $\rhoOut$ take smaller
values, $\gamma$ grows more rapidly.

\subsection{Collision}
\label{sec:scalarCollision}

\begin{figure*}
\subfigure[\,$\lambar = 0.84$]{\includegraphics[width=0.485\textwidth,clip]{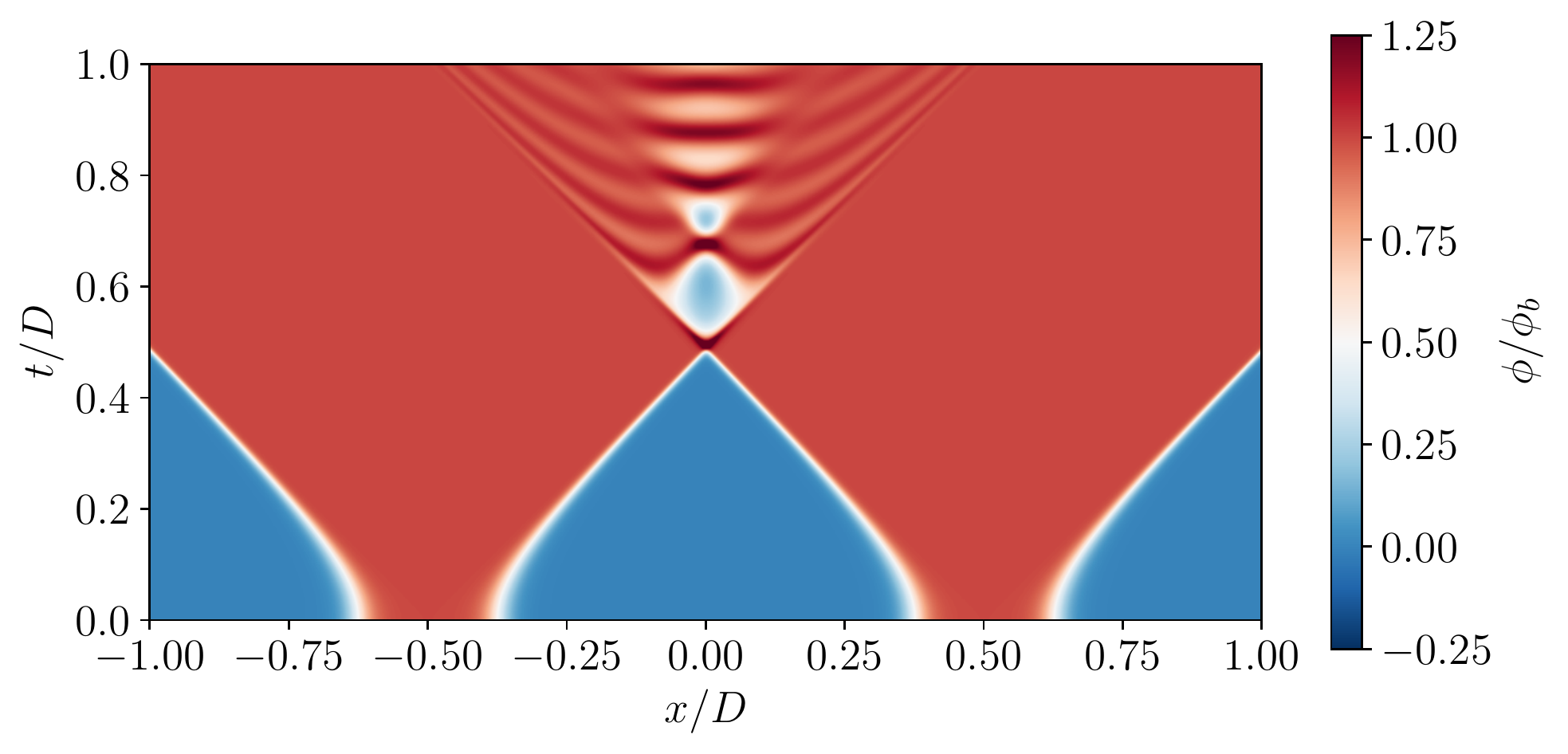}}
\hfill
\subfigure[\,$\lambar = 0.07$]{\includegraphics[width=0.485\textwidth,clip]{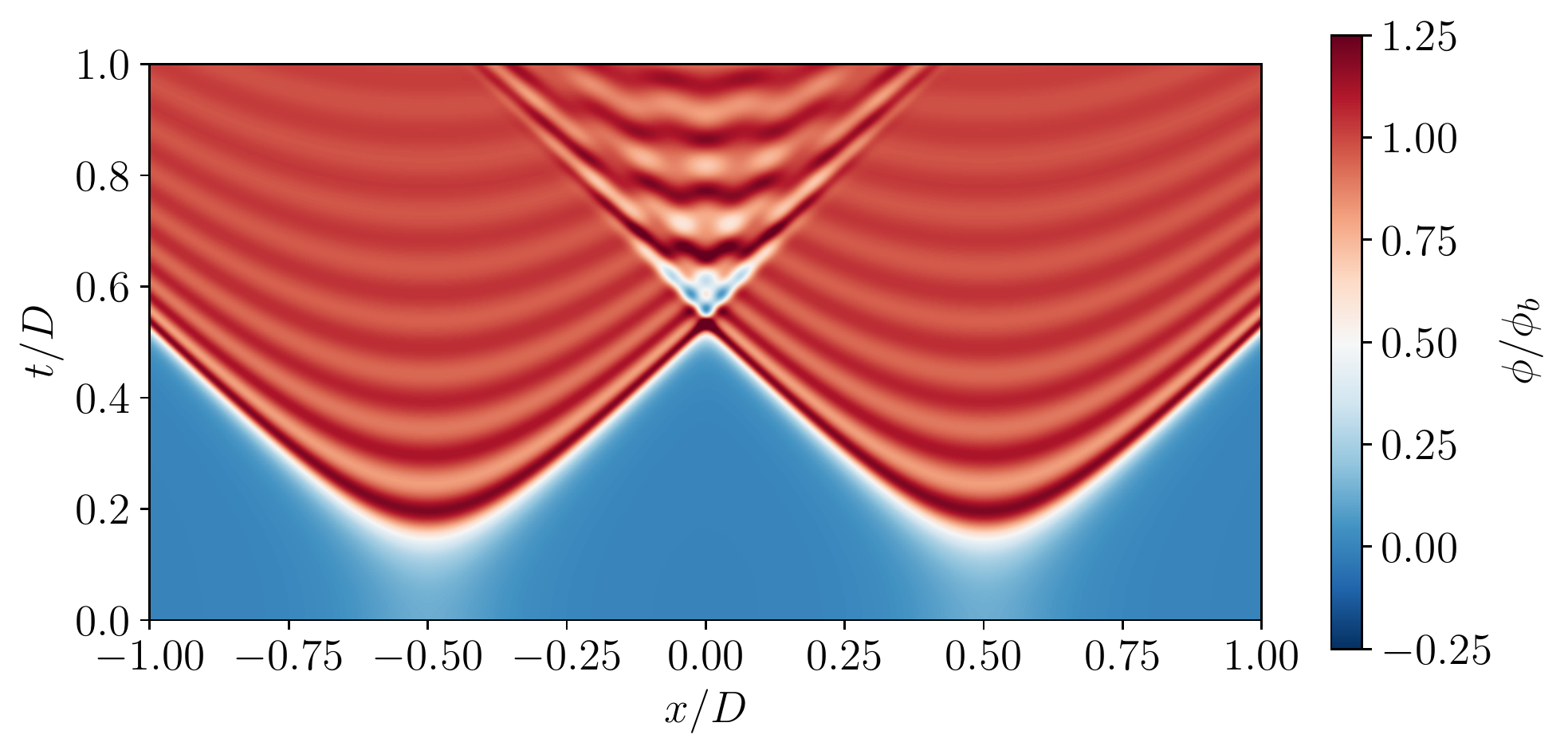}}
\caption{The collision of two bubbles of the true vacuum plotted for a
  thin wall (a) and thick wall (b) potential. The $x$ axis corresponds
to the line joining the two bubble centres, with $D$ being the
separation between bubbles. On the $y$ axis we plot the time $t$ since
the bubbles were nucleated. For both these simulations, the bubbles
collide when the Lorentz factors of the bubble walls are $\gamma = 4.0$. }
\label{fig:collisionAxis}
\end{figure*}

When two true vacuum bubbles collide, the scalar field begins to
oscillate in the region where the bubbles overlap. During this
oscillation the scalar field will rebound towards the false
vacuum~\cite{Hawking:1982ga,Braden:2014cra}. For thin wall potentials
with $\lambar$ closer to 1, the scalar field in the overlap region can
rebound over the potential barrier and return to the false
vacuum. This corresponds to the trapping discussed in
Ref.~\cite{Jinno:2019bxw}. On the other hand, for thick wall
potentials with smaller $\lambar$, the scalar field in the overlap
region will instead oscillate around the true vacuum state. According
to Ref.~\cite{Jinno:2019bxw}, this is where we would expect the bulk
flow model to apply. The value of $\lambar$ separating these
behaviours has been demonstrated to depend on $\gamma$, see Fig.~13
of Ref.~\cite{Jinno:2019bxw} for more details. We show both these
behaviours in Fig.~\ref{fig:collisionAxis}. In both cases, the
oscillations produce scalar field radiation that is emitted at close
to the speed of light. Neither of these effects is accounted for in
the envelope approximation which instead assumes that all shear-stress
disappears in the overlap region.

For all values of $\lambar$, the scalar field will continue to
oscillate around the true vacuum after the true vacuum bubbles have
finished colliding. It is known that the thermalisation of scalar
fields is a long-lasting process in the absence of other
interactions~\cite{Aarts:2000mg,Micha:2002ey,Arrizabalaga:2005tf}.

\section{Gravitational waves from a vacuum transition}
\label{sec:intrograv}

To calculate
the gravitational-wave power spectrum, we need to find the transverse traceless (TT) metric perturbations $h^{TT}_{ij}$ where
\begin{equation}
\Box h^{TT}_{ij} = -16\pi G T^{TT}_{ij}\text,
\end{equation}
and $T^{TT}_{ij}$ is the transverse traceless projection of the energy-momentum tensor,
\begin{equation}
T_{\mu\nu} = \partial_\mu\phi \partial_\nu\phi - \eta_{\mu\nu} \left(  \half (\partial\phi)^2 + V(\phi) \right),
\end{equation}
where $\eta_{\mu\nu}$ is the Minkowski metric.
The energy density in the gravitational waves can be defined as
\begin{equation}
\label{e:GWEnDen}
\rGW(\bx,t) = \frac{1}{64\pi G} \left(\dot
  h^{TT}_{ij}\dot h^{TT}_{ij} + (\nabla h^{TT}_{ij} )(\nabla h^{TT}_{ij}) \right) ,
\end{equation}
Note that an average over many wavelengths and periods may be needed
in order to reduce fluctuations in this quantity.
In general, while gravitational waves are being sourced $\langle (\dot
h^{TT}_{ij})^2 \rangle \neq \langle (\nabla h^{TT}_{ij})^2 \rangle$.

We introduce an auxiliary tensor $u_{ij}$ which satisfies
\cite{GarciaBellido:2007af}
\begin{equation}
\label{e:uEqu}
\Box u_{ij} = -16\pi G (\partial_i \phi)( \partial_j \phi)\text.
\end{equation}
To obtain $h^{TT}_{ij}$ we use the projector $\Lambda_{ij,lm}$ on $u_{ij}$ in momentum space,
\begin{equation}\label{eq:ProjMetric}
    h^{TT}_{ij}(\mathbf{k},t) = \Lambda_{ij,lm}(\mathbf{k})u_{lm}(\mathbf{k},t)\text,
\end{equation}
where
\begin{equation}
\Lambda_{ij,lm}(\mathbf{k})=P_{im}(\mathbf{k})P_{jl}(\mathbf{k})-\frac{1}{2}P_{ij}(\mathbf{k})P_{lm}(\mathbf{k})\text,
\end{equation}
and
\begin{equation}
 P_{ij}(\mathbf{k})=\delta_{ij}-\hat{k}_i\hat{k}_j\text.
\end{equation}
We then define the spectral density of the metric perturbations as
\begin{equation}
\langle h^{TT}_{ij}(\mathbf{k},t) h^{TT}_{ij}(\mathbf{k'},t) \rangle =P_{h}(\mathbf{k},t) (2\pi)^3 \delta(\mathbf{k}+\mathbf{k'})\text.
\end{equation}
Therefore, the power spectrum of gravitational wave energy density is
\begin{equation}
\frac{d\rGW}{d\mathrm{ln}(k)}=\frac{1}{64\pi G}
\frac{k^3}{2\pi^2}\left(P_{\dot{h}}(\mathbf{k},t) +  k^2 P_{h}(\mathbf{k},t) \right)\text,
\end{equation}
and by dividing through by the critical energy density $\rho_c$ we obtain
\begin{equation}
  \label{eq:gwps}
\frac{d\OmGW}{d\mathrm{ln}(k)}=\frac{1}{64\pi G \rho_c}
\frac{k^3}{2\pi^2}\left(P_{\dot{h}}(\mathbf{k},t) + k^2 P_{h}(\mathbf{k},t)\right)\text,
\end{equation}
which we refer to as the gravitational-wave power spectrum.

\subsection{Collision phase}
\label{sec:gwcoll}
Upon collision, the spherical symmetry of isolated bubbles is broken,
and gravitational waves become sourced by the shear-stress located at the bubble walls. During the collision phase, gravitational waves are generated at large wavelengths associated with the scale of the bubble sizes at collision time.

This period has been studied both using lattice field theory
simulations~\cite{Child:2012qg,Cutting:2018tjt} and using simplifying
assumptions such as the envelope
approximation~\cite{Kosowsky:1992vn,Huber:2008,Konstandin:2017sat} and
bulk flow model~\cite{Jinno:2017fby,Konstandin:2017sat}.

The envelope approximation~\cite{Kosowsky:1992vn} is based on two key assumptions. The
first is that the shear-stress in the scalar field is entirely located
in an infinitesimally thin shell located at the bubble wall. The
second approximation is that when bubbles collide, the shear-stress is
removed in the overlap region. Hence, to compute the
transverse traceless shear-stress sourcing the gravitational waves, it
is sufficient to consider the envelope from expanding bubbles. The
gravitational-wave spectrum has been calculated for exponential
nucleation rates using numerical simulations in
Refs.~\cite{Kosowsky:1992vn, Huber:2008,Konstandin:2017sat}.

The gravitational-wave power spectrum is well approximated by a
broken power law
\begin{equation}
  \label{eq:envfit}
\frac{d\OmGW^\text{env}}{d\mathrm{ln}(k)}= {\Om}^\text{env}_\text{p} \frac{(a+b)\tilde{k}^bk^a}{b\tilde{k}^{(a+b)}+ak^{(a+b)}}\text,
\end{equation}
with power law exponents $a$ and $b$, peak amplitude
${\Om}^\text{env}_\text{p}$ and peak wavenumber $\tilde{k}$.

For a vacuum phase transition where the wall velocity approaches the
speed of light, the power law exponents were found to
be $a= 2.9$ and $b= 0.9$~\cite{Konstandin:2017sat}. The peak amplitude
was given by
\begin{equation}
{\Om}^\text{env}_\text{p}\simeq 4.7 \times 10^{-2}\,\left(\frac{H_*}{\beta}\right)^2 \Om_\mathrm{vac}^2\text,
\end{equation}
where $\OmVac=\rVac/\rho_c$ is the vacuum energy density parameter.
The peak wavenumber was estimated to be
\begin{equation}
\tilde{k}/\beta\simeq 1.07\text.
\end{equation}

Note that the value of $\Rstar$ that is expected for a vacuum transition with exponential
nucleation rate is
\begin{equation}
\Rstar  = \frac{(8\pi)^{1/3}}{\beta}\text.
\end{equation}

Analytical investigations using the envelope
approximation~\cite{Jinno:2016vai}, have shown that the two-point
correlator of the energy-momentum tensor can be expressed as a
1-dimensional integral. This then results in the gravitational-wave
power spectrum being given by a broken power law with exponents $a=3$
and $b=1$.

In the bulk flow model, the envelope approximation is
modified~\cite{Jinno:2017fby,Konstandin:2017sat}. The shear-stress
during the transition is still considered to be located in an
infinitesimally thin shell located at the bubble wall. However, in the
bulk flow model, the shear-stress in the bubble wall is not assumed
to disappear upon collision. Instead, the
bubble wall continues to propagate but is no longer driven by
the latent heat of the transition. The bubble wall energy
density per surface element then decays as
$e^{-(t-t_\mathrm{coll})/\tau}/R^2$, where $R$ refers to the bubble
radius and $t_\mathrm{coll}$ the time of collision. The value
of $\tau$ indicates the typical damping timescale of the wall and
should be determined from the particle physics model. The value of
$\tau=0$ corresponds to the envelope approximation, whereas
$\tau=\infty$ corresponds to free propagation of the wall after
collision. Analytical treatments for ultra-relativistic bubbles have shown that as
$\tau\rightarrow\infty$, the IR power law in the gravitational-wave
power spectrum becomes shallower than $k^{3}$, tending towards
$k^{1}$~\cite{Jinno:2017fby}.

The gravitational-wave power spectrum in the bulk flow model with
$\tau=\infty$ and an exponential nucleation rate has also been studied with numerical
simulations~\cite{Konstandin:2017sat}. The resulting fit for
ultra-relativistic wall velocities was given in the same form as
Eq.~\ref{eq:envfit} with power law exponents $a=0.9$ and $b=2.1$. The
peak amplitude was given as
\begin{equation}
{\Om}^\text{bf}_\text{p}\simeq 6.4 \times 10^{-2}\,\left(\frac{H_*}{\beta}\right)^2 \Om_\mathrm{vac}^2\text,
\end{equation}
and peak wave number
\begin{equation}
\tilde{k}/\beta\simeq 0.809\text.
\end{equation}

In Ref.~\cite{Cutting:2018tjt}, full lattice field theory simulations
of colliding vacuum bubbles were conducted. The authors simulated the
gravitational-wave power spectrum produced by colliding thin-wall
bubbles, with $\lambar \geq 0.84$. The bubbles were separated on
average by a distance $\Rstar$, which is then the typical diameter of
bubbles when they collide. The value of $\Rstar$ in each simulation
was chosen so that the Lorentz factor of a bubble with diameter
$\Rstar$ was $\gmStar = 4$. A number of different nucleation scenarios
were investigated, which did not have a significant effect on the
resulting spectrum.

In the aforementioned work, a fit for the spectrum resulting from
bubble collisions was provided in the form
\begin{equation}\label{Eq:OldFit}
\frac{d\OmGW^\text{fit}}{d\text{ln}k}=\Omega^\text{fit}_\text{p}\frac{(a+b)^{c}\tilde{k}^b k^a}{(b \tilde{k}^{(a+b)/c}+ak^{(a+b)/c})^c}\text,
\end{equation}
where the value of $a$ was fixed to $a=3$. From the simulations
conducted it was found that
\begin{align}
\Omega^\text{fit}_\text{p}&= (3.22\pm0.04) \times 10^{-3}\, (\Hc \Rstar \OmVac)^2\text, \\
\tilde{k}\Rstar&=3.20\pm0.04\text, \\
b &= 1.51\pm0.04,  \quad c = 2.18\pm0.15\text,
\end{align}
with $\Hc$ the Hubble parameter at the time of the transition. This
corresponds to a slightly reduced total gravitational-wave power
compared to the envelope approximation, and furthermore a slightly
steeper UV power law. It was suggested that the deviation from the
envelope approximation was due to the behaviour of the scalar field in
the overlap regions. While the envelope approximation assumes that the
shear-stress in the bubble wall disappears upon collision, lattice
field theory simulations indicate that the scalar field oscillates in
the overlap region during bubble collisions.

The fits provided for the gravitational-wave power spectrum arising
from the bulk flow model and the envelope approximation are both
taken from simulations using an exponential nucleation rate. Caution
should be used when comparing them to the simulations in this paper
which correspond to simultaneous nucleation scenario. While the
gravitational-wave power spectrum from lattice simulations did not
show a strong dependence on the nucleation scenario in
Ref.~\cite{Cutting:2018tjt}, it has been shown that the envelope
approximation peak frequency can be shifted by up to a factor of
$\sim1.5$ and the peak amplitude by a factor of $\sim 3$ when changing
between exponential and simultaneous nucleation~\cite{Weir:2016tov}.
It has also been demonstrated that varying the nucleation rate in the
envelope approximation can affect the shape of the power spectrum
around the peak, with simultaneous nucleation creating a sharper peak
than exponential nucleation~\cite{Jinno:2017ixd}.

In this work we intend to extend the results in
Ref.~\cite{Cutting:2018tjt} to potentials with much smaller $\lambar$.
The behaviour of the scalar field in the overlap region during bubble
collisions varies depending on the value of $\lambar$, as described in
Section \ref{sec:scalarCollision}. If it is true that the deviation
from the envelope approximation corresponds to the structure in
overlap regions, the form of the power spectrum may depend on the
value of $\lambar$. We will pay particular attention to whether there
is a change in the total gravitational-wave power or the UV broken
power law exponent due to a variation in $\lambar$.

\subsection{Oscillation phase}

Once the bubbles have finished colliding in a vacuum first-order phase
transition, the scalar field is left in an excited state. In this
state, $\phi$ oscillates around the true vacuum value, $\phiAtMin$, and
as such we refer to this period as the oscillation phase.
Eventually, these oscillations are expected to subside as the scalar
field thermalises and Hubble friction damps away gradients in the
field. In previous lattice field theory simulations, it has been shown
that gravitational waves continue to be sourced during this
period~\cite{Child:2012qg, Cutting:2018tjt}.

In Ref.~\cite{Child:2012qg}, this phase was referred to as a
coalescence phase. It was posited that the gravitational
waves produced during this period would dominate over those produced
from bubble collisions, and will furthermore shift the peak of the
power spectrum towards the UV. However, in Ref.~\cite{Cutting:2018tjt}
it was shown that the peak frequency of the gravitational-wave power spectrum
generated during this phase was associated with the
microphysics of the system, namely $\lw$, rather than the cosmological
scales that correspond to $\Rstar$. When the separation
of these two scales was extrapolated from the simulations up to realistic values, the
gravitational-wave power of the collision phase is expected to
dominate. Furthermore, the peak frequency corresponding to the
oscillation phase would be firmly out of range of any upcoming
gravitational-wave detectors for any realistic early universe phase transition.

In this study we aim to resolve whether the result found in
Ref.~\cite{Cutting:2018tjt} extends to a wider range of $\lambar$, or
whether the gravitational-wave power or peak frequency changes for thick
wall bubbles.

\section{Methods}\label{sec:Methods}

To conduct our simulations in this paper, we employ an updated
version of the code used in Ref.~\cite{Cutting:2018tjt}. This is a 3D
classical lattice field theory code built using the \texttt{LATfield2}
library in \texttt{C++}~\cite{David:2015eya}.

For each simulation, the fields are evolved on a lattice of $L^3$
points using a Crank-Nicholson leapfrog algorithm. We impose periodic
boundary conditions, which corresponds to the approximation that the
universe is isotropic and homogeneous at the scale of the simulation
box. We use a 7 point stencil for the Laplacian operator. We pick an
appropriate lattice spacing $\Delta x$ and fix the timestep
$\Delta t = \Delta x / 5$. The final simulation time is
$t_\mathrm{fin}$.

To understand how the gravitational-wave power spectrum
changes between thin and thick wall bubbles, we perform simulations
with four different values of $\lambar$. These, along with various
corresponding parameters derived from the potential, are given in
Table~\ref{table:potentials}.

When choosing a lattice spacing, we perform a series of
convergence tests for each set of simulations, which we detail in
App.~\ref{app:convergence}. For the largest simulation for each choice
of $\lambar$ we perform a low, medium and high resolution run with a
factor of two smaller lattice spacing for each increase in
resolution. The lattice spacing for the high resolution run is then
used in the rest of the paper. For the gravitational wave power spectrum
we take the uncertainty for each bin to be given by the difference of
the power found in the high and medium resolution run. For the
number of bubbles used in the largest simulations, the uncertainty in
each bin arising from performing multiple realisations is very small
for all but the most infrared modes, and can be neglected compared to
the lattice uncertainty. For more information on the convergence rate,
see App.~\ref{app:convergence}.

To compare our choice of potential parameters with the quartic
potential in Ref.~\cite{Jinno:2019bxw}, we use the conversion between
$\lambar$ and $\epsilon$ given below, \begin{equation}
  \label{eq:epsilonconversion}
 \epsilon = \dfrac{\left(\sqrt{9-8\lambar} -3\right)^2\left(4\lambar +
     \sqrt{9-8\lambar} -3\right)}{8(9-8\lambar)^{3/2}}\text.
\end{equation}
We list the corresponding values of $\epsilon$ for each $\lambar$ in
Table~\ref{table:potentials}. In Fig.~13 of Ref.~\cite{Jinno:2019bxw}
it can be seen that trapping is exhibited for $\epsilon\gtrsim 0.6$ when
$\gamma\simeq 4$. From this we infer that $\lambar=0.84$ and
$\lambar=0.50$ exhibit trapping behaviour for $\gamma = 4.0$, whereas
$\lambar =0.18$ and $\lambar=0.07$ do not.

In Ref.~\cite{Cutting:2018tjt} a range of different nucleation rates
were used. The nucleation rate did not appear to have a detectable
effect on the gravitational-wave power spectrum. To
limit the computational cost, we choose to study only simultaneous
nucleation, where we nucleate all bubbles at the start of the
simulation on the zeroth timestep. Bubbles are nucleated randomly in
the symmetric phase, providing that for all $n<N$, the distance between the $N$th and $n$th bubble centres $r_n$ obeys the following relation
\begin{equation}
r_n^\text{sep}>\Rc+\sqrt{\Rc^2+(t-t_n)^2}\text,
\end{equation}
where $t_{n}$ is the time since nucleation of the $n$th bubble. For
simultaneous nucleation $t_{n}=0$ for all $n$. We nucleate a total of
$\Nb$ bubbles in each simulation.

Bubbles are nucleated into the simulation with the corresponding
critical profile. The critical profile is found by using a shooting
algorithm to determine the bounce solution for a given potential.

The average separation between bubbles is
$\Rstar=\left(\mathcal{V}/\Nb\right)^{1/3}$, where
$\mathcal{V} = \left(L \Delta x\right)^3$ is the volume of the
simulation. When bubbles collide, they will on average have a diameter
of $\Rstar$, and so this quantity sets the length scale of the peak of
the gravitational-wave power spectrum. In our simulations, we choose
$\Rstar$ such that the value of the Lorentz factor for a bubble of
diameter $\Rstar$ is $\gmStar = 4$.

Once $\gmStar$ and $\lambar$ are fixed, the combination $\Rstar \Mb$
is also determined uniquely. This is important as $\Rstar \Mb$
effectively dictates the separation between the length scales of the
physics from bubble collisions and the microscopic physics from
oscillations about the true vacuum. In a true vacuum phase transition,
these scales would be separated by many orders of magnitude as
$\gmStar \rightarrow \infty$, but achieving this separation of
scales numerically is not possible.

To supplement our 3D simulations, we also perform a series of
spherically symmetric 1D simulations of isolated bubbles. This enables
us to study the effect of the lattice on the evolution of $\rIn$,
$\rOut$, $\rMid$ and $\gamma$. This analysis is provided in
Appendix~\ref{app:convergence}. We evaluate $\gmStar$ for an isolated
bubble in both the 1D code and 3D code and list these values in
Tables~\ref{table:simGW} and \ref{table:late-sim}.

We also perform a series of simulations in order to understand the
gravitational waves sourced by the oscillation phase of the scalar
field. To do this we perform long-lasting simulations where the
evolution of the metric perturbations is only turned on after the
phase transition has completed, around $t/\Rstar = 2.0$. We list the
simulations that we conduct to understand the gravitational waves
sourced by the collision phase in Table~\ref{table:simGW}, and for the
oscillation phase in Table~\ref{table:late-sim}.

The simulations studying the collision phase all finish at
$t/\Rstar = 8.0$, with the exception of the largest simulation with
$\lambar = 0.84$ and $\Nb=512$ which terminates at $t/\Rstar = 7.0$
due to time limits imposed by the computing facilities utilised. The
smaller but longer lasting simulations studying the oscillation phase
terminate at $t/\Rstar =40$.

\begin{table}
  \centering
  \resizebox{0.48 \textwidth}{!}{
  \begin{tabular}{D{.}{.}{1.2} D{.}{.}{1.3} D{.}{.}{1.2}  D{.}{.}{1.3} D{.}{.}{1.2} D{.}{.}{1.3} D{.}{.}{1.3}
    D{.}{.}{1.7} D{.}{.}{1.2} D{.}{.}{1.6}}
    \hline \hline
    \multicolumn{1}{c}{$\lambar$}\Tstrut & \multicolumn{1}{c}{$\Rc M$}
    & \multicolumn{1}{c}{$\lw M$} & \multicolumn{1}{c}{$\Rc^\mathrm{tw} M$}
    & \multicolumn{1}{c}{$\lw^\mathrm{tw} M$}  & \multicolumn{1}{c}{$\phi_0/\phiAtMin$} & \multicolumn{1}{c}{$\phiPeak/\phiAtMin$}
    & \multicolumn{1}{c}{$\VPeak/ \rVac$} &
                                            \multicolumn{1}{c}{$\Mb^2/M^2$} & \multicolumn{1}{c}{$\epsilon$}\\
    \hline
    0.84         & 7.15   & 1.71 & 4.04  & 1.42 & 0.981  & 0.334 &
                                                                   1.87 \times 10^{-1} & 1.99 & 1.6\times 10^{-1}\\
    0.50         & 2.07   & 1.24 & 0.36  & 0.83 & 0.570  & 0.146 &
                                                                   8.18
                                                                   \times
                                                                   10^{-3}
                                          & 5.84  & 8.1 \times 10^{-3}
    \\
    0.18         & 1.16   & 0.89 & 0.026 & 0.43 & 0.183  & 0.045 &
                                                                   1.90
                                                                   \times
                                                                   10^{-4}
                                          & 21.46 & 1.9\times 10^{-4} \\
    0.07 \Bstrut & 0.996  & 0.80 & 0.0031& 0.25 & 0.066  & 0.016 &
                                                                   8.21
                                                                   \times
                                                                   10^{-6}
                                          & 61.96 & 8.2 \times 10^{-6}\\
    \hline \hline
  \end{tabular}
}
\caption{The values of $\lambar$ used in our simulations. For each of
  these we give the critical radii, $\Rc$, and wall thicknesses,
  $\lw$, that are used in our simulations, as well as their estimates
  in the thin wall approximation. We also supply the value of the
  scalar field at the centre of the bubble, $\phi_0$, and the value of
  scalar field at the peak of the potential barrier, $\phiPeak$, both
  in terms of the broken phase value, $\phiAtMin$. We also give the
  ratio of the height of the potential barrier, $\VPeak$, compared to
  the potential energy difference, $\rVac$, and the mass of the field
  in the broken phase, $\Mb$, compared to the symmetric phase mass,
  $M$. Finally we give the corresponding value of $\epsilon$ for
  comparison with the quartic potential in Ref.~\cite{Jinno:2019bxw}.}
\label{table:potentials}
\end{table}

\begin{table}
  \centering
  \resizebox{0.48 \textwidth}{!}{
  \begin{tabular}{D{.}{.}{1.2} D{.}{.}{4.0} D{.}{.}{2.2}  D{.}{.}{1.1} D{.}{.}{4.0} D{.}{.}{1.4} D{.}{.}{1.3}
    D{.}{.}{1.3} D{.}{.}{1.3} }
    \hline \hline
    \multicolumn{1}{c}{$\lambar$}\Tstrut & \multicolumn{1}{c}{$\Nb$}
    & \multicolumn{1}{c}{$\Rstar \Mb$} & \multicolumn{1}{c}{$t_\text{fin}/\Rstar$} & \multicolumn{1}{c}{$L$}
    & \multicolumn{1}{c}{$\Delta xM$}& \multicolumn{1}{c}{$\gmStar$} & \multicolumn{1}{c}{$\gmStarOneD$}
    &  \multicolumn{1}{c}{$\gmStarThreeD$}\\
    \hline
    0.84         & 8    & 80.66 & 8.0 & 1200 & 0.0952 & 4.000  & 3.958 & 3.984  \\
    0.84         & 64   & 80.66 & 8.0 & 2400 & 0.0952 & 4.000  & 3.958 & 3.984  \\
    0.84         & 512  & 80.66 & 7.0 & 4800 & 0.0952 & 4.000  & 3.958 & 3.984  \\
    \hline
    0.50         & 8    & 40.53 & 8.0 & 800  & 0.0419 & 4.000  & 3.972 &3.988  \\
    0.50         & 64   & 40.53 & 8.0 & 1600 & 0.0419 & 4.000  & 3.972 &3.988 \\
    0.50         & 512  & 40.53 & 8.0 & 3200 & 0.0419 & 4.000  &  3.972 &3.988  \\
    \hline
    0.18         & 8    & 44.69 & 8.0 & 400  & 0.0482 & 4.000  & 3.927 & 3.966  \\
    0.18         & 64   & 44.69 & 8.0 & 800  & 0.0482 & 4.000  & 3.927 & 3.966 \\
    0.18         & 512  & 44.69 & 8.0 & 1600 & 0.0482 & 4.000  & 3.927 &3.966 \\
    0.18         & 4096 & 44.69 & 8.0 & 3200 & 0.0482 & 4.000  & 3.927 & 3.966  \\
    \hline
    0.07         & 8    & 65.54 & 8.0 & 800  & 0.0482 & 4.000  & 4.021 & 4.004 \\
    0.07         & 64   & 65.54 & 8.0 & 1600  & 0.0482 & 4.000  & 4.021 & 4.004  \\
    0.07         & 512  & 65.54 & 8.0 & 3200 & 0.0482 & 4.000  & 4.021 & 4.004 \Bstrut \\
    \hline \hline
  \end{tabular}
}
\caption{Parameters of the simultaneous nucleation simulations used
  within this paper. Listed here for each run are the values of
  $\lambar$, number of bubbles $\Nb$, average bubble separation
  $\Rstar$, final simulation time $t_\text{fin}$, number of
  lattice points $L^3$, lattice spacing $\Delta x$, typical Lorentz factor at
  collision $\gmStar$, the effective
  $\gmStar$ as found on the lattice in a 1D simulation $\gmStarOneD$
  and in a 3D simulation $\gmStarThreeD$. For details of the potential
  parameters for each $\lambar$, see Table~\ref{table:potentials}. Not given here are simulation
  runs where the metric perturbations are turned on after the bubbles
  have finished colliding, see Table~\ref{table:late-sim}. }
\label{table:simGW}
\end{table}

\begin{table}
  \centering
  \resizebox{0.48 \textwidth}{!}{
  \begin{tabular}{D{.}{.}{1.2} D{.}{.}{4.0} D{.}{.}{2.2}  D{.}{.}{1.1} D{.}{.}{4.0} D{.}{.}{1.4} D{.}{.}{1.3}
    D{.}{.}{1.3} D{.}{.}{1.3} }
    \hline \hline
    \multicolumn{1}{c}{$\lambar$}\Tstrut & \multicolumn{1}{c}{$\Nb$}
    & \multicolumn{1}{c}{$\Rstar \Mb$} & \multicolumn{1}{c}{$t_\text{fin}/\Rstar$} & \multicolumn{1}{c}{$L$}
    & \multicolumn{1}{c}{$\Delta xM$}& \multicolumn{1}{c}{$\gmStar$} & \multicolumn{1}{c}{$\gmStarOneD$}
    &  \multicolumn{1}{c}{$\gmStarThreeD$}\\
    \hline
    0.84         & 8    & 80.66 & 40.0 & 1200 & 0.0952 & 4.000  & 3.958 & 3.984  \\
    0.50         & 8    & 40.53 & 40.0 & 800  & 0.0419 & 4.000  & 3.972 &3.988  \\
    0.18         & 8    & 44.69 & 40.0 & 400  & 0.0482 & 4.000  & 3.927 & 3.966  \\
    0.07         & 8    & 65.54 & 40.0 & 800  & 0.0482 & 4.000  & 4.021 & 4.004 \\
    \hline \hline
  \end{tabular}
}
\caption{Parameters of the simultaneous nucleation runs where we turn
  the evolution of
  metric perturbations on well after the bubbles have finished
  colliding at $t/\Rstar = 2.5$. This allows us to
  study the gravitational-wave signal produced from the oscillation
  phase.}
\label{table:late-sim}
\end{table}

\section{Results: scalar field}\label{sec:ResultsScalar}

During a vacuum first-order phase transition, the scalar field
undergoes several phases of evolution. First occurs the nucleation and
expansion of bubbles. Next, the bubbles begin to collide and the field
oscillates in the overlap regions. Finally, the bubbles finish
colliding, and the scalar field oscillates around $\phiAtMin$ as the
field thermalises.

\begin{figure}
\includegraphics[width=0.485\textwidth,clip]{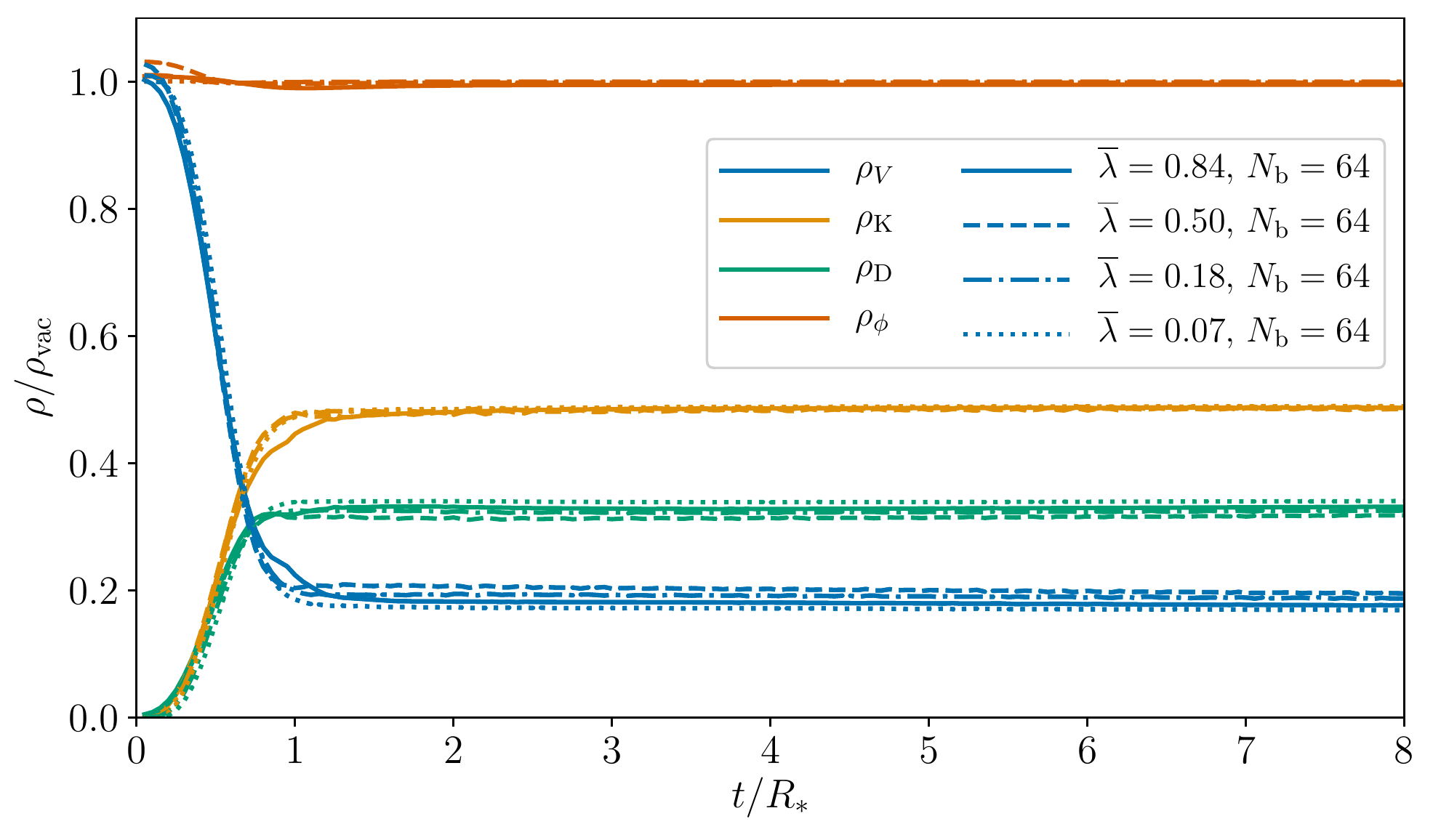}
\caption{ The evolution of mean energy densities corresponding to the scalar field for simulations with varying $\lambar$.}
\label{fig:scalar_globals}
\end{figure}

It is useful to investigate the evolution of the total, kinetic,
gradient and potential energy densities of the scalar field. We show
this for several simulations with a range of $\lambar$ and $\Nb=64$ in
Fig.~\ref{fig:scalar_globals}. There appears to be little variation in
the mean energy densities for different $\lambar$, nor any consistent
trend as it changes. By tracking the evolution of $\rho_{V}$ we can
see that in all cases the bubbles finish colliding shortly after
$t/\Rstar = 1$. Around this time the kinetic, gradient and potential
energy densities settle to constant values. As $\rho_{V}$ does not
tend to zero at the end of the simulation, we know that the scalar
field continues in the oscillation phase after the bubbles finish
colliding. The scalar field does not thermalise during the duration of
our simulations. We can also see that the total energy density in the
scalar field $\rho_{\phi}$ is well conserved, with minimal energy
being lost to the lattice.

Further insight into the behaviour of the scalar field can be deduced
from examining slices through the simulation box. In
Appendix~\ref{app:slices} we show slices of the scalar field for two
simulations with $\lambar=0.07$ and $\lambar=0.84$ and $\Nb=64$. We
plot these slices at three different times,
$t/\Rstar \in \{0.5,\,1.0,\,4.0\} $. These correspond to early on in the bubble
collision phase, towards the end of this phase where most bubbles have
finished colliding, and much later during the oscillation phase. These
simulation slices confirm the behaviour outlined in
Section~\ref{sec:scalarCollision}. When $\lambar$ is small, the
expanding scalar field profile oscillates around $\phiAtMin$ and the
rebound in the overlap region towards the symmetric phase is minimal.
For larger $\lambar$, the rebound can be quite dramatic.

\begin{figure*}
\subfigure[\,$\lambar = 0.84$, $\Nb = 512$]{\includegraphics[width=0.485\textwidth,clip]{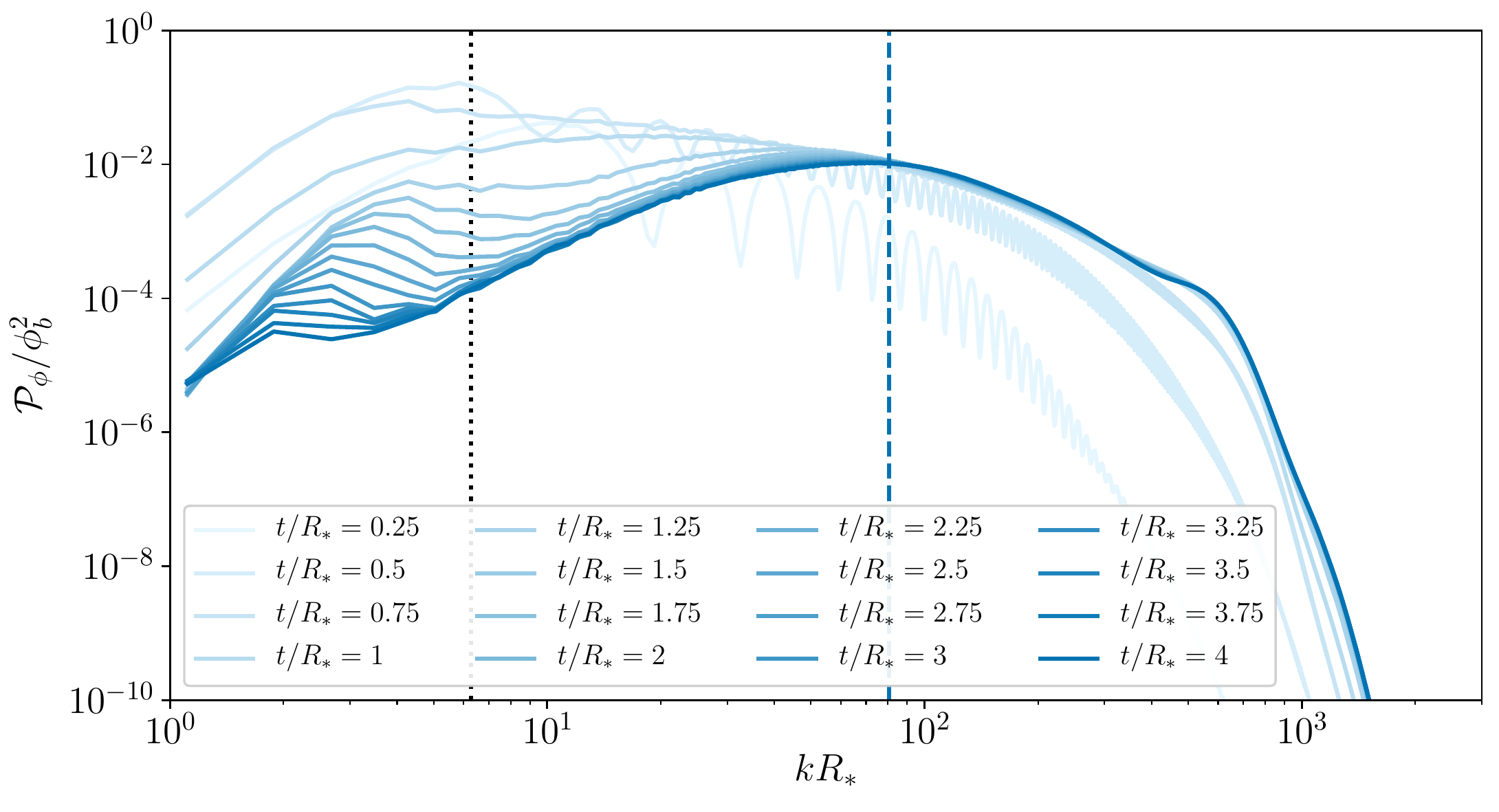}}
\hfill
\subfigure[\,$\lambar = 0.07$,  $\Nb=512$]{\includegraphics[width=0.485\textwidth,clip]{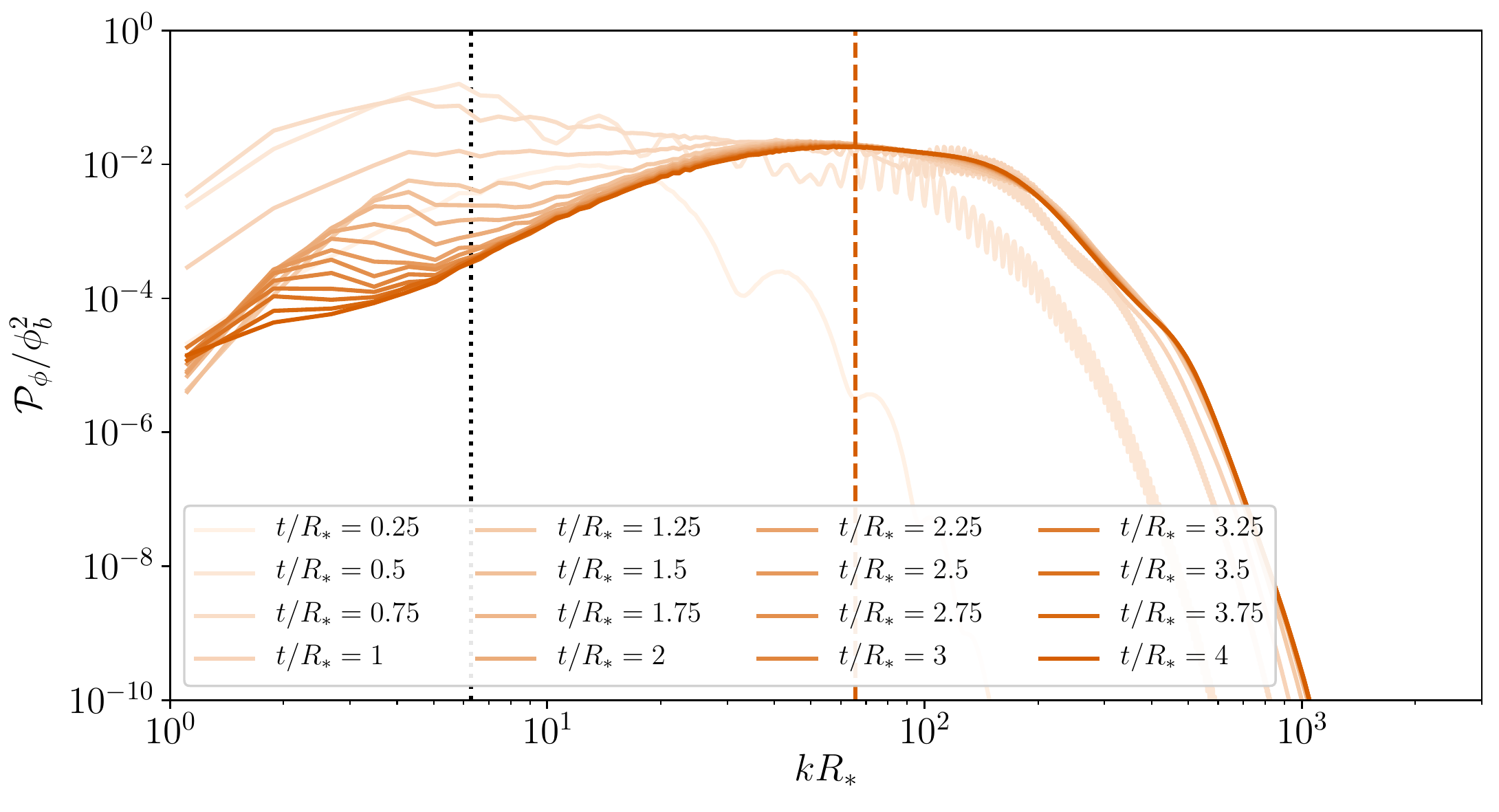}}
\caption{ The power spectrum of the scalar field $\mathcal{P}_{\phi}$. In each plot darker shades indicate later times. The vertical black dotted line shows the location of $k=2\pi/R_*$, whereas the vertical dashed coloured line shows the location of $k=\Mb$. }
\label{fig:cSpec_waterfall}
\end{figure*}

In Fig.~\ref{fig:cSpec_waterfall} we show the power spectrum of the
scalar field, $\mathcal{P}_{\phi}$ for two simulations with
$\lambar =0.84$ and $\lambar=0.07$. We see that at early times while
the bubble is expanding, the power spectrum is peaked around the scale
of $\Rstar$.  At later times, as the scalar field begins to oscillate,
the peak wavenumber for the power spectrum increases, moving further
towards the length scale associated with $\Mb$. It is interesting to
note that the decay of power in the IR is not as rapid as one might
initially expect. Although the bubbles have finished colliding around
$t= \Rstar$, the power in the scalar field in the IR decays slower,
reaching a minimum only after several $t/\Rstar$.

When trying to understand how the gradients in the scalar field source
gravitational waves, it is useful to follow the evolution of
$T^{TT}_{ij}$. The transverse traceless shear-stress tells us about
the instantaneous source of gravitational waves at any given point in
the simulation. By examining slices of the modulus of the transverse
traceless shear-stress, $\sqrt{T^{TT}_{ij} T^{TT}_{ij}}$, we are able
to determine also the location where gravitational waves are being
sourced. We show this alongside the scalar field slices in
Appendix~\ref{app:slices}.

\begin{figure*}
\subfigure[\,$\lambar = 0.84$, $\Nb = 512$]{\includegraphics[width=0.485\textwidth,clip]{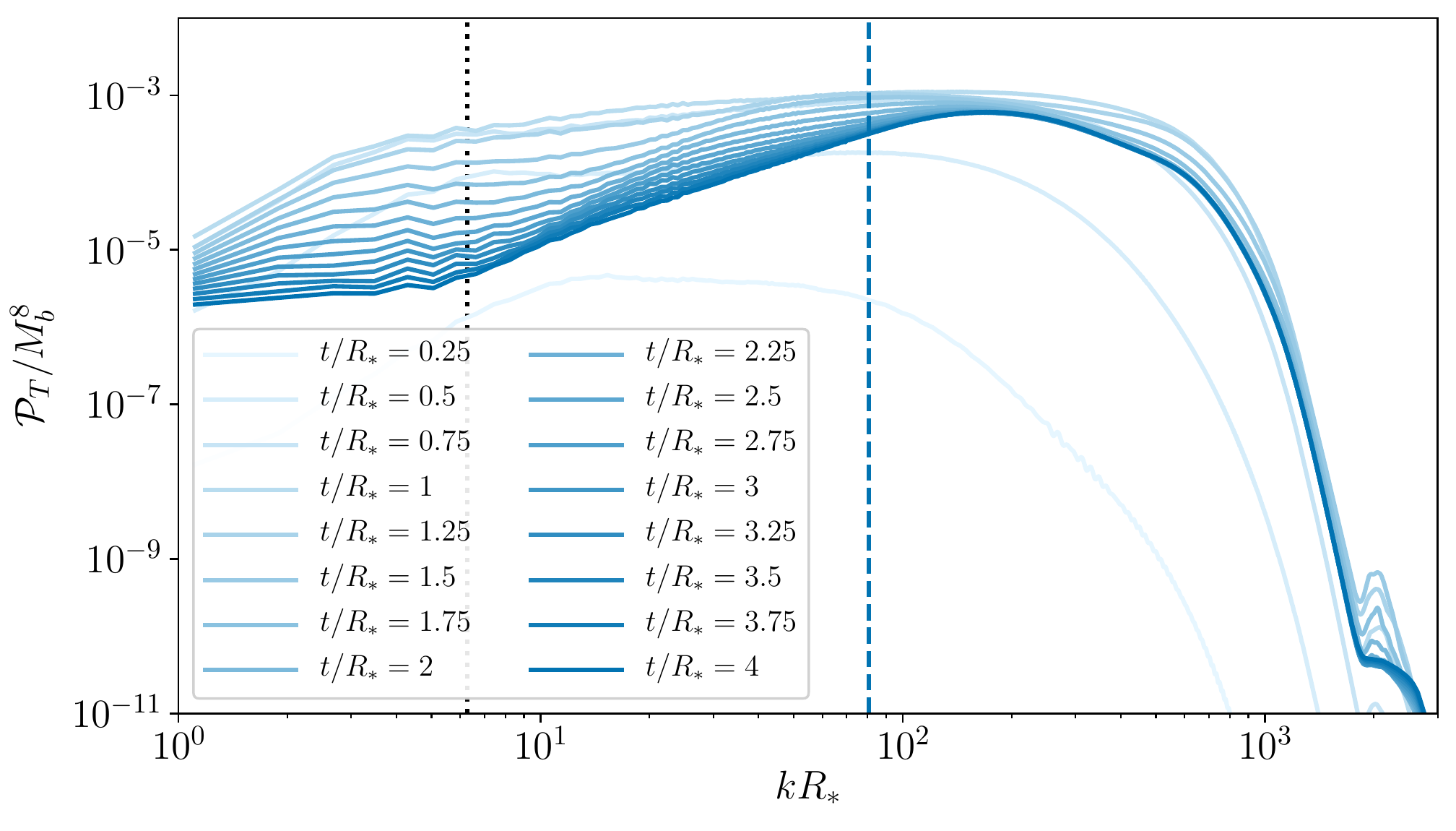}}
\hfill
\subfigure[\,$\lambar = 0.07$,  $\Nb=512$]{\includegraphics[width=0.485\textwidth,clip]{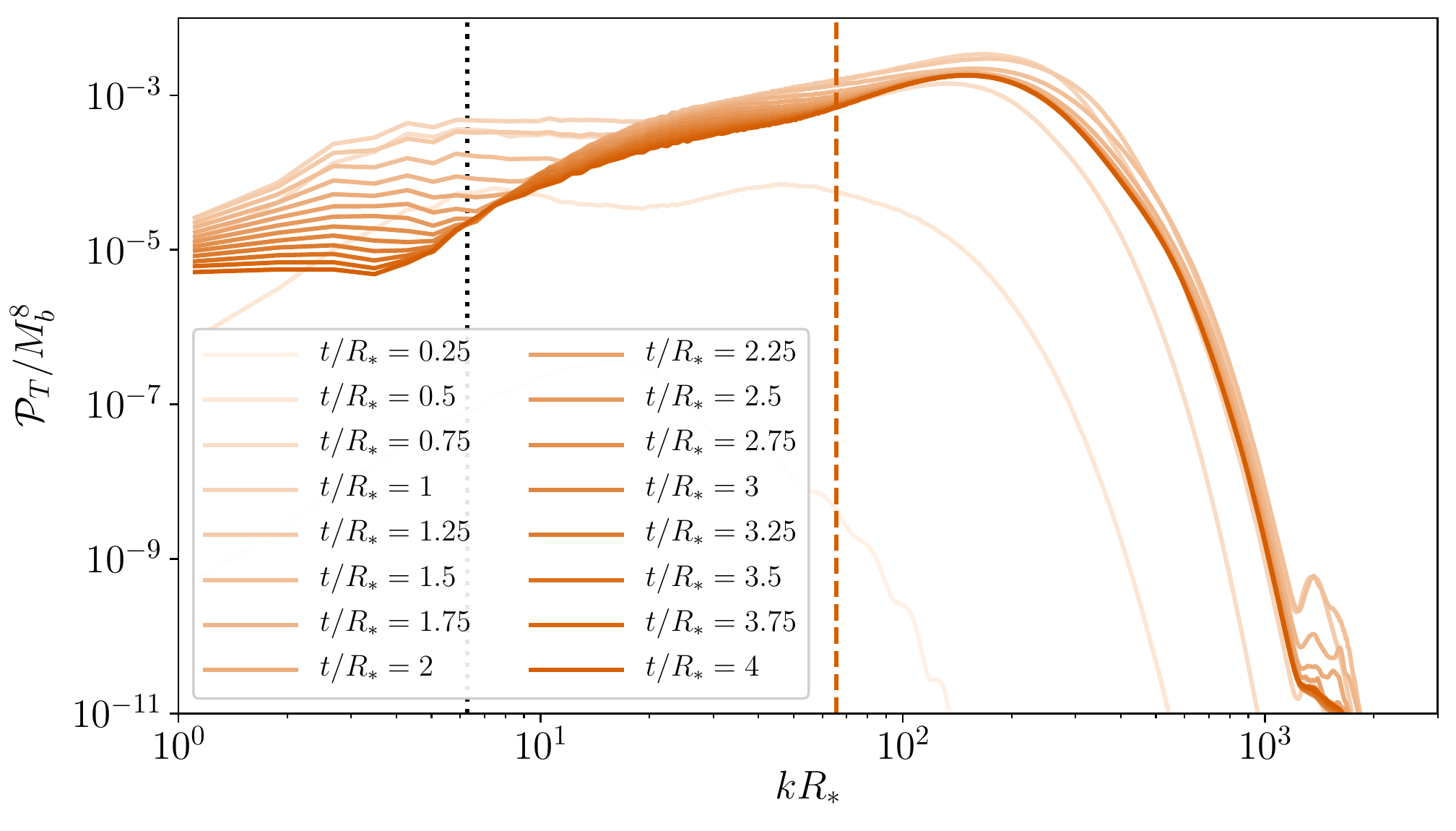}}
\caption{ The power spectrum of the transverse traceless shear-stress
  $T_{ij}^{TT}$. In each plot darker shades indicate later times. The
  vertical black dotted line shows the location of $k=2\pi/R_*$,
  whereas the vertical dashed coloured line shows the location of
  $k=\Mb$. }
\label{fig:emTT_waterfall}
\end{figure*}

From the slices, we can see that, contrary to the prediction of the
envelope approximation, there is substantial shear-stress in the
overlap region of collided bubbles. This appears to be particularly
true for potentials with smaller $\lambar$. Furthermore, even after
the final bubbles have finished colliding, waves of shear-stress
previously associated with the bubble collisions propagate outward
with length scales of $\Rstar$ or larger. This shows some similarity
to that which is predicted in the bulk flow model. At later times the shear
stress appears to have power on much smaller length scales.

It is useful to study the power spectrum of the transverse traceless
shear-stress, $\mathcal{P}_T$. We
plot the evolution of $\mathcal{P}_T $ for two simulations with
$\lambar=0.07$ and $\lambar=0.84$ in Fig.~\ref{fig:emTT_waterfall}.
From $\mathcal{P}_T $, we can see that as the transition
progresses, the shear-stress starts to grow as bubbles start to
collide. Initially, there is substantial power in the IR, corresponding
to typical length scales of the bubbles when they collide. At later
times the power shifts more towards the UV, with a peak developing
close to the scale associated with $\Mb$. This occurs as the scalar
field has entered the oscillation phase of the transition.

Interestingly, we see that the power in the IR does not disappear
immediately after the bubbles finish colliding, around $t/\Rstar = 1$.
Instead the power slowly decreases for several $t/\Rstar$. This
appears to agree with what we saw in the slices of
$\sqrt{T^{TT}_{ij} T^{TT}_{ij}}$ in which the shear-stress associated
with the bubble wall and collision regions continued to propagate for
some time after the bubbles finished colliding, giving further support
to the bulk flow model. Unfortunately, we cannot resolve a sufficient
distance into the IR to see any fall off of the shear-stress
corresponding to the causal interval.

\begin{figure*}
\subfigure[\,$\lambar = 0.84$, $\Nb = 8$]{\includegraphics[width=0.485\textwidth,clip]{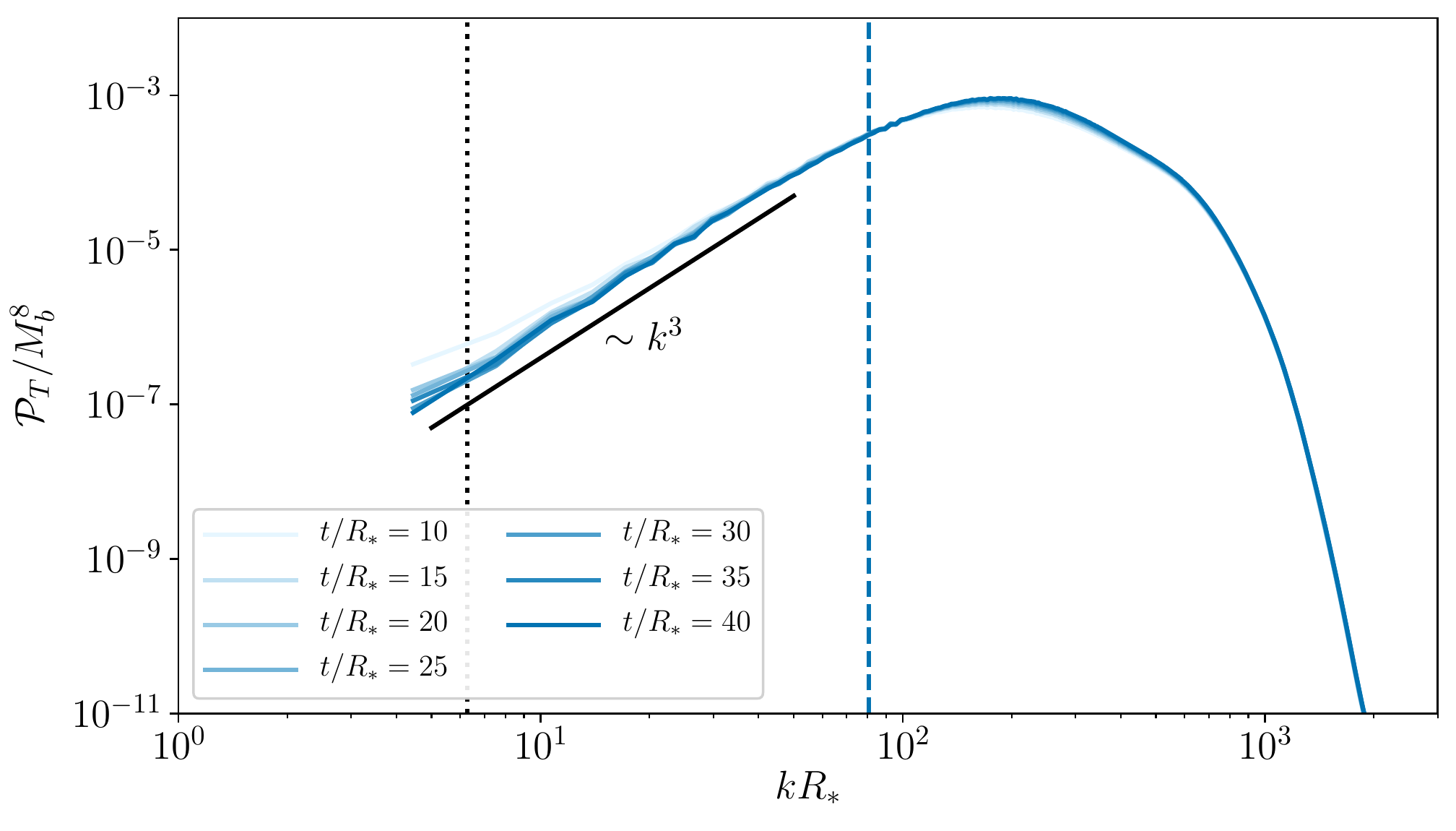}}
\hfill
\subfigure[\,$\lambar = 0.07$,  $\Nb=8$]{\includegraphics[width=0.485\textwidth,clip]{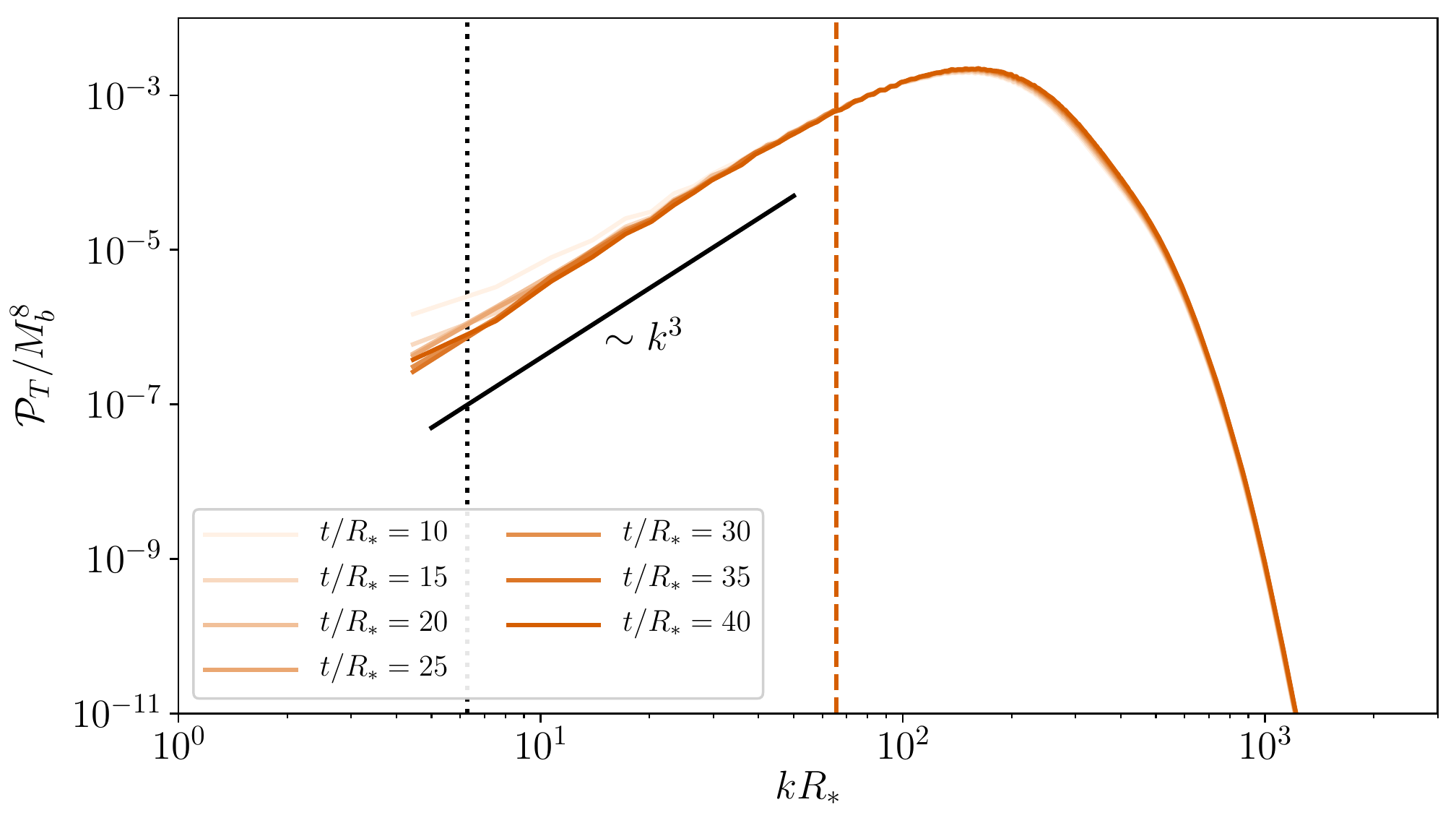}}
\caption{ The power spectrum of the transverse traceless shear-stress
  $T_{ij}^{TT}$ at very late times. In each plot darker shades
  indicate later times. The vertical black dotted line
  shows the location of $k=2\pi/R_*$, whereas the vertical dashed
  coloured line shows the location of $k=\Mb$. The solid black
  line shows a power law of $k^3$.}
\label{fig:lateshst}
\end{figure*}

From our smaller, but much longer simulations outlined in
Table~\ref{table:late-sim} we can show how $\mathcal{P}_T$ behaves at
very late times. We plot the evolution of $\mathcal{P}_T$ up to
$t/\Rstar =40$ for two simulations with $\Nb=8$ and for $\lambar=0.84$
and $\lambar=0.07$ in Fig.~\ref{fig:lateshst}. It can clearly be seen
that at very late times, the power spectrum settles into a shape with
a characteristic power law of $k^3$ rising from the IR. In our
simulations, we do not allow for the decay of the scalar field into
other particles, and we also do not account for the damping of
oscillations of the scalar field due to expansion. Both of these would
reduce the power in $T^{TT}_{ij}$. It is still interesting to note
however that the non-linear behaviour in the scalar field continues to
source gravitational waves long after the collisions phase terminates.

\section{Results: gravitational waves}\label{sec:ResultsGrav}

\begin{figure*}[htbp]
\subfigure[\,$t/\Rstar = 0.60$]{\includegraphics[width=0.485\textwidth,clip]{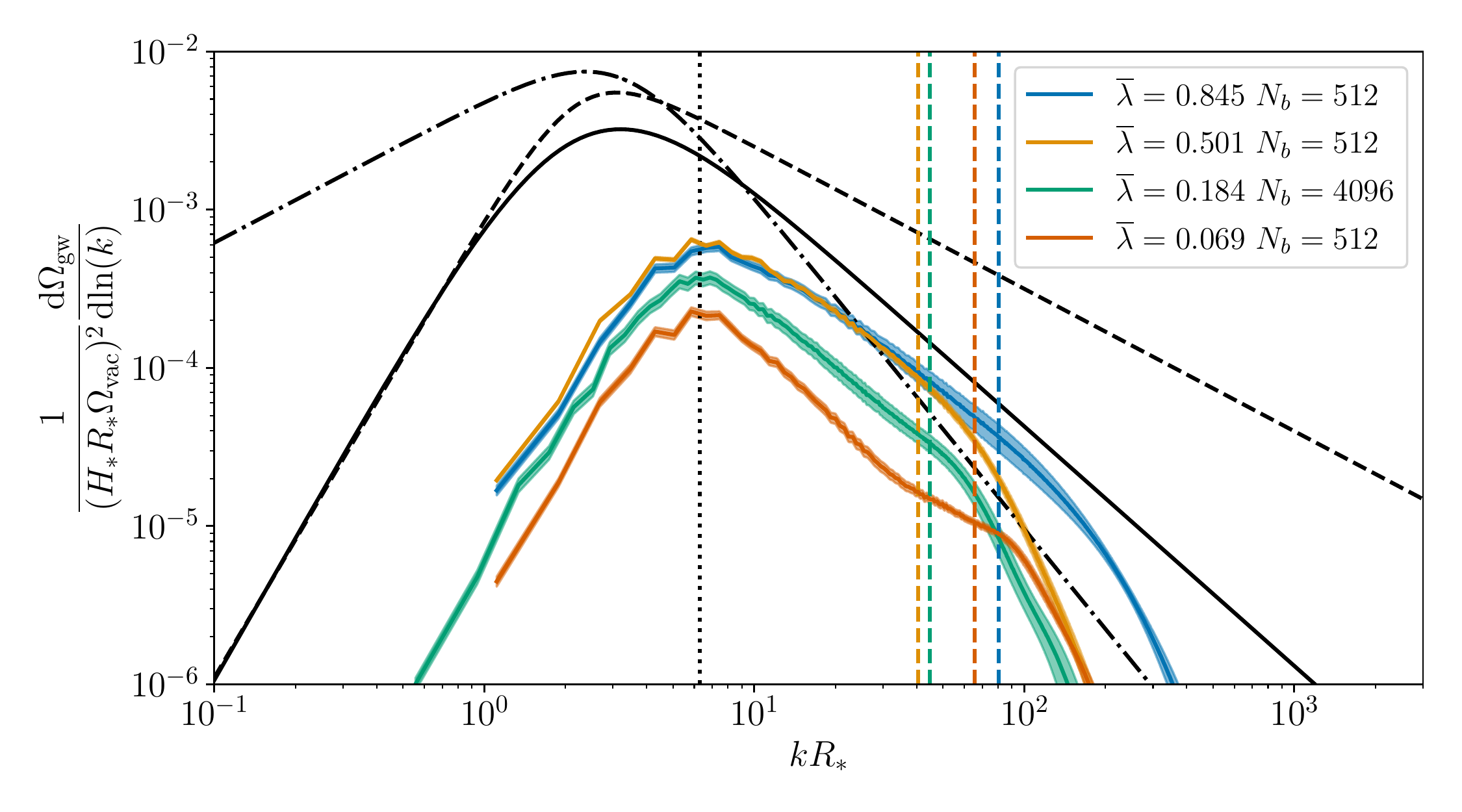}}
\hfill
\subfigure[\,$t/\Rstar = 1.6$]{\includegraphics[width=0.485\textwidth,clip]{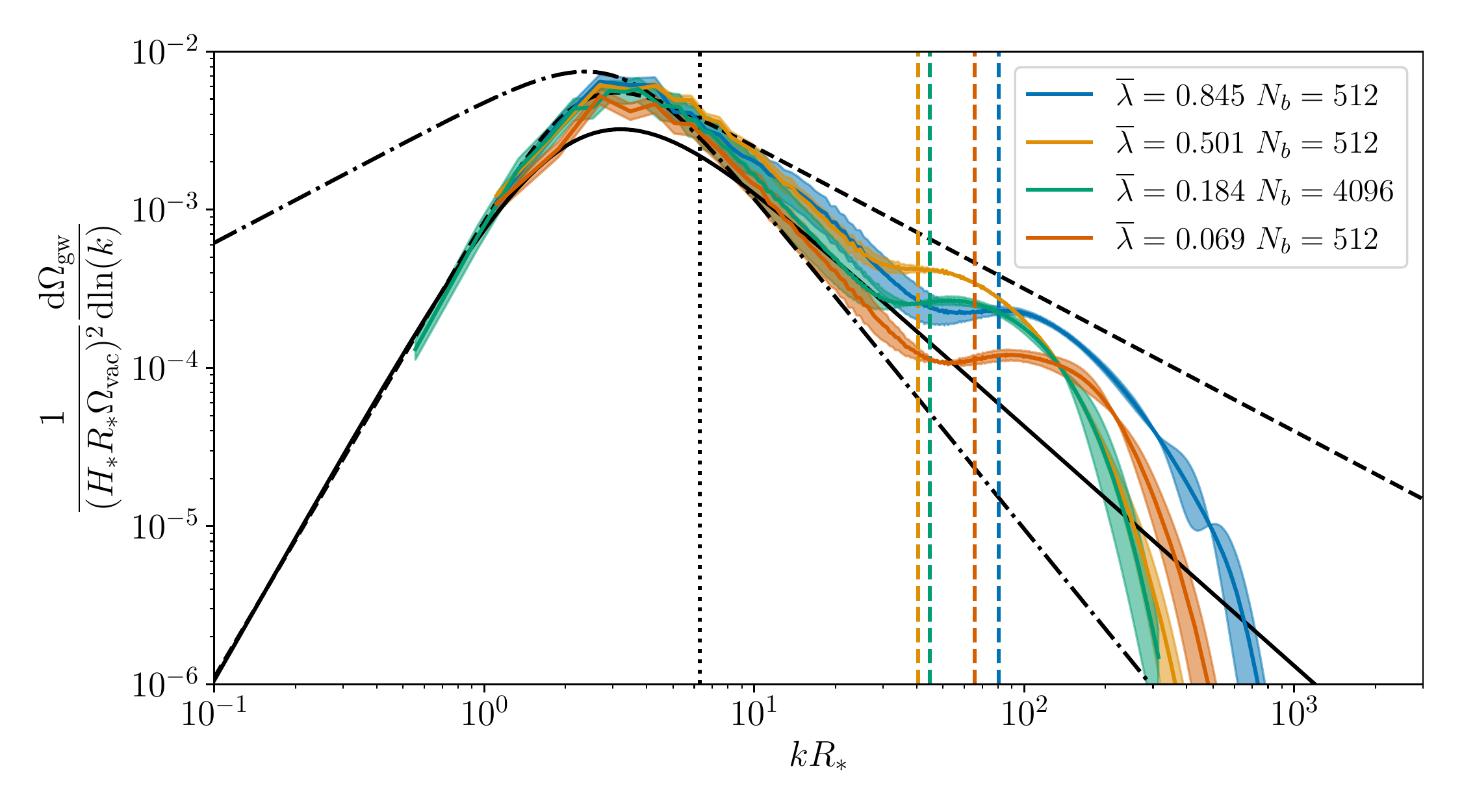}}
\\
\subfigure[\,$t/\Rstar = 2.5$]{\includegraphics[width=0.485\textwidth,clip]{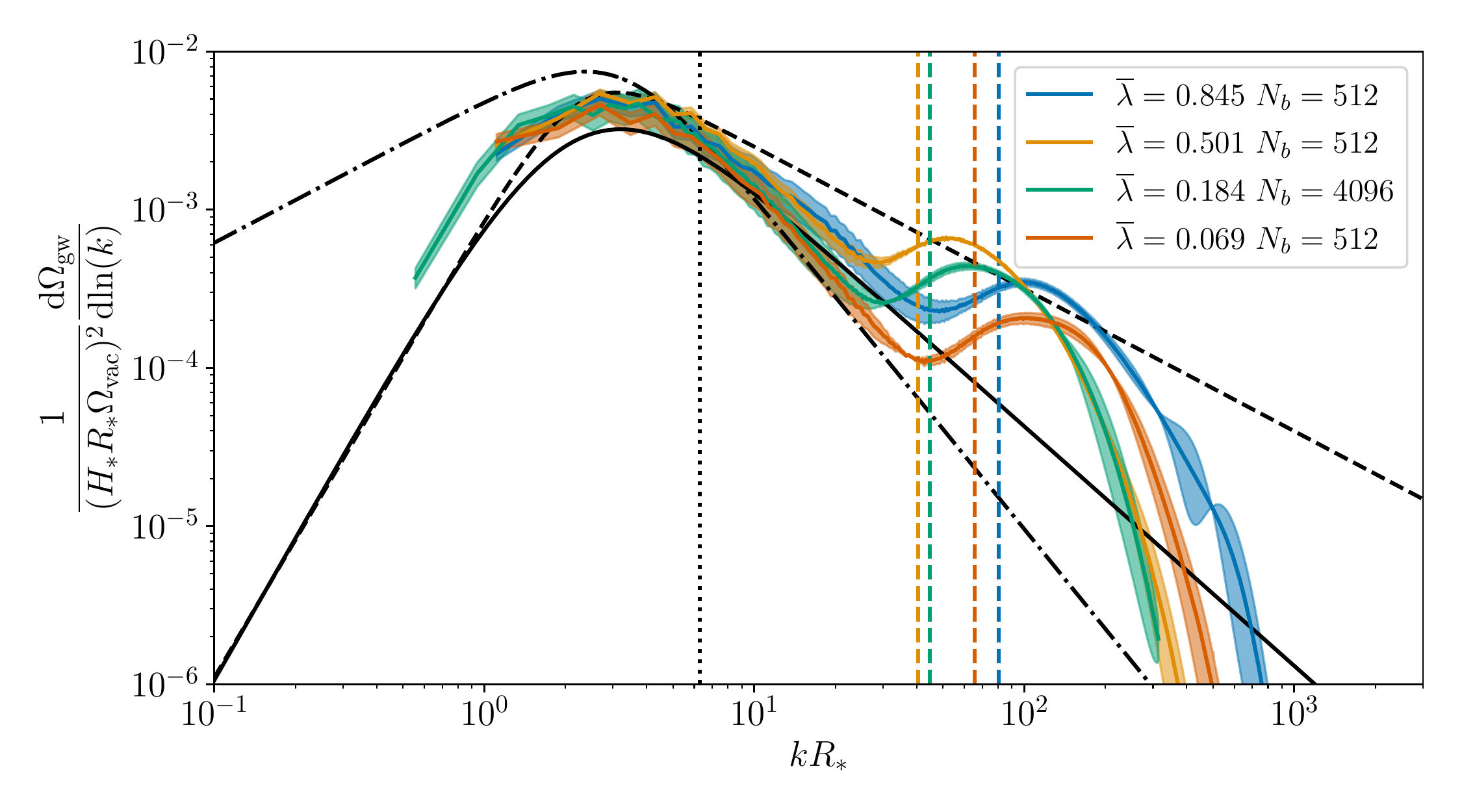}}
\hfill
\subfigure[\,$t/\Rstar = 4.0$]{\includegraphics[width=0.485\textwidth,clip]{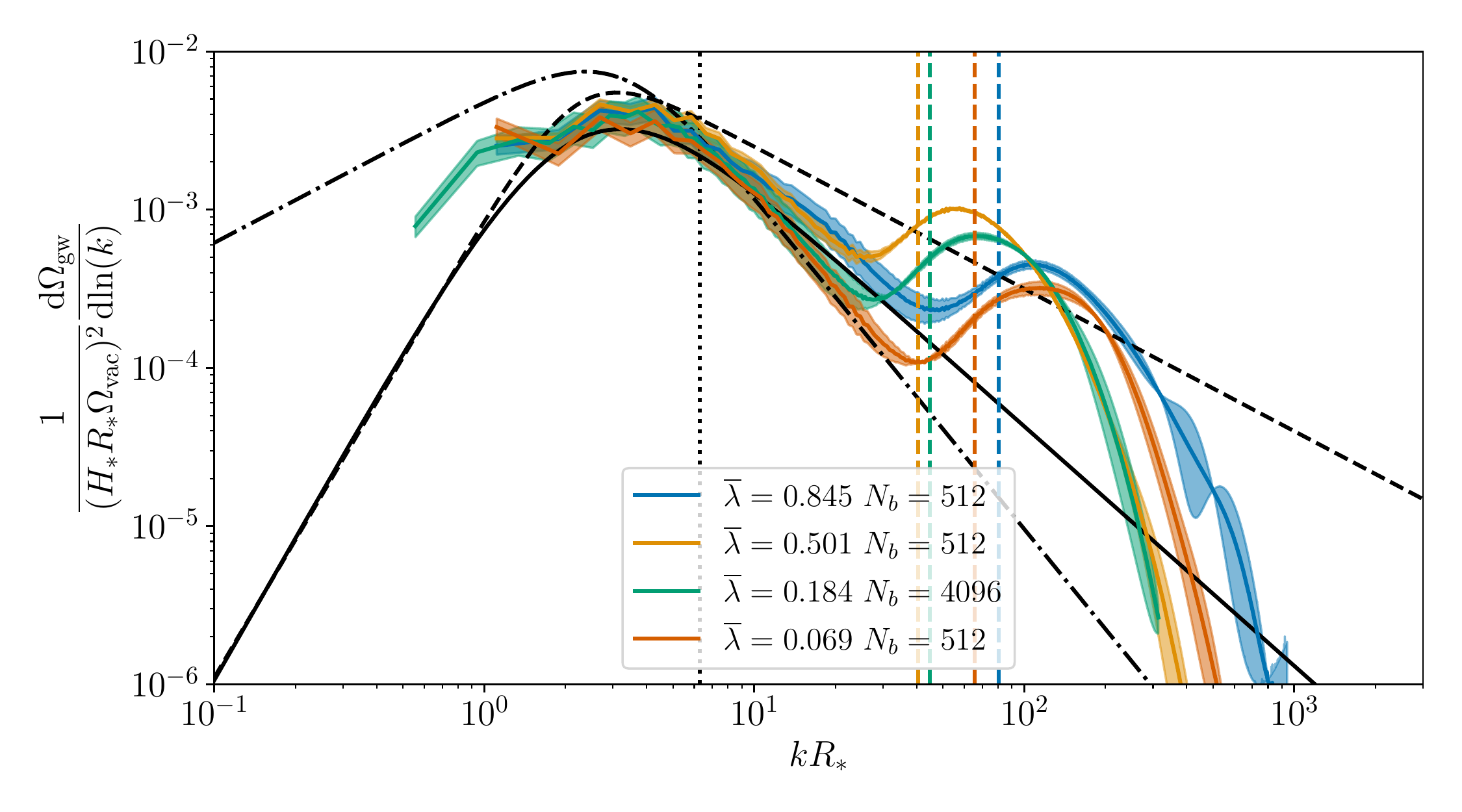}}
\caption{ Evolution of the gravitational-wave power spectrum for the
  largest simulations performed for each $\lambar$. Each simulation
  uses a simultaneous nucleation scenario, and the Lorentz factor of
  the wall of a bubble with diameter $\Rstar$ is $\gmStar=4.0$. We
  plot the power spectra at four different times, early collision
  phase (a), late collision phase (b), early oscillation phase (c),
  and later in the oscillation phase (d). The black dashed line gives the
  result from the envelope approximation~\cite{Konstandin:2017sat},
  the black dash-dot line gives the prediction from the bulk flow
  model~\cite{Konstandin:2017sat}, and the solid black line indicates
  the previous fit provided in Ref.~\cite{Cutting:2018tjt}. The
  envelope approximation and bulk flow model fits are for an
  exponential nucleation rate. The vertical dotted line gives the
  location of $k = 2\pi / \Rstar$, whereas the coloured dashed lines
  indicate where $k=\Mb$. For each simulation we shade a region
  corresponding to $\pm$ the difference in power between our high and
  medium resolution runs. At high wavenumbers the signal is
  overwhelmed by noise arising from single-precision floating point
  numerical errors. This noise is identified by comparing a smaller
  single-precision and double-precision run. We therefore apply a cut off
  in the UV at $k=\pi/2\Delta x$.}
\label{fig:gwallEvol}
\end{figure*}

We measure the gravitational-wave power spectrum produced during each
simulation. We compare the resulting power spectra with the fit
predicted from the previous work in Ref.~\cite{Cutting:2018tjt}, as
well as the envelope approximation and bulk flow model fits using an
exponential nucleation rate, as detailed in
Section~\ref{sec:gwcoll}. In Fig.~\ref{fig:gwallEvol} we plot four
snapshots showing the evolution of the gravitational-wave power
spectrum (Eq.~\ref{eq:gwps}) for the largest simulations performed for
each $\lambar$. These simulations are listed in
Table~\ref{table:simGW}. The uncertainty for each power spectrum bin
is given by the difference between its value in our high and medium
resolution runs.

We see that early on in the collision phase at $t/\Rstar =0.6$, the
power spectrum is growing with a peak at $k \approx 2\pi / \Rstar$. At
early times, for all $\lambar$ there is a characteristic infrared
power law in $k$ with exponent $\sim 3$. Later in the collision phase
at $t/\Rstar = 1.6$, most of the bubbles have finished colliding, and
we see that for all $\lambar$ the peak has shifted towards lower
values of $k$, aligning with the peak locations predicted in
Ref.~\cite{Cutting:2018tjt} and earlier studies of the envelope
approximation. The peak gravitational-wave power at this point seems
very close to that predicted for an exponential nucleation rate in the
envelope approximation for all $\lambar$. With our limited resolution
of the IR power law we see that it appears to still be roughly
consistent with an exponent of $\sim 3$. The UV power laws vary
between different $\lambar$, with thicker potentials having steeper
exponents. At later times in the collision phase, we see a rise in the
first few bins for our gravitational-wave power spectra, consistent
with the slow decay of the IR power in $\mathcal{P}_T$ shown in
Fig.~\ref{fig:emTT_waterfall}. The limited range we have in the IR
makes it difficult to be conclusive about this. The peak location
appears to remain fixed. We also see for each simulation the steady
growth of a bump in the power spectrum towards the UV, associated with
the length scale of $k\sim \Mb$, consistent with that seen in
Ref.~\cite{Cutting:2018tjt}.

We first turn our attention to understanding the evolution of the UV
bump in the power spectrum. This is made up of gravitational waves
sourced during the oscillation phase, where the scalar field is
oscillating around the scale of $\Mb$. To see the shape of
the power spectra produced from these oscillations, we conduct a
series of long-lasting simulations where we only turn on the evolution
of the metric perturbations at $t/\Rstar=2.5$, long after the last
bubbles have collided. These simulation runs are listed in
Table~\ref{table:late-sim}.

\begin{figure*}[htbp]
\subfigure[\,$\lambar = 0.84$, $\Nb = 8$]{\includegraphics[width=0.485\textwidth,clip]{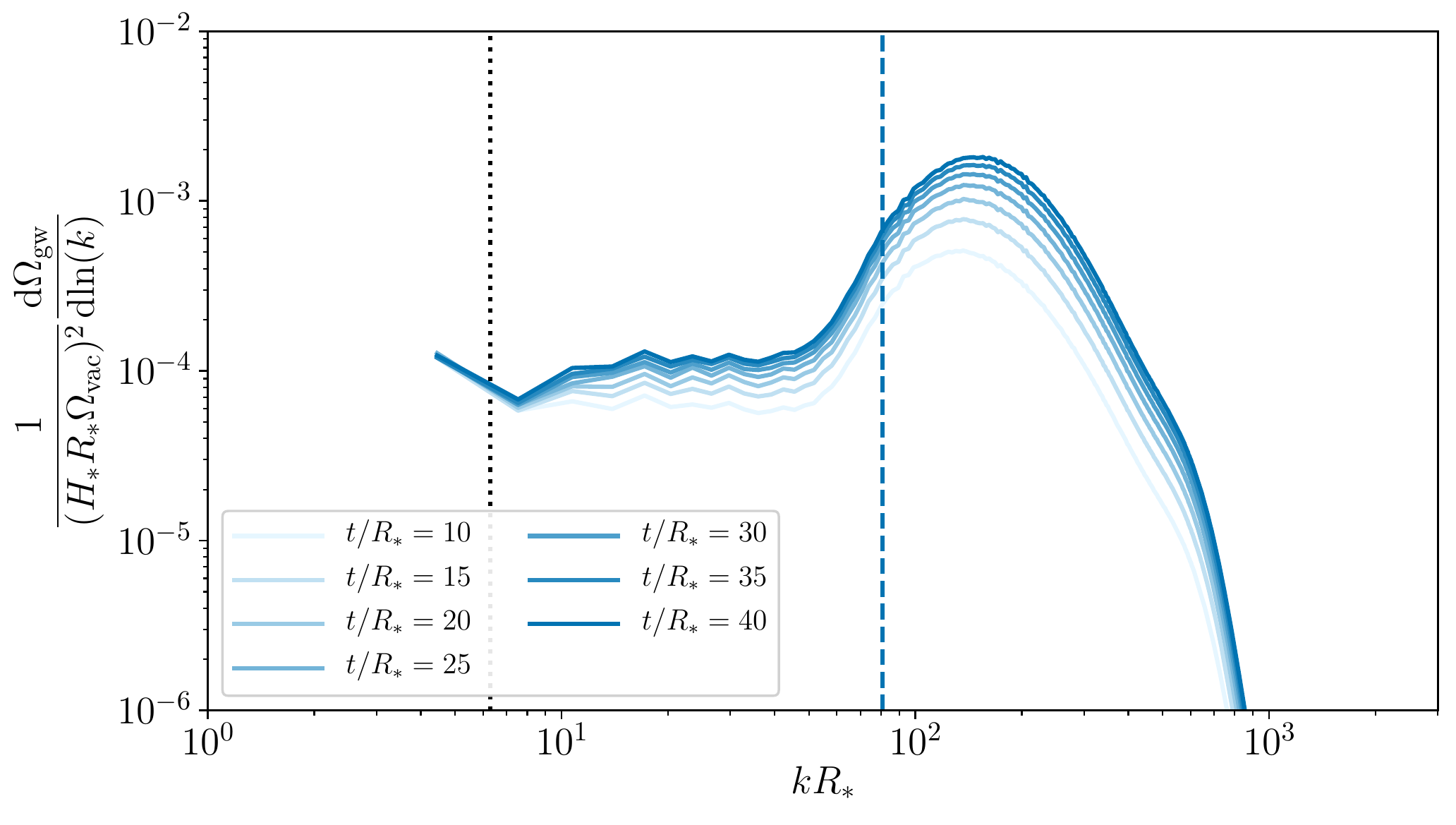}}
\hfill
\subfigure[\,$\lambar = 0.07$,  $\Nb=8$]{\includegraphics[width=0.485\textwidth,clip]{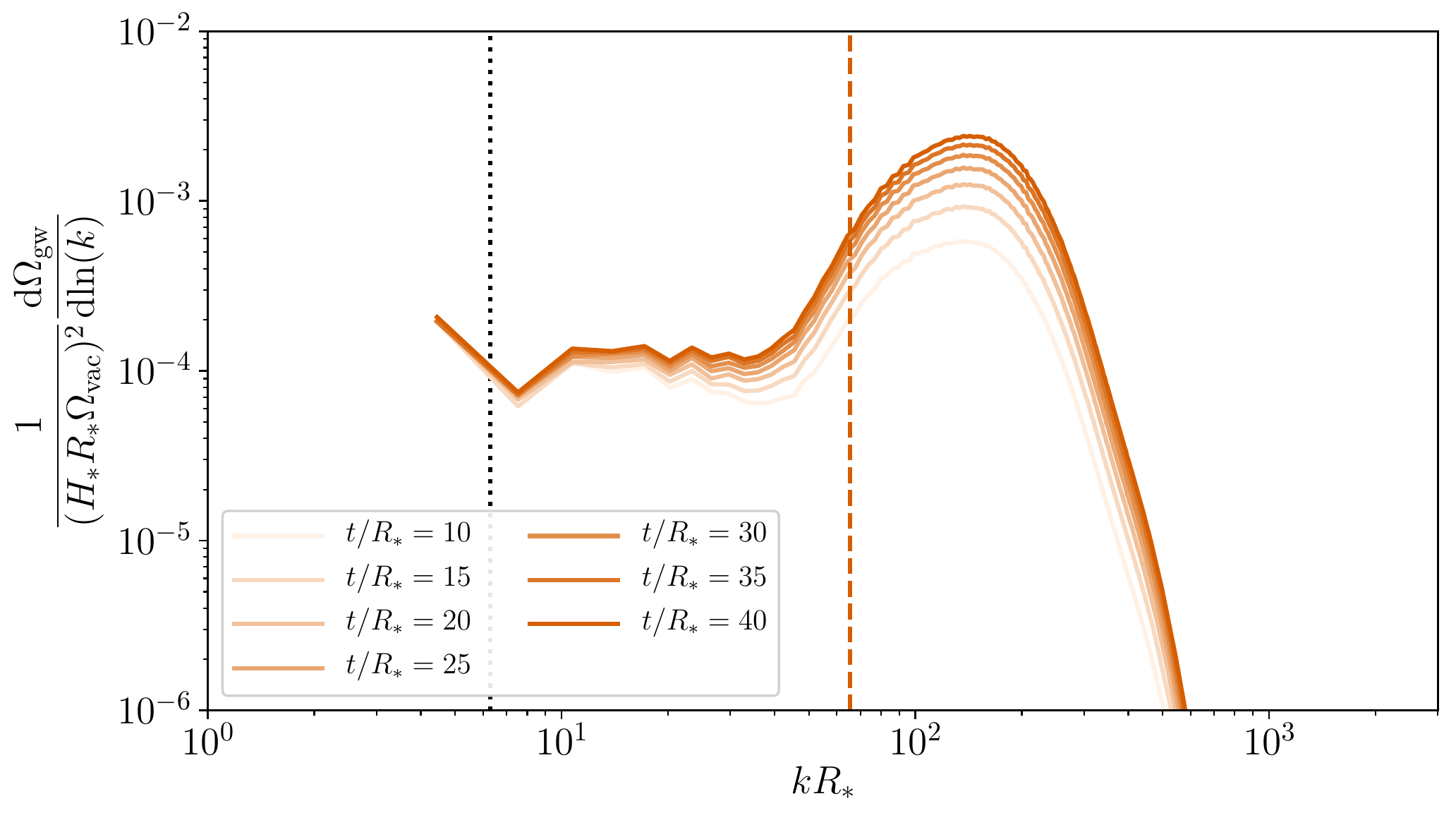}}
\caption{ The power spectrum of the gravitational wave energy density
  parameter for two of the simulations listed in
  Table~\ref{table:late-sim}. In these simulations the metric
  perturbations are only turned on after the bubbles have finished
  colliding, at $t/\Rstar=2.5$. In each plot darker shades indicate
  later times. The vertical black dotted line shows the location of
  $k=2\pi/R_*$, whereas the vertical dashed coloured line shows the
  location of $k=\Mb$. } \label{fig:Omg_late_waterfall} \end{figure*}

We plot the resulting power spectra for $\lambar=0.84$ and
$\lambar=0.07$ in Fig.~\ref{fig:Omg_late_waterfall}. We see that for
both $\lambar$ the power spectra are characterised by a plateau in the
IR, presumably turning over at wavelengths larger than we can access
within our simulations,
and a growing bump at a length scale associated with $\Mb$.

\begin{figure}[htbp]
\includegraphics[width=0.485\textwidth,clip]{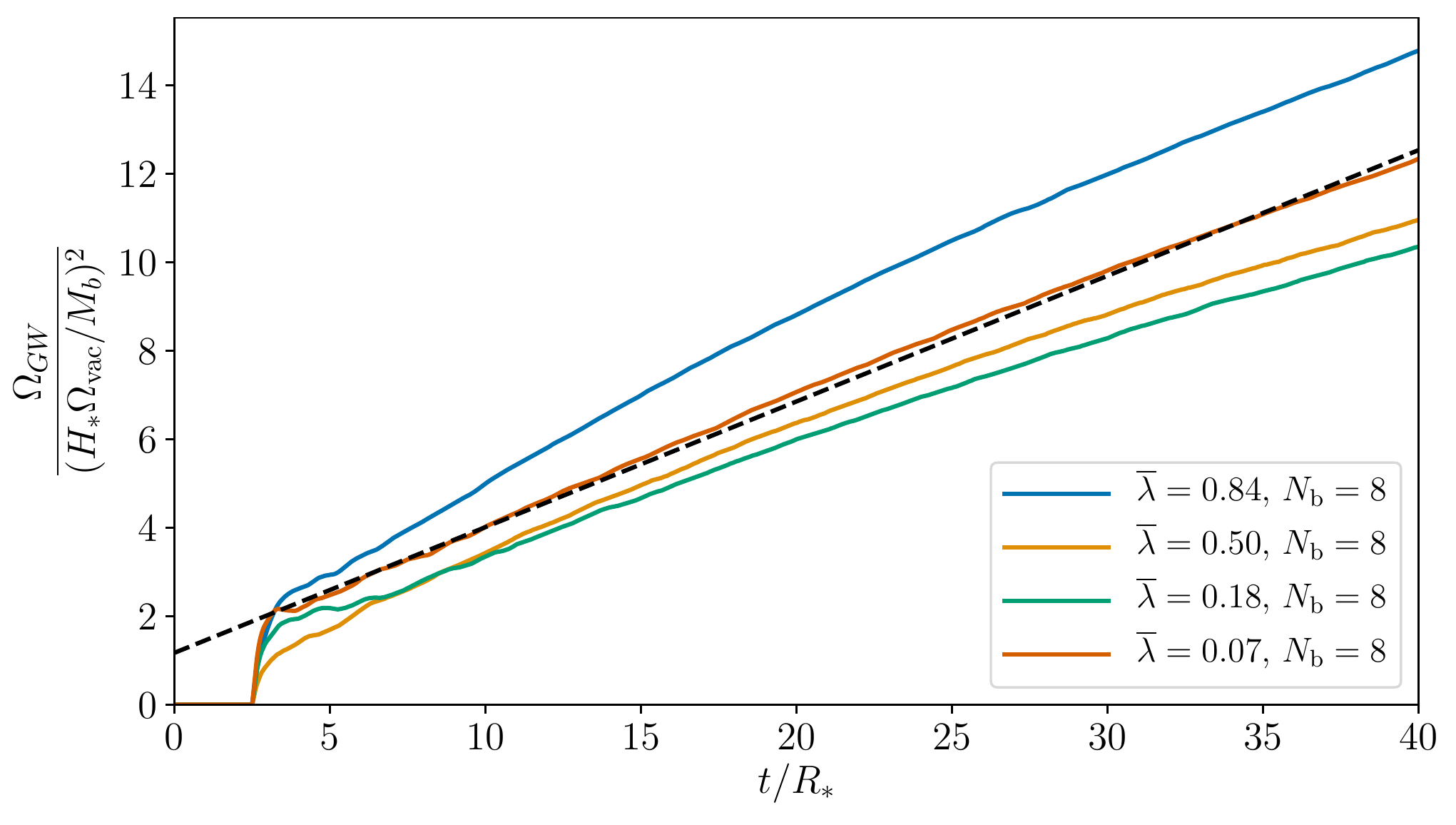}
\caption{ Evolution of the total gravitational wave energy density
  parameter $\OmGW$ for a series of $\lambar$. These are the
  simulations listed in Table~\ref{table:late-sim}, in which the
  evolution of $h^{TT }_{ij}$ is only turned on at $t/\Rstar=2.5$. The
  black dashed line represents a linear fit to the data with slope
  $\frac{d\OmGW}{dt} \sim 0.28 (\Hc \OmVac/\Mb)^2 / \Rstar$.}
\label{fig:lateOmgw}
\end{figure}

We can also use these simulations to calculate the growth rate of
$\OmGW$ during the oscillation phase. We plot $\OmGW$ for our late
time simulations in Fig.~\ref{fig:lateOmgw}. From this plot, we can see
that the growth rate is fairly similar for all $\lambar$. The rate
appears to be slower than linear. Note that the growth of
gravitational waves shown in our simulations is in effect an upper
bound, as in reality other effects will come into play such as the
decay of the scalar field into other particles and damping of the
scalar field gradients due to the effects of expansion.

We find that the calculation of the growth of $\OmGW$ during the
oscillation phase is similar for all $\lambar$ to that found in
Ref.~\cite{Cutting:2018tjt}. Therefore, upon
extrapolation to a realistic separation of scales, the gravitational
wave energy density will be dominated by the production in the
collision phase providing $M_b\ll m_\mathrm{Pl}$, with $m_\mathrm{Pl}$
the Plank mass.

\subsection{Fitting}
\label{sec:fitting}
To attempt to distinguish between the resulting power spectra
for different $\lambar$ we calculate fits for the spectrum. We do this
for the largest simulation performed for each $\lambar$. We choose to
fit according to the following function,
\begin{equation}
  \frac{d\OmGW}{d\mathrm{ln}(k)}= \tilde{\Om}_\mathrm{GW} \frac{(a+b)\tilde{k}^bk^a}{b\tilde{k}^{(a+b)}+ak^{(a+b)}}\text,
\label{eq:thisfit}
\end{equation}
where $a$, $b$, $\tilde{k}$ and $\tilde{\Om}_\mathrm{GW}$ are the
fitting parameters. The fit is calculated using the difference in
power between the high resolution and medium resolution runs as the
one sigma uncertainty for each bin.

We are able to see from Fig.~\ref{fig:gwallEvol} that there appears to
be some indication of time dependence in the power spectra, even
after the bubbles have finished colliding. This is also indicated due
to the evolution of $\mathcal{P}_T$ shown in
Fig.\ref{fig:emTT_waterfall}. We, therefore, choose to perform our fit
throughout the simulation and track how the fitting parameters
evolve. We fit for values of $k$ up to $k = \Mb/2$ in order to
avoid the UV power law being affected by the growing bump associated
with oscillations in the scalar field about the mass scale.

\begin{figure*}[htbp]
\subfigure[]{\includegraphics[width=0.485\textwidth,clip]{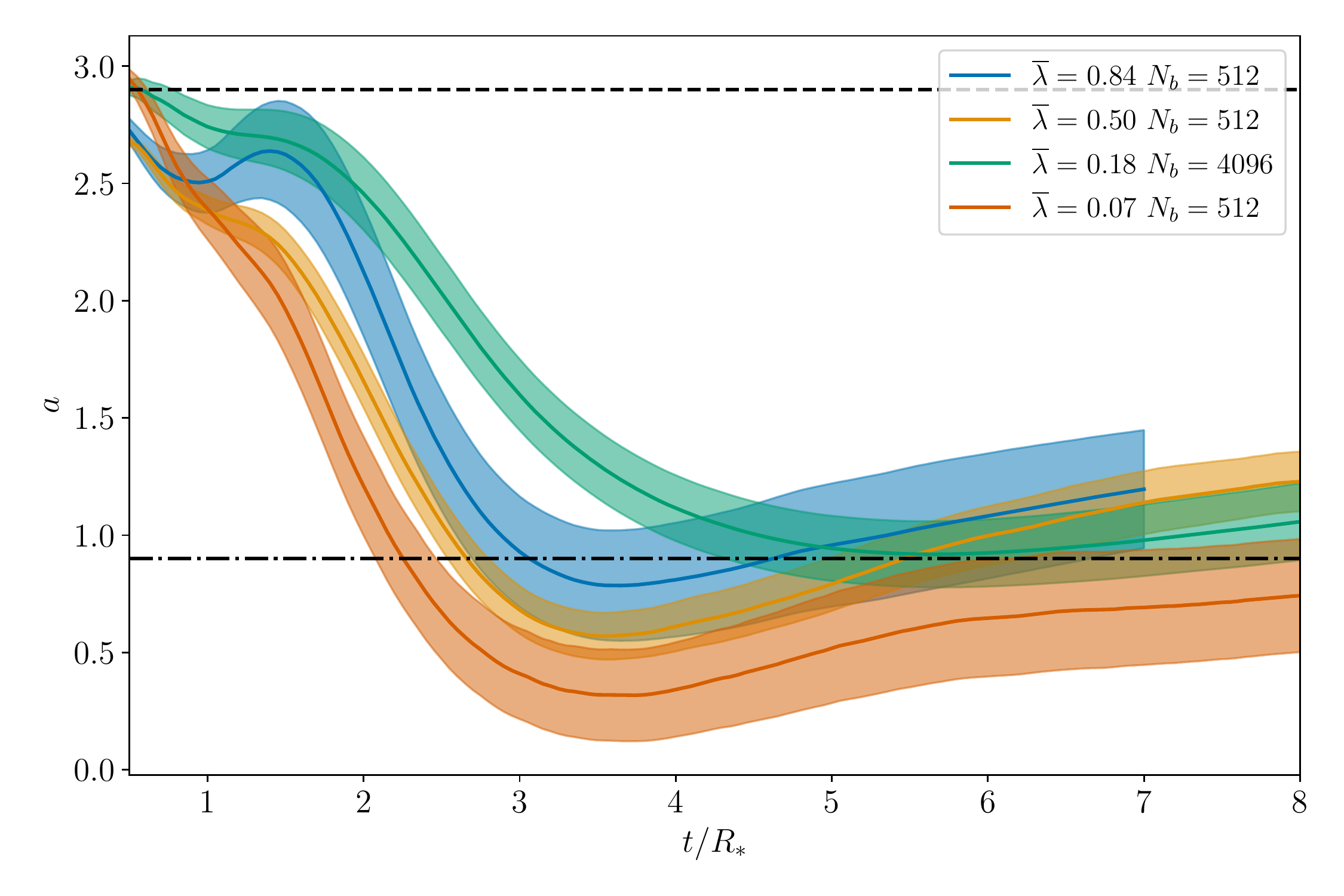}}
\hfill
\subfigure[]{\includegraphics[width=0.485\textwidth,clip]{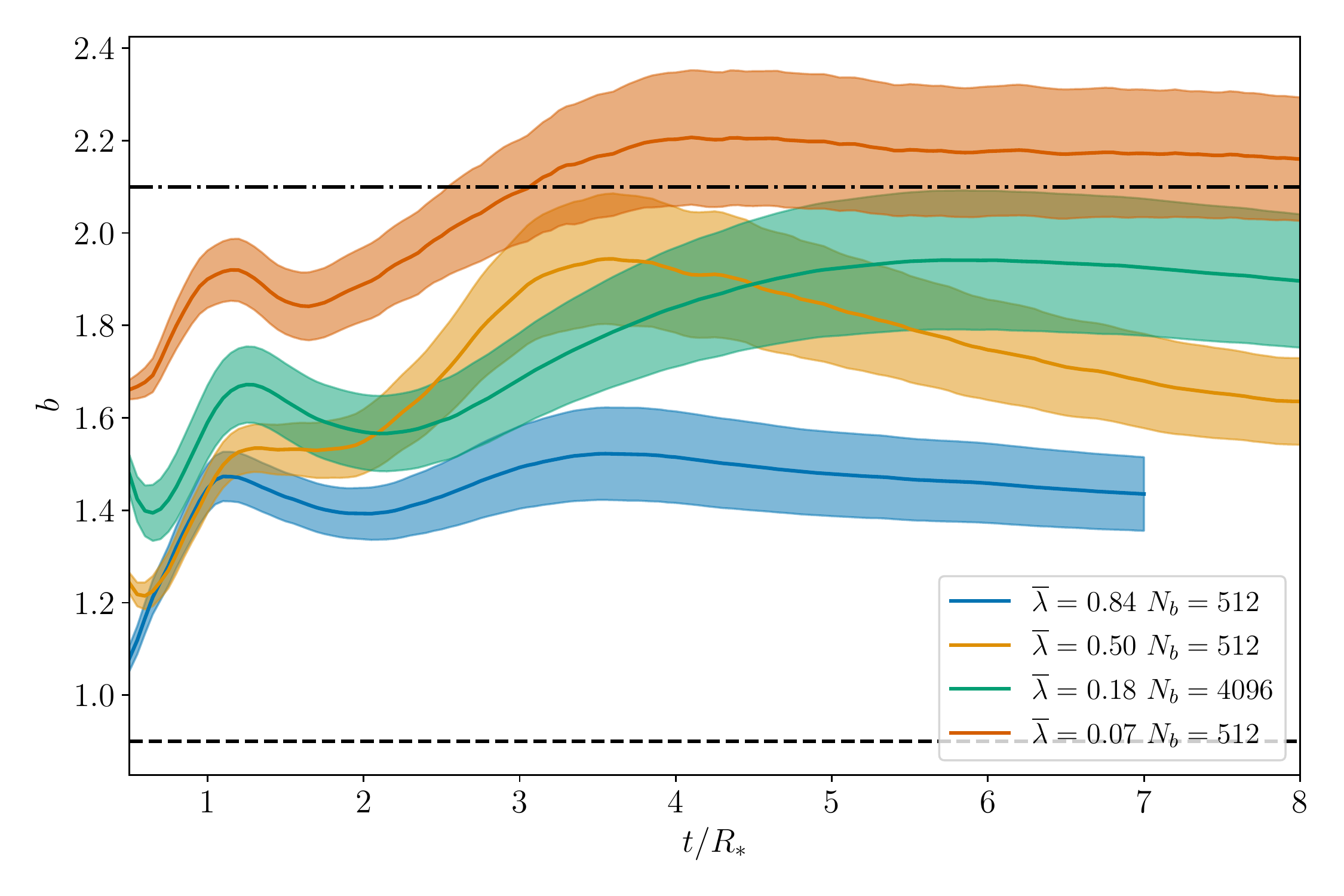}}
\\
\subfigure[]{\includegraphics[width=0.485\textwidth,clip]{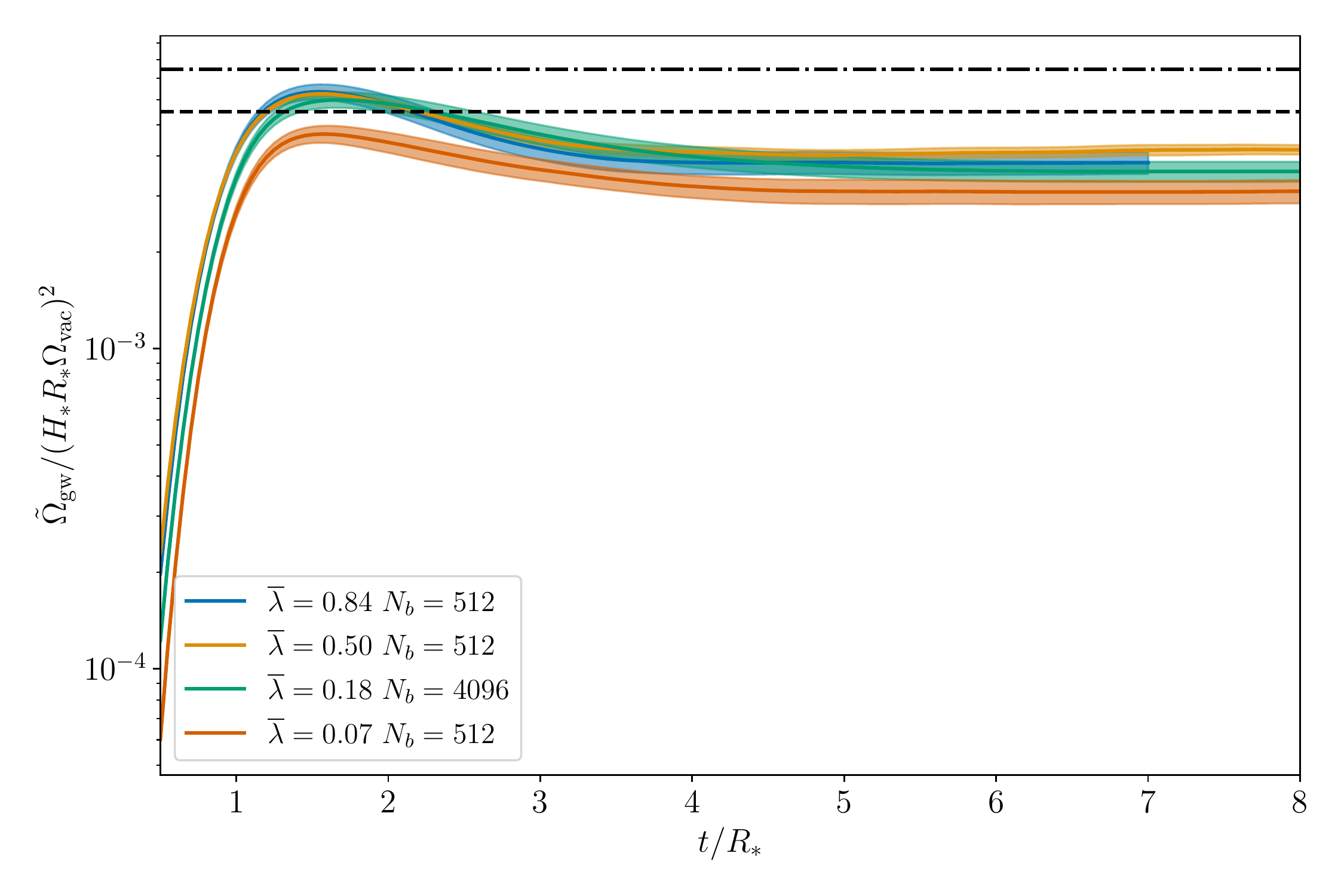}}
\hfill
\subfigure[]{\includegraphics[width=0.485\textwidth,clip]{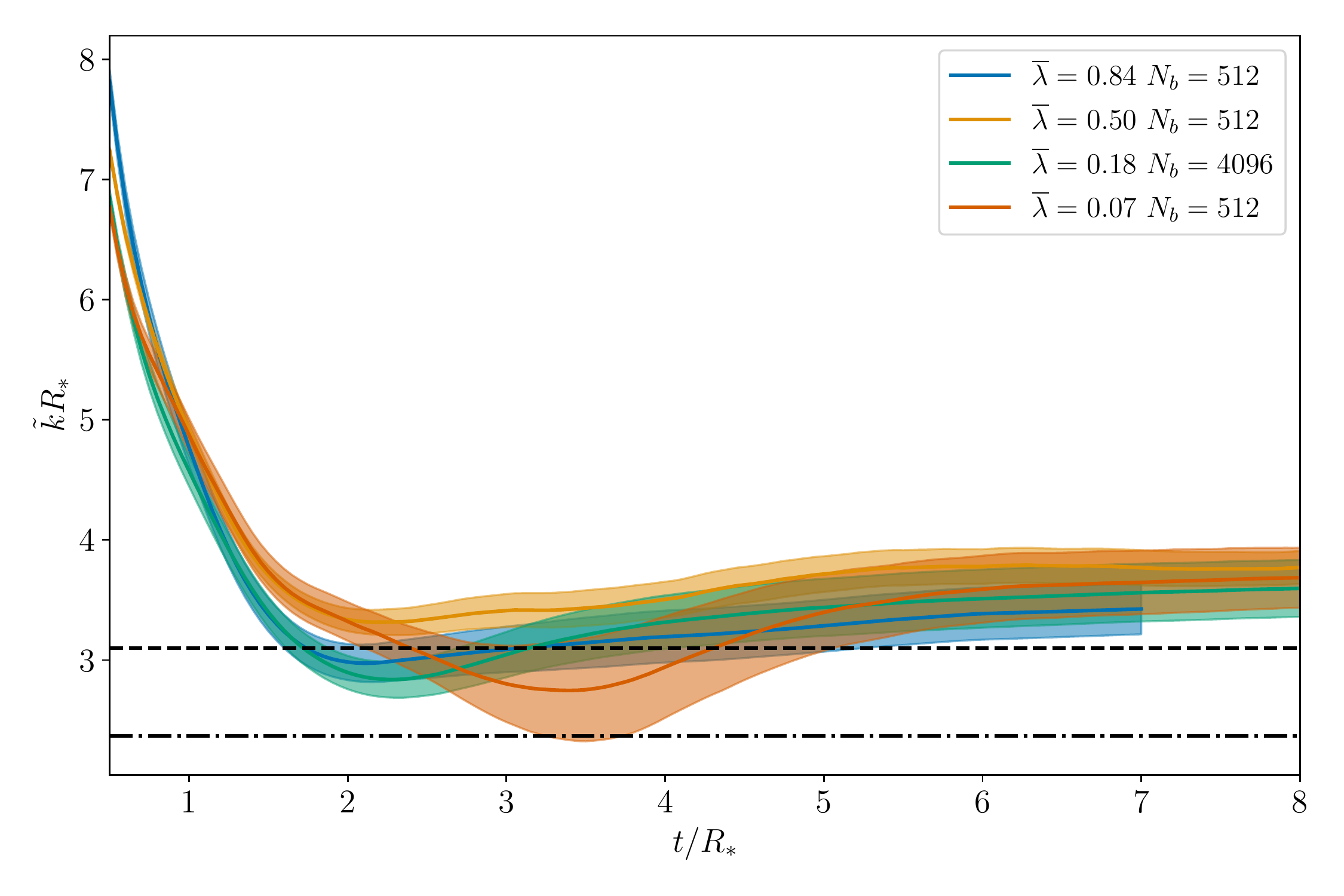}}
\caption{Plot of the values of all the fitting parameters in
  Eq.~\ref{eq:thisfit} for a simultaneous nucleation rate. These have
  been found using the largest simulation for each $\lambar$ in
  Table~\ref{table:simGW}. We plot how these values vary with time
  during the simulations. In (a) we show the IR power law $a$, in (b)
  we show the UV power law $b$, in (c) we plot the peak amplitude
  $\tilde{\Omega}_\mathrm{GW}$ and in (d) we plot the peak frequency
  $\tilde{k}$. The coloured bands show the region corresponding to one
  standard deviation on the fitting parameters.  In each plot we
  highlight the prediction for each parameter for an exponential
  nucleation rate in the envelope approximation by a horizontal dashed
  black line, and in the bulk flow model by a dash-dot black line.  }
\label{fig:Fitting}
\end{figure*}

In Fig.~\ref{fig:Fitting}, we plot how all four fitting parameters
$a$, $b$, $\tilde{k}$ and $\tilde{\Om}_\mathrm{GW}$ evolve for the
largest simulation for each $\lambar$ in Table~\ref{table:simGW}. We
include lines to illustrate the predictions for each parameter by the
envelope approximation and by the bulk flow model.

Note that the envelope and bulk flow predictions are taken from
simulations with an exponential nucleation rate, whereas our
simulations use simultaneous nucleation. This could result in a
discrepancy between the peak frequency and
amplitude~\cite{Weir:2016tov}, though the power law exponents are not
typically affected by the nucleation scenario. In previous lattice
simulations conducted in Ref.~\cite{Cutting:2018tjt}, no strong
dependence on the nucleation rate was seen in the peak amplitude or
frequency.

At early times the peak frequency is slightly more than
$\tilde{k}\sim 2\pi/\Rstar$, but as the bubbles finish colliding this
shifts to smaller values, closer to $\tilde{k}\sim \pi/\Rstar$. This
behaviour is consistent across all $\lambar$. In all cases, the final
value of $\tilde{k}$ is larger than predicted for an exponential rate
in the bulk flow model and slightly larger than the envelope
approximation prediction.

The peak gravitational-wave amplitude is obtained around the time of
$t/\Rstar=1.5$. At later times, the peak amplitude drops as the power
spectrum becomes more broad. We observe that there is some deviation
between $\lambar$ with the peak gravitational-wave power larger for
the two thin wall potentials, and smaller for the two thick wall
potentials. This effect is overall quite small, and
$\tilde{\Omega}_\mathrm{gw}$ is the fitting parameter most sensitive
to lattice effects, see App.~\ref{app:convergence}. The peak amplitude
is smaller than that predicted for an exponential nucleation rate in
the envelope approximation, which in turn is smaller than the
amplitude predicted in the bulk flow model.

We see that at early times, for all $\lambar$, the IR power law is
close to a white noise spectrum of $k^{3}$. After the bubbles finish
colliding, the IR power law decreases. This indicates that
gravitational waves are being sourced on scales larger than $\Rstar$.
This agrees with what we have seen in both $\mathcal{P}_T$ and
Fig.~\ref{fig:gwallEvol}. There is some indication of the IR power law
exponent $a$ growing towards the end of the simulations. Our limited
resolution in the IR and the small number of modes in the bins with
smallest $k$ values mean that we cannot evaluate accurately the value
of $a$, particularly at late times where the peak frequency is
smallest. However, the data that we do have do show a strong
indication of an IR power law that becomes shallower than $k^3$ at
late times after the bubbles have finished colliding. This is
particularly true in the case of $\lambar =0.18$ which has the largest
number of bubbles, $\Nb=4096$. For all $\lambar$, the final value of
the IR power law is close to $a=1$ as predicted by the bulk flow
model. We do see some indication that, as $\lambar$ increases, the
final IR power law becomes steeper.

At early times, the UV power law exponent $b$ grows for all $\lambar$.
At late times we see that there is also a consistent trend in $b$
according to $\lambar$, with the UV power law becoming steeper as
$\lambar$ decreases. The final value of $b$ for $\lambar=0.07$ is
close to that of the bulk flow model prediction. As
$\lambar\rightarrow 1$, the value of $b$ decreases, moving towards the
value predicted in the envelope approximation, though it does not
reach it for the values of $\lambar$ we study.

There is some indication that there is a slow decrease in $b$ at late
times. This is in part because $a$ and $b$ are anti-correlated around
the peak of the spectrum, and as $a$ grows $b$ decreases. The limited
separation of scales we obtain between $\Rstar$ and $1/\Mb$ increases
the influence of $a$ on $b$. This effect is strongest for
$\lambar=0.50$ where the peak in the spectrum from bubble collisions
and that from oscillations in the scalar field are closest together.
Forcing $a$ to be fixed leads to a more stable value of $b$ at the end
of the simulation, though a worse fit overall. To obtain a
more accurate fit for $b$, we need to improve our resolution in the IR
to obtain a better estimate on $a$. Alternatively, we could increase
the separation between $\Rstar$ and $1/\Mb$ by increasing $\gmStar$.
Both of these options require larger simulations and a dynamic range
currently unavailable to us.

In Table~\ref{table:fitting} we provide the late time values of the
fitting parameters for each $\lambar$. These are taken at the end of
the simulation, corresponding to $t/\Rstar=7.0$ for $\lambar =0.84$
and $t/\Rstar=8.0$ for the other $\lambar$. Caution should be taken
when using these values, as from Fig.~\ref{fig:Fitting} it can be seen
that the exponents $a$ and $b$ have not completely settled by the end
of our simulations.

\begin{table}
  \centering
  \resizebox{0.48 \textwidth}{!}{
  \begin{tabular}{D{.}{.}{1.2} D{.}{.}{1.13} D{.}{.}{1.9}  D{.}{.}{1.9} D{.}{.}{1.9} }
    \hline \hline
    \multicolumn{1}{c}{$\lambar$}\Tstrut &
                                           \multicolumn{1}{c}{$\tilde{\Om}_\mathrm{GW}/(\Hc
                                           \Rstar \OmVac)^2$}
    & \multicolumn{1}{c}{$\tilde{k} \Rstar$} & \multicolumn{1}{c}{$a$} & \multicolumn{1}{c}{$b$}\\
    \hline
    0.84         & (3.81 \pm 0.30) \times 10^{-3}   & 3.42 \pm 0.21 &
                                                                      1.20\pm 0.25 & 1.44\pm0.08 \\
    0.50         & (4.18 \pm 0.15) \times 10^{-3}   & 3.77 \pm 0.14 &
                                                                      1.23
                                                                      \pm
                                                                      0.13&
                                                                             1.64
                                                                             \pm 0.09\\
    0.18         & (3.56 \pm 0.26) \times 10^{-3}    & 3.60 \pm 0.24 &
                                                                       1.06
                                                                       \pm
                                                                       0.16&
                                                                         1.90
    \pm 0.14\\
    0.07         & (3.10 \pm 0.26)\times 10^{-3}    & 3.68 \pm 0.25 &
                                                                      0.742
                                                                      \pm
                                                                       0.241&
                                                                         2.16
    \pm 0.13\\
    \hline \hline
  \end{tabular}
}
\caption{Final values of fitting parameter values in
  Eq.~\ref{eq:thisfit} which gives gravitational-wave power spectrum
  arising from bubble collisions. These are calculated for the largest
  simulation for all $\lambar$ given in Table~\ref{table:simGW}. The
  values supplied here are taken at the end of the simulations, which
  corresponds to $t/\Rstar=7.0$ for $\lambar=0.84$ and $t/\Rstar=8.0$
  for the other values of $\lambar$. Uncertainties on the fitting
  parameters are calculated taking the one sigma uncertainty on each
  power spectrum bin to be given by the difference between its value in our
  medium and high resolution runs.}
\label{table:fitting}
\end{table}

\section{Conclusions}\label{sec:Conclusions}

In this work, we have investigated whether the underlying potential for
a vacuum phase transition can affect the resulting gravitational-wave
signal. We note that, for a quartic effective potential with a cubic
term, the effect of the potential on the dynamics of the scalar field
is determined through a single parameter, $\lambar$. When
$\lambar \rightarrow 1$, we are in the thin wall limit, and the
critical profile can be approximated by a $\mathrm{tanh}$ function.
The thick wall limit is approached for $\lambar \rightarrow 0$, and in
this case, the critical profile is approximated well by a Gaussian.

The dynamics of the scalar field in the overlap region between
colliding bubbles depends on the value of $\lambar$. When $\lambar$ is
close to one, after bubbles collide, the scalar field rebounds in the
overlap region towards the symmetric phase. The rebound is reduced as
$\lambar\rightarrow 0$ for fixed $\gmStar$.

We have explored a range of $\lambar$ in a series of simulations with
up to $4800^3$ lattice sites in which as many as $4096$ bubbles are
nucleated simultaneously. From these, we evaluate the transverse
traceless shear-stress $T^{TT}_{ij}$ and compute the power spectrum
$\mathcal{P}_T$. We find evidence that even after the bubbles have
finished colliding, gravitational waves continue to be sourced at
scales larger than $\Rstar$. This could be as a result of energy
density in the bubble walls continuing to propagate after collision.
Continued propagation of shells of energy density after collision is a
violation of one of the assumptions of the envelope approximation and
matches closer to the bulk flow model.

After the bubbles have finished colliding, we enter an oscillation
phase during which the scalar field is oscillating around $\phi_b$.
This produces a peak in $\mathcal{P}_T$ around $k\sim\Mb$, and at very
late times this develops a white noise IR power law of $k^{3}$. While
this feature is very long-lasting within our simulations, we would
expect that in reality, the amplitude would decay as the scalar
field gradients decrease due to quantum processes and Hubble friction.

During the oscillation phase, gravitational waves are sourced by a
feature in $\mathcal{P}_{{T}^{TT}}$ with a peak around $k\sim\Mb$. This
behaviour was already noted in
Refs.~\cite{Child:2012qg,Cutting:2018tjt}. This feature produces a
bump in the gravitational-wave power spectrum around $k\sim\Mb$ for
all $\lambar$. The growth rate of the $\OmGW$ during the oscillation
phase is slightly slower than linear. As our simulations do not
account for damping from Hubble friction or allow for the scalar field
to decay via quantum processes, this should be taken as an upper bound
on the growth rate for $\OmGW$. Our results on the growth rate are
consistent with those in Ref.~\cite{Cutting:2018tjt}, where it was
shown that the total gravitational-wave power from the oscillation
phase will be suppressed compared to that arising from bubble
collisions providing that $\Mb\ll\mpl$.

We also compute the gravitational wave energy density parameter power
spectrum $d\OmGW/d\mathrm{ln}(k)$. We perform a fit for the spectrum
arising from bubble collisions using Eq.~\ref{eq:thisfit}. We
calculate how the fitting parameters vary during our simulations.

There are some indications that $\lambar$ can affect the resulting
gravitational-wave power spectrum. The peak power of the
gravitational-wave power spectrum varies according to $\lambar$,
though the variation is probably not significant enough to be
observable. The values of $\tilde{\Om}_\mathrm{GW}$ found at the end
of our simulations are larger for $\lambar>0.5$, and decreases for
smaller $\lambar$. The peak amplitude
$\tilde{\Om}_\mathrm{GW}/(\Hc R_* \OmVac)^2$ varies between
$4.2 \times 10^{-3}$ for $\lambar=0.50$, and $3.1 \times 10^{-3}$ for
$\lambar =0.07$.

More hopeful is the possibility that we could distinguish vacuum
transitions with different potentials due to the UV power law. The UV power laws
we find at the end of the simulations become steeper as $\lambar$
decreases. The gravitational-wave power spectrum falls as
$k^{-1.4}$ for $\lambar =0.84$ and $k^{-2.2}$ for $\lambar= 0.07$.

The IR power law is close to $k^3$ when bubbles start to collide, with
a peak in the spectrum around $k\sim 2\pi/\Rstar$. At later times the peak shifts
slightly towards the IR. The section of the IR power law that we can
resolve appears to become shallower with an exponent $<3$. Our limited
resolution in the IR means that we can only infer the power law from
the first few bins in our power spectrum. We find that at the end of
our simulations the IR power law is shallower for smaller $\lambar$,
varying between $k^{1.2}$ for $\lambar=0.84$ and $k^{0.7}$ for
$\lambar=0.07$. Presumably, at larger scales than we can resolve
within our simulations, the power law turns over to a white noise
spectrum as causality dictates.

We find that neither the envelope approximation or the bulk flow model
correctly predict the final gravitational-wave power spectrum. For all $\lambar$, the
peak power is slightly smaller than predicted by the envelope
approximation which is itself smaller than the bulk flow model
predicts. The peak location is closer to that predicted by the
envelope approximation. The UV power law is similar to the bulk flow
model for small $\lambar$, and moves towards the envelope
approximation prediction as $\lambar$ increases. The value of the
IR power law also seems to be closer to the bulk flow model, though as
$\lambar$ increases, it does become steeper, shifting towards the
envelope approximation value. This roughly follows the picture
proposed in Ref.~\cite{Jinno:2019bxw}. It remains to be determined if
simulations with larger $\lambar$ become even closer to the envelope
approximation, and whether at larger $\gmStar$ the proposal of
Ref.~\cite{Jinno:2019bxw} becomes more exact.

Overall, we have shown that for vacuum phase transitions, the
underlying effective potential can affect the resulting gravitational
wave power spectrum. In particular, we have seen that, for the
quartic potential that we investigated, the UV power law appears
sensitive to $\lambar$. The IR power law is challenging to resolve
with our simulations, but we see some indication that it may be
shallower than a $k^{3}$ spectrum, and it appears to be evolving long
after the bubbles have finished colliding. Further explorations into
the IR power law and behaviour of the gravitational-wave power spectrum as
we extrapolate to larger $\gmStar$ will require new techniques as we
have reached the limit of the computing resources available to us.

\begin{acknowledgments}
  We thank Oliver Gould, Rysuke Jinno, Thomas Konstandin, Eugene Lim,
  Satumaaria Sukuvaara and Essi Vilhonen for useful discussions. DC,
  DJW, and MH would like to thank Nordita for their hospitality during
  the ``Gravitational Waves from the Early Universe'' workshop. Our
  simulations were carried out at the Finnish Centre for Scientific
  Computing CSC. DC (ORCID ID 0000-0002-7395-7802) is supported by a
  Science and Technology Facilities Council Studentship and the
  University of Sussex. EGE was supported by the Academy of Finland,
  grant 320123. EGE and DJW were supported by the Research Funds of
  the University of Helsinki. DJW (ORCID ID 0000-0001-6986-0517) was
  supported by a Science and Technology Facilities Council Ernest
  Rutherford Fellowship, grant no. ST/R003904/1, and by the Academy of
  Finland, grants 324882 and 328958. MH was supported by the Science
  and Technology Facilities Council (grant number ST/P000819/1) and
  the Academy of Finland (grant number 286769).
\end{acknowledgments}

\bibliography{thick-wall-scalar}

\begin{thebibliography}{78}%
\makeatletter
\providecommand \@ifxundefined [1]{%
 \@ifx{#1\undefined}
}%
\providecommand \@ifnum [1]{%
 \ifnum #1\expandafter \@firstoftwo
 \else \expandafter \@secondoftwo
 \fi
}%
\providecommand \@ifx [1]{%
 \ifx #1\expandafter \@firstoftwo
 \else \expandafter \@secondoftwo
 \fi
}%
\providecommand \natexlab [1]{#1}%
\providecommand \enquote  [1]{``#1''}%
\providecommand \bibnamefont  [1]{#1}%
\providecommand \bibfnamefont [1]{#1}%
\providecommand \citenamefont [1]{#1}%
\providecommand \href@noop [0]{\@secondoftwo}%
\providecommand \href [0]{\begingroup \@sanitize@url \@href}%
\providecommand \@href[1]{\@@startlink{#1}\@@href}%
\providecommand \@@href[1]{\endgroup#1\@@endlink}%
\providecommand \@sanitize@url [0]{\catcode `\\12\catcode `\$12\catcode
  `\&12\catcode `\#12\catcode `\^12\catcode `\_12\catcode `\%12\relax}%
\providecommand \@@startlink[1]{}%
\providecommand \@@endlink[0]{}%
\providecommand \url  [0]{\begingroup\@sanitize@url \@url }%
\providecommand \@url [1]{\endgroup\@href {#1}{\urlprefix }}%
\providecommand \urlprefix  [0]{URL }%
\providecommand \Eprint [0]{\href }%
\providecommand \doibase [0]{http://dx.doi.org/}%
\providecommand \selectlanguage [0]{\@gobble}%
\providecommand \bibinfo  [0]{\@secondoftwo}%
\providecommand \bibfield  [0]{\@secondoftwo}%
\providecommand \translation [1]{[#1]}%
\providecommand \BibitemOpen [0]{}%
\providecommand \bibitemStop [0]{}%
\providecommand \bibitemNoStop [0]{.\EOS\space}%
\providecommand \EOS [0]{\spacefactor3000\relax}%
\providecommand \BibitemShut  [1]{\csname bibitem#1\endcsname}%
\let\auto@bib@innerbib\@empty
%</preamble>
\bibitem [{\citenamefont {Audley}\ \emph {et~al.}(2017)\citenamefont {Audley}
  \emph {et~al.}}]{Audley:2017drz}%
  \BibitemOpen
  \bibfield  {author} {\bibinfo {author} {\bibfnamefont {H.}~\bibnamefont
  {Audley}} \emph {et~al.},\ }\href@noop {} {\  (\bibinfo {year} {2017})},\
  \Eprint {http://arxiv.org/abs/1702.00786} {arXiv:1702.00786 [astro-ph.IM]}
  \BibitemShut {NoStop}%
%%CITATION = ARXIV:1702.00786;%%
\bibitem [{\citenamefont {Caprini}\ and\ \citenamefont
  {Figueroa}(2018)}]{Caprini:2018mtu}%
  \BibitemOpen
  \bibfield  {author} {\bibinfo {author} {\bibfnamefont {C.}~\bibnamefont
  {Caprini}}\ and\ \bibinfo {author} {\bibfnamefont {D.~G.}\ \bibnamefont
  {Figueroa}},\ }\href@noop {} {\  (\bibinfo {year} {2018})},\ \Eprint
  {http://arxiv.org/abs/1801.04268} {arXiv:1801.04268 [astro-ph.CO]}
  \BibitemShut {NoStop}%
%%CITATION = ARXIV:1801.04268;%%
\bibitem [{\citenamefont {Caprini}\ \emph {et~al.}(2016)\citenamefont {Caprini}
  \emph {et~al.}}]{Caprini:2015zlo}%
  \BibitemOpen
  \bibfield  {author} {\bibinfo {author} {\bibfnamefont {C.}~\bibnamefont
  {Caprini}} \emph {et~al.},\ }\href {\doibase 10.1088/1475-7516/2016/04/001}
  {\bibfield  {journal} {\bibinfo  {journal} {JCAP}\ }\textbf {\bibinfo
  {volume} {1604}},\ \bibinfo {pages} {001} (\bibinfo {year} {2016})},\ \Eprint
  {http://arxiv.org/abs/1512.06239} {arXiv:1512.06239 [astro-ph.CO]}
  \BibitemShut {NoStop}%
%%CITATION = ARXIV:1512.06239;%%
\bibitem [{\citenamefont {Caprini}\ \emph {et~al.}(2020)\citenamefont {Caprini}
  \emph {et~al.}}]{Caprini:2019egz}%
  \BibitemOpen
  \bibfield  {author} {\bibinfo {author} {\bibfnamefont {C.}~\bibnamefont
  {Caprini}} \emph {et~al.},\ }\href {\doibase 10.1088/1475-7516/2020/03/024}
  {\bibfield  {journal} {\bibinfo  {journal} {JCAP}\ }\textbf {\bibinfo
  {volume} {2003}},\ \bibinfo {pages} {024} (\bibinfo {year} {2020})},\ \Eprint
  {http://arxiv.org/abs/1910.13125} {arXiv:1910.13125 [astro-ph.CO]}
  \BibitemShut {NoStop}%
%%CITATION = ARXIV:1910.13125;%%
\bibitem [{\citenamefont {Kajantie}\ \emph {et~al.}(1996)\citenamefont
  {Kajantie}, \citenamefont {Laine}, \citenamefont {Rummukainen},\ and\
  \citenamefont {Shaposhnikov}}]{Kajantie:1996mn}%
  \BibitemOpen
  \bibfield  {author} {\bibinfo {author} {\bibfnamefont {K.}~\bibnamefont
  {Kajantie}}, \bibinfo {author} {\bibfnamefont {M.}~\bibnamefont {Laine}},
  \bibinfo {author} {\bibfnamefont {K.}~\bibnamefont {Rummukainen}}, \ and\
  \bibinfo {author} {\bibfnamefont {M.~E.}\ \bibnamefont {Shaposhnikov}},\
  }\href {\doibase 10.1103/PhysRevLett.77.2887} {\bibfield  {journal} {\bibinfo
   {journal} {Phys.Rev.Lett.}\ }\textbf {\bibinfo {volume} {77}},\ \bibinfo
  {pages} {2887} (\bibinfo {year} {1996})},\ \Eprint
  {http://arxiv.org/abs/hep-ph/9605288} {arXiv:hep-ph/9605288 [hep-ph]}
  \BibitemShut {NoStop}%
%%CITATION = HEP-PH/9605288;%%
\bibitem [{\citenamefont {Kajantie}\ \emph {et~al.}(1997)\citenamefont
  {Kajantie}, \citenamefont {Laine}, \citenamefont {Rummukainen},\ and\
  \citenamefont {Shaposhnikov}}]{Kajantie:1996qd}%
  \BibitemOpen
  \bibfield  {author} {\bibinfo {author} {\bibfnamefont {K.}~\bibnamefont
  {Kajantie}}, \bibinfo {author} {\bibfnamefont {M.}~\bibnamefont {Laine}},
  \bibinfo {author} {\bibfnamefont {K.}~\bibnamefont {Rummukainen}}, \ and\
  \bibinfo {author} {\bibfnamefont {M.~E.}\ \bibnamefont {Shaposhnikov}},\
  }\href {\doibase 10.1016/S0550-3213(97)00164-8} {\bibfield  {journal}
  {\bibinfo  {journal} {Nucl.Phys.}\ }\textbf {\bibinfo {volume} {B493}},\
  \bibinfo {pages} {413} (\bibinfo {year} {1997})},\ \Eprint
  {http://arxiv.org/abs/hep-lat/9612006} {arXiv:hep-lat/9612006 [hep-lat]}
  \BibitemShut {NoStop}%
%%CITATION = HEP-LAT/9612006;%%
\bibitem [{\citenamefont {Profumo}\ \emph {et~al.}(2007)\citenamefont
  {Profumo}, \citenamefont {Ramsey-Musolf},\ and\ \citenamefont
  {Shaughnessy}}]{Profumo:2007wc}%
  \BibitemOpen
  \bibfield  {author} {\bibinfo {author} {\bibfnamefont {S.}~\bibnamefont
  {Profumo}}, \bibinfo {author} {\bibfnamefont {M.~J.}\ \bibnamefont
  {Ramsey-Musolf}}, \ and\ \bibinfo {author} {\bibfnamefont {G.}~\bibnamefont
  {Shaughnessy}},\ }\href {\doibase 10.1088/1126-6708/2007/08/010} {\bibfield
  {journal} {\bibinfo  {journal} {JHEP}\ }\textbf {\bibinfo {volume} {08}},\
  \bibinfo {pages} {010} (\bibinfo {year} {2007})},\ \Eprint
  {http://arxiv.org/abs/0705.2425} {arXiv:0705.2425 [hep-ph]} \BibitemShut
  {NoStop}%
%%CITATION = ARXIV:0705.2425;%%
\bibitem [{\citenamefont {Espinosa}\ \emph {et~al.}(2012)\citenamefont
  {Espinosa}, \citenamefont {Konstandin},\ and\ \citenamefont
  {Riva}}]{Espinosa:2011ax}%
  \BibitemOpen
  \bibfield  {author} {\bibinfo {author} {\bibfnamefont {J.~R.}\ \bibnamefont
  {Espinosa}}, \bibinfo {author} {\bibfnamefont {T.}~\bibnamefont
  {Konstandin}}, \ and\ \bibinfo {author} {\bibfnamefont {F.}~\bibnamefont
  {Riva}},\ }\href {\doibase 10.1016/j.nuclphysb.2011.09.010} {\bibfield
  {journal} {\bibinfo  {journal} {Nucl. Phys.}\ }\textbf {\bibinfo {volume}
  {B854}},\ \bibinfo {pages} {592} (\bibinfo {year} {2012})},\ \Eprint
  {http://arxiv.org/abs/1107.5441} {arXiv:1107.5441 [hep-ph]} \BibitemShut
  {NoStop}%
%%CITATION = ARXIV:1107.5441;%%
\bibitem [{\citenamefont {Cline}\ and\ \citenamefont
  {Kainulainen}(2013)}]{Cline:2012hg}%
  \BibitemOpen
  \bibfield  {author} {\bibinfo {author} {\bibfnamefont {J.~M.}\ \bibnamefont
  {Cline}}\ and\ \bibinfo {author} {\bibfnamefont {K.}~\bibnamefont
  {Kainulainen}},\ }\href {\doibase 10.1088/1475-7516/2013/01/012} {\bibfield
  {journal} {\bibinfo  {journal} {JCAP}\ }\textbf {\bibinfo {volume} {1301}},\
  \bibinfo {pages} {012} (\bibinfo {year} {2013})},\ \Eprint
  {http://arxiv.org/abs/1210.4196} {arXiv:1210.4196 [hep-ph]} \BibitemShut
  {NoStop}%
%%CITATION = ARXIV:1210.4196;%%
\bibitem [{\citenamefont {Profumo}\ \emph {et~al.}(2015)\citenamefont
  {Profumo}, \citenamefont {Ramsey-Musolf}, \citenamefont {Wainwright},\ and\
  \citenamefont {Winslow}}]{Profumo:2014opa}%
  \BibitemOpen
  \bibfield  {author} {\bibinfo {author} {\bibfnamefont {S.}~\bibnamefont
  {Profumo}}, \bibinfo {author} {\bibfnamefont {M.~J.}\ \bibnamefont
  {Ramsey-Musolf}}, \bibinfo {author} {\bibfnamefont {C.~L.}\ \bibnamefont
  {Wainwright}}, \ and\ \bibinfo {author} {\bibfnamefont {P.}~\bibnamefont
  {Winslow}},\ }\href {\doibase 10.1103/PhysRevD.91.035018} {\bibfield
  {journal} {\bibinfo  {journal} {Phys. Rev.}\ }\textbf {\bibinfo {volume}
  {D91}},\ \bibinfo {pages} {035018} (\bibinfo {year} {2015})},\ \Eprint
  {http://arxiv.org/abs/1407.5342} {arXiv:1407.5342 [hep-ph]} \BibitemShut
  {NoStop}%
%%CITATION = ARXIV:1407.5342;%%
\bibitem [{\citenamefont {Beniwal}\ \emph {et~al.}(2019)\citenamefont
  {Beniwal}, \citenamefont {Lewicki}, \citenamefont {White},\ and\
  \citenamefont {Williams}}]{Beniwal:2018hyi}%
  \BibitemOpen
  \bibfield  {author} {\bibinfo {author} {\bibfnamefont {A.}~\bibnamefont
  {Beniwal}}, \bibinfo {author} {\bibfnamefont {M.}~\bibnamefont {Lewicki}},
  \bibinfo {author} {\bibfnamefont {M.}~\bibnamefont {White}}, \ and\ \bibinfo
  {author} {\bibfnamefont {A.~G.}\ \bibnamefont {Williams}},\ }\href {\doibase
  10.1007/JHEP02(2019)183} {\bibfield  {journal} {\bibinfo  {journal} {JHEP}\
  }\textbf {\bibinfo {volume} {02}},\ \bibinfo {pages} {183} (\bibinfo {year}
  {2019})},\ \Eprint {http://arxiv.org/abs/1810.02380} {arXiv:1810.02380
  [hep-ph]} \BibitemShut {NoStop}%
%%CITATION = ARXIV:1810.02380;%%
\bibitem [{\citenamefont {Kakizaki}\ \emph {et~al.}(2015)\citenamefont
  {Kakizaki}, \citenamefont {Kanemura},\ and\ \citenamefont
  {Matsui}}]{Kakizaki:2015wua}%
  \BibitemOpen
  \bibfield  {author} {\bibinfo {author} {\bibfnamefont {M.}~\bibnamefont
  {Kakizaki}}, \bibinfo {author} {\bibfnamefont {S.}~\bibnamefont {Kanemura}},
  \ and\ \bibinfo {author} {\bibfnamefont {T.}~\bibnamefont {Matsui}},\ }\href
  {\doibase 10.1103/PhysRevD.92.115007} {\bibfield  {journal} {\bibinfo
  {journal} {Phys. Rev.}\ }\textbf {\bibinfo {volume} {D92}},\ \bibinfo {pages}
  {115007} (\bibinfo {year} {2015})},\ \Eprint
  {http://arxiv.org/abs/1509.08394} {arXiv:1509.08394 [hep-ph]} \BibitemShut
  {NoStop}%
%%CITATION = ARXIV:1509.08394;%%
\bibitem [{\citenamefont {Dorsch}\ \emph {et~al.}(2017)\citenamefont {Dorsch},
  \citenamefont {Huber}, \citenamefont {Konstandin},\ and\ \citenamefont
  {No}}]{Dorsch:2016nrg}%
  \BibitemOpen
  \bibfield  {author} {\bibinfo {author} {\bibfnamefont {G.~C.}\ \bibnamefont
  {Dorsch}}, \bibinfo {author} {\bibfnamefont {S.~J.}\ \bibnamefont {Huber}},
  \bibinfo {author} {\bibfnamefont {T.}~\bibnamefont {Konstandin}}, \ and\
  \bibinfo {author} {\bibfnamefont {J.~M.}\ \bibnamefont {No}},\ }\href
  {\doibase 10.1088/1475-7516/2017/05/052} {\bibfield  {journal} {\bibinfo
  {journal} {JCAP}\ }\textbf {\bibinfo {volume} {1705}},\ \bibinfo {pages}
  {052} (\bibinfo {year} {2017})},\ \Eprint {http://arxiv.org/abs/1611.05874}
  {arXiv:1611.05874 [hep-ph]} \BibitemShut {NoStop}%
%%CITATION = ARXIV:1611.05874;%%
\bibitem [{\citenamefont {Basler}\ \emph {et~al.}(2017)\citenamefont {Basler},
  \citenamefont {Krause}, \citenamefont {Muhlleitner}, \citenamefont
  {Wittbrodt},\ and\ \citenamefont {Wlotzka}}]{Basler:2016obg}%
  \BibitemOpen
  \bibfield  {author} {\bibinfo {author} {\bibfnamefont {P.}~\bibnamefont
  {Basler}}, \bibinfo {author} {\bibfnamefont {M.}~\bibnamefont {Krause}},
  \bibinfo {author} {\bibfnamefont {M.}~\bibnamefont {Muhlleitner}}, \bibinfo
  {author} {\bibfnamefont {J.}~\bibnamefont {Wittbrodt}}, \ and\ \bibinfo
  {author} {\bibfnamefont {A.}~\bibnamefont {Wlotzka}},\ }\href {\doibase
  10.1007/JHEP02(2017)121} {\bibfield  {journal} {\bibinfo  {journal} {JHEP}\
  }\textbf {\bibinfo {volume} {02}},\ \bibinfo {pages} {121} (\bibinfo {year}
  {2017})},\ \Eprint {http://arxiv.org/abs/1612.04086} {arXiv:1612.04086
  [hep-ph]} \BibitemShut {NoStop}%
%%CITATION = ARXIV:1612.04086;%%
\bibitem [{\citenamefont {Randall}\ and\ \citenamefont
  {Servant}(2007)}]{Randall:2006py}%
  \BibitemOpen
  \bibfield  {author} {\bibinfo {author} {\bibfnamefont {L.}~\bibnamefont
  {Randall}}\ and\ \bibinfo {author} {\bibfnamefont {G.}~\bibnamefont
  {Servant}},\ }\href {\doibase 10.1088/1126-6708/2007/05/054} {\bibfield
  {journal} {\bibinfo  {journal} {JHEP}\ }\textbf {\bibinfo {volume} {05}},\
  \bibinfo {pages} {054} (\bibinfo {year} {2007})},\ \Eprint
  {http://arxiv.org/abs/hep-ph/0607158} {arXiv:hep-ph/0607158 [hep-ph]}
  \BibitemShut {NoStop}%
%%CITATION = HEP-PH/0607158;%%
\bibitem [{\citenamefont {Konstandin}\ \emph {et~al.}(2010)\citenamefont
  {Konstandin}, \citenamefont {Nardini},\ and\ \citenamefont
  {Quiros}}]{PhysRevD.82.083513}%
  \BibitemOpen
  \bibfield  {author} {\bibinfo {author} {\bibfnamefont {T.}~\bibnamefont
  {Konstandin}}, \bibinfo {author} {\bibfnamefont {G.}~\bibnamefont {Nardini}},
  \ and\ \bibinfo {author} {\bibfnamefont {M.}~\bibnamefont {Quiros}},\ }\href
  {\doibase 10.1103/PhysRevD.82.083513} {\bibfield  {journal} {\bibinfo
  {journal} {Phys. Rev. D}\ }\textbf {\bibinfo {volume} {82}},\ \bibinfo
  {pages} {083513} (\bibinfo {year} {2010})}\BibitemShut {NoStop}%
\bibitem [{\citenamefont {Konstandin}\ and\ \citenamefont
  {Servant}(2011)}]{Konstandin:2011dr}%
  \BibitemOpen
  \bibfield  {author} {\bibinfo {author} {\bibfnamefont {T.}~\bibnamefont
  {Konstandin}}\ and\ \bibinfo {author} {\bibfnamefont {G.}~\bibnamefont
  {Servant}},\ }\href {\doibase 10.1088/1475-7516/2011/12/009} {\bibfield
  {journal} {\bibinfo  {journal} {JCAP}\ }\textbf {\bibinfo {volume} {1112}},\
  \bibinfo {pages} {009} (\bibinfo {year} {2011})},\ \Eprint
  {http://arxiv.org/abs/1104.4791} {arXiv:1104.4791 [hep-ph]} \BibitemShut
  {NoStop}%
%%CITATION = ARXIV:1104.4791;%%
\bibitem [{\citenamefont {von Harling}\ and\ \citenamefont
  {Servant}(2018)}]{vonHarling:2017yew}%
  \BibitemOpen
  \bibfield  {author} {\bibinfo {author} {\bibfnamefont {B.}~\bibnamefont {von
  Harling}}\ and\ \bibinfo {author} {\bibfnamefont {G.}~\bibnamefont
  {Servant}},\ }\href {\doibase 10.1007/JHEP01(2018)159} {\bibfield  {journal}
  {\bibinfo  {journal} {JHEP}\ }\textbf {\bibinfo {volume} {01}},\ \bibinfo
  {pages} {159} (\bibinfo {year} {2018})},\ \Eprint
  {http://arxiv.org/abs/1711.11554} {arXiv:1711.11554 [hep-ph]} \BibitemShut
  {NoStop}%
%%CITATION = ARXIV:1711.11554;%%
\bibitem [{\citenamefont {Dillon}\ \emph {et~al.}(2018)\citenamefont {Dillon},
  \citenamefont {El-Menoufi}, \citenamefont {Huber},\ and\ \citenamefont
  {Manuel}}]{Dillon:2017ctw}%
  \BibitemOpen
  \bibfield  {author} {\bibinfo {author} {\bibfnamefont {B.~M.}\ \bibnamefont
  {Dillon}}, \bibinfo {author} {\bibfnamefont {B.~K.}\ \bibnamefont
  {El-Menoufi}}, \bibinfo {author} {\bibfnamefont {S.~J.}\ \bibnamefont
  {Huber}}, \ and\ \bibinfo {author} {\bibfnamefont {J.~P.}\ \bibnamefont
  {Manuel}},\ }\href {\doibase 10.1103/PhysRevD.98.086005} {\bibfield
  {journal} {\bibinfo  {journal} {Phys. Rev.}\ }\textbf {\bibinfo {volume}
  {D98}},\ \bibinfo {pages} {086005} (\bibinfo {year} {2018})},\ \Eprint
  {http://arxiv.org/abs/1708.02953} {arXiv:1708.02953 [hep-th]} \BibitemShut
  {NoStop}%
%%CITATION = ARXIV:1708.02953;%%
\bibitem [{\citenamefont {Meg{\'\i}as}\ \emph {et~al.}(2018)\citenamefont
  {Meg{\'\i}as}, \citenamefont {Nardini},\ and\ \citenamefont
  {Quir{\'o}s}}]{Megias:2018sxv}%
  \BibitemOpen
  \bibfield  {author} {\bibinfo {author} {\bibfnamefont {E.}~\bibnamefont
  {Meg{\'\i}as}}, \bibinfo {author} {\bibfnamefont {G.}~\bibnamefont
  {Nardini}}, \ and\ \bibinfo {author} {\bibfnamefont {M.}~\bibnamefont
  {Quir{\'o}s}},\ }\href {\doibase 10.1007/JHEP09(2018)095} {\bibfield
  {journal} {\bibinfo  {journal} {JHEP}\ }\textbf {\bibinfo {volume} {09}},\
  \bibinfo {pages} {095} (\bibinfo {year} {2018})},\ \Eprint
  {http://arxiv.org/abs/1806.04877} {arXiv:1806.04877 [hep-ph]} \BibitemShut
  {NoStop}%
%%CITATION = ARXIV:1806.04877;%%
\bibitem [{\citenamefont {Bruggisser}\ \emph {et~al.}(2018)\citenamefont
  {Bruggisser}, \citenamefont {Von~Harling}, \citenamefont {Matsedonskyi},\
  and\ \citenamefont {Servant}}]{Bruggisser:2018mrt}%
  \BibitemOpen
  \bibfield  {author} {\bibinfo {author} {\bibfnamefont {S.}~\bibnamefont
  {Bruggisser}}, \bibinfo {author} {\bibfnamefont {B.}~\bibnamefont
  {Von~Harling}}, \bibinfo {author} {\bibfnamefont {O.}~\bibnamefont
  {Matsedonskyi}}, \ and\ \bibinfo {author} {\bibfnamefont {G.}~\bibnamefont
  {Servant}},\ }\href {\doibase 10.1007/JHEP12(2018)099} {\bibfield  {journal}
  {\bibinfo  {journal} {JHEP}\ }\textbf {\bibinfo {volume} {12}},\ \bibinfo
  {pages} {099} (\bibinfo {year} {2018})},\ \Eprint
  {http://arxiv.org/abs/1804.07314} {arXiv:1804.07314 [hep-ph]} \BibitemShut
  {NoStop}%
%%CITATION = ARXIV:1804.07314;%%
\bibitem [{\citenamefont {Schwaller}(2015)}]{Schwaller:2015tja}%
  \BibitemOpen
  \bibfield  {author} {\bibinfo {author} {\bibfnamefont {P.}~\bibnamefont
  {Schwaller}},\ }\href {\doibase 10.1103/PhysRevLett.115.181101} {\bibfield
  {journal} {\bibinfo  {journal} {Phys. Rev. Lett.}\ }\textbf {\bibinfo
  {volume} {115}},\ \bibinfo {pages} {181101} (\bibinfo {year} {2015})},\
  \Eprint {http://arxiv.org/abs/1504.07263} {arXiv:1504.07263 [hep-ph]}
  \BibitemShut {NoStop}%
%%CITATION = ARXIV:1504.07263;%%
\bibitem [{\citenamefont {Addazi}\ and\ \citenamefont
  {Marciano}(2018)}]{Addazi:2017gpt}%
  \BibitemOpen
  \bibfield  {author} {\bibinfo {author} {\bibfnamefont {A.}~\bibnamefont
  {Addazi}}\ and\ \bibinfo {author} {\bibfnamefont {A.}~\bibnamefont
  {Marciano}},\ }\href {\doibase 10.1088/1674-1137/42/2/023107} {\bibfield
  {journal} {\bibinfo  {journal} {Chin. Phys.}\ }\textbf {\bibinfo {volume}
  {C42}},\ \bibinfo {pages} {023107} (\bibinfo {year} {2018})},\ \Eprint
  {http://arxiv.org/abs/1703.03248} {arXiv:1703.03248 [hep-ph]} \BibitemShut
  {NoStop}%
%%CITATION = ARXIV:1703.03248;%%
\bibitem [{\citenamefont {Aoki}\ \emph {et~al.}(2017)\citenamefont {Aoki},
  \citenamefont {Goto},\ and\ \citenamefont {Kubo}}]{Aoki:2017aws}%
  \BibitemOpen
  \bibfield  {author} {\bibinfo {author} {\bibfnamefont {M.}~\bibnamefont
  {Aoki}}, \bibinfo {author} {\bibfnamefont {H.}~\bibnamefont {Goto}}, \ and\
  \bibinfo {author} {\bibfnamefont {J.}~\bibnamefont {Kubo}},\ }\href {\doibase
  10.1103/PhysRevD.96.075045} {\bibfield  {journal} {\bibinfo  {journal} {Phys.
  Rev.}\ }\textbf {\bibinfo {volume} {D96}},\ \bibinfo {pages} {075045}
  (\bibinfo {year} {2017})},\ \Eprint {http://arxiv.org/abs/1709.07572}
  {arXiv:1709.07572 [hep-ph]} \BibitemShut {NoStop}%
%%CITATION = ARXIV:1709.07572;%%
\bibitem [{\citenamefont {Croon}\ \emph {et~al.}(2018)\citenamefont {Croon},
  \citenamefont {Sanz},\ and\ \citenamefont {White}}]{Croon:2018erz}%
  \BibitemOpen
  \bibfield  {author} {\bibinfo {author} {\bibfnamefont {D.}~\bibnamefont
  {Croon}}, \bibinfo {author} {\bibfnamefont {V.}~\bibnamefont {Sanz}}, \ and\
  \bibinfo {author} {\bibfnamefont {G.}~\bibnamefont {White}},\ }\href
  {\doibase 10.1007/JHEP08(2018)203} {\bibfield  {journal} {\bibinfo  {journal}
  {JHEP}\ }\textbf {\bibinfo {volume} {08}},\ \bibinfo {pages} {203} (\bibinfo
  {year} {2018})},\ \Eprint {http://arxiv.org/abs/1806.02332} {arXiv:1806.02332
  [hep-ph]} \BibitemShut {NoStop}%
%%CITATION = ARXIV:1806.02332;%%
\bibitem [{\citenamefont {Breitbach}\ \emph {et~al.}(2018)\citenamefont
  {Breitbach}, \citenamefont {Kopp}, \citenamefont {Madge}, \citenamefont
  {Opferkuch},\ and\ \citenamefont {Schwaller}}]{Breitbach:2018ddu}%
  \BibitemOpen
  \bibfield  {author} {\bibinfo {author} {\bibfnamefont {M.}~\bibnamefont
  {Breitbach}}, \bibinfo {author} {\bibfnamefont {J.}~\bibnamefont {Kopp}},
  \bibinfo {author} {\bibfnamefont {E.}~\bibnamefont {Madge}}, \bibinfo
  {author} {\bibfnamefont {T.}~\bibnamefont {Opferkuch}}, \ and\ \bibinfo
  {author} {\bibfnamefont {P.}~\bibnamefont {Schwaller}},\ }\href@noop {} {\
  (\bibinfo {year} {2018})},\ \Eprint {http://arxiv.org/abs/1811.11175}
  {arXiv:1811.11175 [hep-ph]} \BibitemShut {NoStop}%
%%CITATION = ARXIV:1811.11175;%%
\bibitem [{\citenamefont {Okada}\ and\ \citenamefont
  {Seto}(2018)}]{Okada:2018xdh}%
  \BibitemOpen
  \bibfield  {author} {\bibinfo {author} {\bibfnamefont {N.}~\bibnamefont
  {Okada}}\ and\ \bibinfo {author} {\bibfnamefont {O.}~\bibnamefont {Seto}},\
  }\href {\doibase 10.1103/PhysRevD.98.063532} {\bibfield  {journal} {\bibinfo
  {journal} {Phys. Rev.}\ }\textbf {\bibinfo {volume} {D98}},\ \bibinfo {pages}
  {063532} (\bibinfo {year} {2018})},\ \Eprint
  {http://arxiv.org/abs/1807.00336} {arXiv:1807.00336 [hep-ph]} \BibitemShut
  {NoStop}%
%%CITATION = ARXIV:1807.00336;%%
\bibitem [{\citenamefont {Hasegawa}\ \emph {et~al.}(2019)\citenamefont
  {Hasegawa}, \citenamefont {Okada},\ and\ \citenamefont
  {Seto}}]{Hasegawa:2019amx}%
  \BibitemOpen
  \bibfield  {author} {\bibinfo {author} {\bibfnamefont {T.}~\bibnamefont
  {Hasegawa}}, \bibinfo {author} {\bibfnamefont {N.}~\bibnamefont {Okada}}, \
  and\ \bibinfo {author} {\bibfnamefont {O.}~\bibnamefont {Seto}},\ }\href
  {\doibase 10.1103/PhysRevD.99.095039} {\bibfield  {journal} {\bibinfo
  {journal} {Phys. Rev.}\ }\textbf {\bibinfo {volume} {D99}},\ \bibinfo {pages}
  {095039} (\bibinfo {year} {2019})},\ \Eprint
  {http://arxiv.org/abs/1904.03020} {arXiv:1904.03020 [hep-ph]} \BibitemShut
  {NoStop}%
%%CITATION = ARXIV:1904.03020;%%
\bibitem [{\citenamefont {Hall}\ \emph {et~al.}(2019)\citenamefont {Hall},
  \citenamefont {Konstandin}, \citenamefont {McGehee},\ and\ \citenamefont
  {Murayama}}]{Hall:2019rld}%
  \BibitemOpen
  \bibfield  {author} {\bibinfo {author} {\bibfnamefont {E.}~\bibnamefont
  {Hall}}, \bibinfo {author} {\bibfnamefont {T.}~\bibnamefont {Konstandin}},
  \bibinfo {author} {\bibfnamefont {R.}~\bibnamefont {McGehee}}, \ and\
  \bibinfo {author} {\bibfnamefont {H.}~\bibnamefont {Murayama}},\ }\href@noop
  {} {\  (\bibinfo {year} {2019})},\ \Eprint {http://arxiv.org/abs/1911.12342}
  {arXiv:1911.12342 [hep-ph]} \BibitemShut {NoStop}%
\bibitem [{\citenamefont {Hall}\ \emph {et~al.}(2020)\citenamefont {Hall},
  \citenamefont {Konstandin}, \citenamefont {McGehee}, \citenamefont
  {Murayama},\ and\ \citenamefont {Servant}}]{Hall:2019ank}%
  \BibitemOpen
  \bibfield  {author} {\bibinfo {author} {\bibfnamefont {E.}~\bibnamefont
  {Hall}}, \bibinfo {author} {\bibfnamefont {T.}~\bibnamefont {Konstandin}},
  \bibinfo {author} {\bibfnamefont {R.}~\bibnamefont {McGehee}}, \bibinfo
  {author} {\bibfnamefont {H.}~\bibnamefont {Murayama}}, \ and\ \bibinfo
  {author} {\bibfnamefont {G.}~\bibnamefont {Servant}},\ }\href {\doibase
  10.1007/JHEP04(2020)042} {\bibfield  {journal} {\bibinfo  {journal} {JHEP}\
  }\textbf {\bibinfo {volume} {04}},\ \bibinfo {pages} {042} (\bibinfo {year}
  {2020})},\ \Eprint {http://arxiv.org/abs/1910.08068} {arXiv:1910.08068
  [hep-ph]} \BibitemShut {NoStop}%
\bibitem [{\citenamefont {Coleman}(1977)}]{Coleman:1977py}%
  \BibitemOpen
  \bibfield  {author} {\bibinfo {author} {\bibfnamefont {S.~R.}\ \bibnamefont
  {Coleman}},\ }\href {\doibase 10.1103/PhysRevD.15.2929} {\bibfield  {journal}
  {\bibinfo  {journal} {Phys. Rev.}\ }\textbf {\bibinfo {volume} {D15}},\
  \bibinfo {pages} {2929} (\bibinfo {year} {1977})},\ \bibinfo {note}
  {[Erratum: Phys. Rev.D16,1248(1977)]}\BibitemShut {NoStop}%
%%CITATION = PHRVA,D15,2929;%%
\bibitem [{\citenamefont {Linde}(1983)}]{Linde:1981zj}%
  \BibitemOpen
  \bibfield  {author} {\bibinfo {author} {\bibfnamefont {A.~D.}\ \bibnamefont
  {Linde}},\ }\href {\doibase 10.1016/0550-3213(83)90293-6} {\bibfield
  {journal} {\bibinfo  {journal} {Nucl.Phys.}\ }\textbf {\bibinfo {volume}
  {B216}},\ \bibinfo {pages} {421} (\bibinfo {year} {1983})}\BibitemShut
  {NoStop}%
%%CITATION = NUPHA,B216,421;%%
\bibitem [{\citenamefont {Steinhardt}(1982)}]{Steinhardt:1981ct}%
  \BibitemOpen
  \bibfield  {author} {\bibinfo {author} {\bibfnamefont {P.~J.}\ \bibnamefont
  {Steinhardt}},\ }\href {\doibase 10.1103/PhysRevD.25.2074} {\bibfield
  {journal} {\bibinfo  {journal} {Phys.Rev.}\ }\textbf {\bibinfo {volume}
  {D25}},\ \bibinfo {pages} {2074} (\bibinfo {year} {1982})}\BibitemShut
  {NoStop}%
%%CITATION = PHRVA,D25,2074;%%
\bibitem [{\citenamefont {Witten}(1984)}]{Witten:1984rs}%
  \BibitemOpen
  \bibfield  {author} {\bibinfo {author} {\bibfnamefont {E.}~\bibnamefont
  {Witten}},\ }\href {\doibase 10.1103/PhysRevD.30.272} {\bibfield  {journal}
  {\bibinfo  {journal} {Phys.Rev.}\ }\textbf {\bibinfo {volume} {D30}},\
  \bibinfo {pages} {272} (\bibinfo {year} {1984})}\BibitemShut {NoStop}%
%%CITATION = PHRVA,D30,272;%%
\bibitem [{\citenamefont {{Hogan}}(1986)}]{1986MNRAS.218..629H}%
  \BibitemOpen
  \bibfield  {author} {\bibinfo {author} {\bibfnamefont {C.~J.}\ \bibnamefont
  {{Hogan}}},\ }\href@noop {} {\bibfield  {journal} {\bibinfo  {journal}
  {MNRAS}\ }\textbf {\bibinfo {volume} {218}},\ \bibinfo {pages} {629}
  (\bibinfo {year} {1986})}\BibitemShut {NoStop}%
\bibitem [{\citenamefont {Kurki-Suonio}(1985)}]{KurkiSuonio:1984ba}%
  \BibitemOpen
  \bibfield  {author} {\bibinfo {author} {\bibfnamefont {H.}~\bibnamefont
  {Kurki-Suonio}},\ }\href {\doibase 10.1016/0550-3213(85)90135-X} {\bibfield
  {journal} {\bibinfo  {journal} {Nucl.Phys.}\ }\textbf {\bibinfo {volume}
  {B255}},\ \bibinfo {pages} {231} (\bibinfo {year} {1985})}\BibitemShut
  {NoStop}%
%%CITATION = NUPHA,B255,231;%%
\bibitem [{\citenamefont {Bodeker}\ and\ \citenamefont
  {Moore}(2009)}]{Bodeker:2009qy}%
  \BibitemOpen
  \bibfield  {author} {\bibinfo {author} {\bibfnamefont {D.}~\bibnamefont
  {Bodeker}}\ and\ \bibinfo {author} {\bibfnamefont {G.~D.}\ \bibnamefont
  {Moore}},\ }\href {\doibase 10.1088/1475-7516/2009/05/009} {\bibfield
  {journal} {\bibinfo  {journal} {JCAP}\ }\textbf {\bibinfo {volume} {0905}},\
  \bibinfo {pages} {009} (\bibinfo {year} {2009})},\ \Eprint
  {http://arxiv.org/abs/0903.4099} {arXiv:0903.4099 [hep-ph]} \BibitemShut
  {NoStop}%
%%CITATION = ARXIV:0903.4099;%%
\bibitem [{\citenamefont {Bodeker}\ and\ \citenamefont
  {Moore}(2017)}]{Bodeker:2017cim}%
  \BibitemOpen
  \bibfield  {author} {\bibinfo {author} {\bibfnamefont {D.}~\bibnamefont
  {Bodeker}}\ and\ \bibinfo {author} {\bibfnamefont {G.~D.}\ \bibnamefont
  {Moore}},\ }\href@noop {} {\  (\bibinfo {year} {2017})},\ \Eprint
  {http://arxiv.org/abs/1703.08215} {arXiv:1703.08215 [hep-ph]} \BibitemShut
  {NoStop}%
%%CITATION = ARXIV:1703.08215;%%
\bibitem [{\citenamefont {Ellis}\ \emph
  {et~al.}(2019{\natexlab{a}})\citenamefont {Ellis}, \citenamefont {Lewicki},
  \citenamefont {No},\ and\ \citenamefont {Vaskonen}}]{Ellis:2019oqb}%
  \BibitemOpen
  \bibfield  {author} {\bibinfo {author} {\bibfnamefont {J.}~\bibnamefont
  {Ellis}}, \bibinfo {author} {\bibfnamefont {M.}~\bibnamefont {Lewicki}},
  \bibinfo {author} {\bibfnamefont {J.~M.}\ \bibnamefont {No}}, \ and\ \bibinfo
  {author} {\bibfnamefont {V.}~\bibnamefont {Vaskonen}},\ }\href@noop {} {\
  (\bibinfo {year} {2019}{\natexlab{a}})},\ \Eprint
  {http://arxiv.org/abs/1903.09642} {arXiv:1903.09642 [hep-ph]} \BibitemShut
  {NoStop}%
%%CITATION = ARXIV:1903.09642;%%
\bibitem [{\citenamefont {Ellis}\ \emph
  {et~al.}(2019{\natexlab{b}})\citenamefont {Ellis}, \citenamefont {Lewicki},\
  and\ \citenamefont {No}}]{Ellis:2018mja}%
  \BibitemOpen
  \bibfield  {author} {\bibinfo {author} {\bibfnamefont {J.}~\bibnamefont
  {Ellis}}, \bibinfo {author} {\bibfnamefont {M.}~\bibnamefont {Lewicki}}, \
  and\ \bibinfo {author} {\bibfnamefont {J.~M.}\ \bibnamefont {No}},\ }\href
  {\doibase 10.1088/1475-7516/2019/04/003} {\bibfield  {journal} {\bibinfo
  {journal} {JCAP}\ }\textbf {\bibinfo {volume} {04}},\ \bibinfo {pages} {003}
  (\bibinfo {year} {2019}{\natexlab{b}})},\ \Eprint
  {http://arxiv.org/abs/1809.08242} {arXiv:1809.08242 [hep-ph]} \BibitemShut
  {NoStop}%
\bibitem [{\citenamefont {Kosowsky}\ and\ \citenamefont
  {Turner}(1993)}]{Kosowsky:1992vn}%
  \BibitemOpen
  \bibfield  {author} {\bibinfo {author} {\bibfnamefont {A.}~\bibnamefont
  {Kosowsky}}\ and\ \bibinfo {author} {\bibfnamefont {M.~S.}\ \bibnamefont
  {Turner}},\ }\href {\doibase 10.1103/PhysRevD.47.4372} {\bibfield  {journal}
  {\bibinfo  {journal} {Phys.Rev.}\ }\textbf {\bibinfo {volume} {D47}},\
  \bibinfo {pages} {4372} (\bibinfo {year} {1993})},\ \Eprint
  {http://arxiv.org/abs/astro-ph/9211004} {arXiv:astro-ph/9211004 [astro-ph]}
  \BibitemShut {NoStop}%
%%CITATION = ASTRO-PH/9211004;%%
\bibitem [{\citenamefont {Kamionkowski}\ \emph {et~al.}(1994)\citenamefont
  {Kamionkowski}, \citenamefont {Kosowsky},\ and\ \citenamefont
  {Turner}}]{Kamionkowski:1993fg}%
  \BibitemOpen
  \bibfield  {author} {\bibinfo {author} {\bibfnamefont {M.}~\bibnamefont
  {Kamionkowski}}, \bibinfo {author} {\bibfnamefont {A.}~\bibnamefont
  {Kosowsky}}, \ and\ \bibinfo {author} {\bibfnamefont {M.~S.}\ \bibnamefont
  {Turner}},\ }\href {\doibase 10.1103/PhysRevD.49.2837} {\bibfield  {journal}
  {\bibinfo  {journal} {Phys.Rev.}\ }\textbf {\bibinfo {volume} {D49}},\
  \bibinfo {pages} {2837} (\bibinfo {year} {1994})},\ \Eprint
  {http://arxiv.org/abs/astro-ph/9310044} {arXiv:astro-ph/9310044 [astro-ph]}
  \BibitemShut {NoStop}%
%%CITATION = ASTRO-PH/9310044;%%
\bibitem [{\citenamefont {Huber}\ and\ \citenamefont
  {Konstandin}(2008)}]{Huber:2008}%
  \BibitemOpen
  \bibfield  {author} {\bibinfo {author} {\bibfnamefont {S.~J.}\ \bibnamefont
  {Huber}}\ and\ \bibinfo {author} {\bibfnamefont {T.}~\bibnamefont
  {Konstandin}},\ }\href {\doibase 10.1088/1475-7516/2008/09/022} {\bibfield
  {journal} {\bibinfo  {journal} {JCAP}\ }\textbf {\bibinfo {volume} {0809}},\
  \bibinfo {pages} {022} (\bibinfo {year} {2008})},\ \Eprint
  {http://arxiv.org/abs/0806.1828} {arXiv:0806.1828 [hep-ph]} \BibitemShut
  {NoStop}%
%%CITATION = ARXIV:0806.1828;%%
\bibitem [{\citenamefont {Konstandin}(2018)}]{Konstandin:2017sat}%
  \BibitemOpen
  \bibfield  {author} {\bibinfo {author} {\bibfnamefont {T.}~\bibnamefont
  {Konstandin}},\ }\href {\doibase 10.1088/1475-7516/2018/03/047} {\bibfield
  {journal} {\bibinfo  {journal} {JCAP}\ }\textbf {\bibinfo {volume} {1803}},\
  \bibinfo {pages} {047} (\bibinfo {year} {2018})},\ \Eprint
  {http://arxiv.org/abs/1712.06869} {arXiv:1712.06869 [astro-ph.CO]}
  \BibitemShut {NoStop}%
%%CITATION = ARXIV:1712.06869;%%
\bibitem [{\citenamefont {Jinno}\ and\ \citenamefont
  {Takimoto}(2017)}]{Jinno:2016vai}%
  \BibitemOpen
  \bibfield  {author} {\bibinfo {author} {\bibfnamefont {R.}~\bibnamefont
  {Jinno}}\ and\ \bibinfo {author} {\bibfnamefont {M.}~\bibnamefont
  {Takimoto}},\ }\href {\doibase 10.1103/PhysRevD.95.024009} {\bibfield
  {journal} {\bibinfo  {journal} {Phys. Rev.}\ }\textbf {\bibinfo {volume}
  {D95}},\ \bibinfo {pages} {024009} (\bibinfo {year} {2017})},\ \Eprint
  {http://arxiv.org/abs/1605.01403} {arXiv:1605.01403 [astro-ph.CO]}
  \BibitemShut {NoStop}%
%%CITATION = ARXIV:1605.01403;%%
\bibitem [{\citenamefont {Jinno}\ and\ \citenamefont
  {Takimoto}(2019)}]{Jinno:2017fby}%
  \BibitemOpen
  \bibfield  {author} {\bibinfo {author} {\bibfnamefont {R.}~\bibnamefont
  {Jinno}}\ and\ \bibinfo {author} {\bibfnamefont {M.}~\bibnamefont
  {Takimoto}},\ }\href {\doibase 10.1088/1475-7516/2019/01/060} {\bibfield
  {journal} {\bibinfo  {journal} {JCAP}\ }\textbf {\bibinfo {volume} {01}},\
  \bibinfo {pages} {060} (\bibinfo {year} {2019})},\ \Eprint
  {http://arxiv.org/abs/1707.03111} {arXiv:1707.03111 [hep-ph]} \BibitemShut
  {NoStop}%
\bibitem [{\citenamefont {Hindmarsh}\ \emph {et~al.}(2014)\citenamefont
  {Hindmarsh}, \citenamefont {Huber}, \citenamefont {Rummukainen},\ and\
  \citenamefont {Weir}}]{Hindmarsh:2013xza}%
  \BibitemOpen
  \bibfield  {author} {\bibinfo {author} {\bibfnamefont {M.}~\bibnamefont
  {Hindmarsh}}, \bibinfo {author} {\bibfnamefont {S.~J.}\ \bibnamefont
  {Huber}}, \bibinfo {author} {\bibfnamefont {K.}~\bibnamefont {Rummukainen}},
  \ and\ \bibinfo {author} {\bibfnamefont {D.~J.}\ \bibnamefont {Weir}},\
  }\href {\doibase 10.1103/PhysRevLett.112.041301} {\bibfield  {journal}
  {\bibinfo  {journal} {Phys.Rev.Lett.}\ }\textbf {\bibinfo {volume} {112}},\
  \bibinfo {pages} {041301} (\bibinfo {year} {2014})},\ \Eprint
  {http://arxiv.org/abs/1304.2433} {arXiv:1304.2433 [hep-ph]} \BibitemShut
  {NoStop}%
%%CITATION = ARXIV:1304.2433;%%
\bibitem [{\citenamefont {Giblin}\ and\ \citenamefont
  {Mertens}(2014)}]{Giblin:2014qia}%
  \BibitemOpen
  \bibfield  {author} {\bibinfo {author} {\bibfnamefont {J.~T.}\ \bibnamefont
  {Giblin}}\ and\ \bibinfo {author} {\bibfnamefont {J.~B.}\ \bibnamefont
  {Mertens}},\ }\href {\doibase 10.1103/PhysRevD.90.023532} {\bibfield
  {journal} {\bibinfo  {journal} {Phys.Rev.}\ }\textbf {\bibinfo {volume}
  {D90}},\ \bibinfo {pages} {023532} (\bibinfo {year} {2014})},\ \Eprint
  {http://arxiv.org/abs/1405.4005} {arXiv:1405.4005 [astro-ph.CO]} \BibitemShut
  {NoStop}%
%%CITATION = ARXIV:1405.4005;%%
\bibitem [{\citenamefont {Hindmarsh}\ \emph {et~al.}(2015)\citenamefont
  {Hindmarsh}, \citenamefont {Huber}, \citenamefont {Rummukainen},\ and\
  \citenamefont {Weir}}]{Hindmarsh:2015qta}%
  \BibitemOpen
  \bibfield  {author} {\bibinfo {author} {\bibfnamefont {M.}~\bibnamefont
  {Hindmarsh}}, \bibinfo {author} {\bibfnamefont {S.~J.}\ \bibnamefont
  {Huber}}, \bibinfo {author} {\bibfnamefont {K.}~\bibnamefont {Rummukainen}},
  \ and\ \bibinfo {author} {\bibfnamefont {D.~J.}\ \bibnamefont {Weir}},\
  }\href {\doibase 10.1103/PhysRevD.92.123009} {\bibfield  {journal} {\bibinfo
  {journal} {Phys. Rev.}\ }\textbf {\bibinfo {volume} {D92}},\ \bibinfo {pages}
  {123009} (\bibinfo {year} {2015})},\ \Eprint
  {http://arxiv.org/abs/1504.03291} {arXiv:1504.03291 [astro-ph.CO]}
  \BibitemShut {NoStop}%
%%CITATION = ARXIV:1504.03291;%%
\bibitem [{\citenamefont {Hindmarsh}\ \emph {et~al.}(2017)\citenamefont
  {Hindmarsh}, \citenamefont {Huber}, \citenamefont {Rummukainen},\ and\
  \citenamefont {Weir}}]{Hindmarsh:2017gnf}%
  \BibitemOpen
  \bibfield  {author} {\bibinfo {author} {\bibfnamefont {M.}~\bibnamefont
  {Hindmarsh}}, \bibinfo {author} {\bibfnamefont {S.~J.}\ \bibnamefont
  {Huber}}, \bibinfo {author} {\bibfnamefont {K.}~\bibnamefont {Rummukainen}},
  \ and\ \bibinfo {author} {\bibfnamefont {D.~J.}\ \bibnamefont {Weir}},\
  }\href {\doibase 10.1103/PhysRevD.96.103520} {\bibfield  {journal} {\bibinfo
  {journal} {Phys. Rev.}\ }\textbf {\bibinfo {volume} {D96}},\ \bibinfo {pages}
  {103520} (\bibinfo {year} {2017})},\ \Eprint
  {http://arxiv.org/abs/1704.05871} {arXiv:1704.05871 [astro-ph.CO]}
  \BibitemShut {NoStop}%
%%CITATION = ARXIV:1704.05871;%%
\bibitem [{\citenamefont {Weir}(2016)}]{Weir:2016tov}%
  \BibitemOpen
  \bibfield  {author} {\bibinfo {author} {\bibfnamefont {D.~J.}\ \bibnamefont
  {Weir}},\ }\href {\doibase 10.1103/PhysRevD.93.124037} {\bibfield  {journal}
  {\bibinfo  {journal} {Phys. Rev.}\ }\textbf {\bibinfo {volume} {D93}},\
  \bibinfo {pages} {124037} (\bibinfo {year} {2016})},\ \Eprint
  {http://arxiv.org/abs/1604.08429} {arXiv:1604.08429 [astro-ph.CO]}
  \BibitemShut {NoStop}%
%%CITATION = ARXIV:1604.08429;%%
\bibitem [{\citenamefont {Hindmarsh}(2018)}]{Hindmarsh:2016lnk}%
  \BibitemOpen
  \bibfield  {author} {\bibinfo {author} {\bibfnamefont {M.}~\bibnamefont
  {Hindmarsh}},\ }\href {\doibase 10.1103/PhysRevLett.120.071301} {\bibfield
  {journal} {\bibinfo  {journal} {Phys. Rev. Lett.}\ }\textbf {\bibinfo
  {volume} {120}},\ \bibinfo {pages} {071301} (\bibinfo {year} {2018})},\
  \Eprint {http://arxiv.org/abs/1608.04735} {arXiv:1608.04735 [astro-ph.CO]}
  \BibitemShut {NoStop}%
%%CITATION = ARXIV:1608.04735;%%
\bibitem [{\citenamefont {Hindmarsh}\ and\ \citenamefont
  {Hijazi}(2019)}]{Hindmarsh:2019phv}%
  \BibitemOpen
  \bibfield  {author} {\bibinfo {author} {\bibfnamefont {M.}~\bibnamefont
  {Hindmarsh}}\ and\ \bibinfo {author} {\bibfnamefont {M.}~\bibnamefont
  {Hijazi}},\ }\href {\doibase 10.1088/1475-7516/2019/12/062} {\bibfield
  {journal} {\bibinfo  {journal} {JCAP}\ }\textbf {\bibinfo {volume} {12}},\
  \bibinfo {pages} {062} (\bibinfo {year} {2019})},\ \Eprint
  {http://arxiv.org/abs/1909.10040} {arXiv:1909.10040 [astro-ph.CO]}
  \BibitemShut {NoStop}%
\bibitem [{\citenamefont {Cutting}\ \emph {et~al.}(2019)\citenamefont
  {Cutting}, \citenamefont {Hindmarsh},\ and\ \citenamefont
  {Weir}}]{Cutting:2019zws}%
  \BibitemOpen
  \bibfield  {author} {\bibinfo {author} {\bibfnamefont {D.}~\bibnamefont
  {Cutting}}, \bibinfo {author} {\bibfnamefont {M.}~\bibnamefont {Hindmarsh}},
  \ and\ \bibinfo {author} {\bibfnamefont {D.~J.}\ \bibnamefont {Weir}},\
  }\href@noop {} {\  (\bibinfo {year} {2019})},\ \Eprint
  {http://arxiv.org/abs/1906.00480} {arXiv:1906.00480 [hep-ph]} \BibitemShut
  {NoStop}%
\bibitem [{\citenamefont {Jinno}\ \emph
  {et~al.}(2019{\natexlab{a}})\citenamefont {Jinno}, \citenamefont {Seong},
  \citenamefont {Takimoto},\ and\ \citenamefont {Um}}]{Jinno:2019jhi}%
  \BibitemOpen
  \bibfield  {author} {\bibinfo {author} {\bibfnamefont {R.}~\bibnamefont
  {Jinno}}, \bibinfo {author} {\bibfnamefont {H.}~\bibnamefont {Seong}},
  \bibinfo {author} {\bibfnamefont {M.}~\bibnamefont {Takimoto}}, \ and\
  \bibinfo {author} {\bibfnamefont {C.~M.}\ \bibnamefont {Um}},\ }\href
  {\doibase 10.1088/1475-7516/2019/10/033} {\bibfield  {journal} {\bibinfo
  {journal} {JCAP}\ }\textbf {\bibinfo {volume} {10}},\ \bibinfo {pages} {033}
  (\bibinfo {year} {2019}{\natexlab{a}})},\ \Eprint
  {http://arxiv.org/abs/1905.00899} {arXiv:1905.00899 [astro-ph.CO]}
  \BibitemShut {NoStop}%
\bibitem [{\citenamefont {Ellis}\ \emph {et~al.}(2020)\citenamefont {Ellis},
  \citenamefont {Lewicki},\ and\ \citenamefont {No}}]{Ellis:2020awk}%
  \BibitemOpen
  \bibfield  {author} {\bibinfo {author} {\bibfnamefont {J.}~\bibnamefont
  {Ellis}}, \bibinfo {author} {\bibfnamefont {M.}~\bibnamefont {Lewicki}}, \
  and\ \bibinfo {author} {\bibfnamefont {J.~M.}\ \bibnamefont {No}},\
  }\href@noop {} {\  (\bibinfo {year} {2020})},\ \Eprint
  {http://arxiv.org/abs/2003.07360} {arXiv:2003.07360 [hep-ph]} \BibitemShut
  {NoStop}%
\bibitem [{\citenamefont {Caprini}\ \emph {et~al.}(2008)\citenamefont
  {Caprini}, \citenamefont {Durrer},\ and\ \citenamefont
  {Servant}}]{Caprini:2007xq}%
  \BibitemOpen
  \bibfield  {author} {\bibinfo {author} {\bibfnamefont {C.}~\bibnamefont
  {Caprini}}, \bibinfo {author} {\bibfnamefont {R.}~\bibnamefont {Durrer}}, \
  and\ \bibinfo {author} {\bibfnamefont {G.}~\bibnamefont {Servant}},\ }\href
  {\doibase 10.1103/PhysRevD.77.124015} {\bibfield  {journal} {\bibinfo
  {journal} {Phys.Rev.}\ }\textbf {\bibinfo {volume} {D77}},\ \bibinfo {pages}
  {124015} (\bibinfo {year} {2008})},\ \Eprint {http://arxiv.org/abs/0711.2593}
  {arXiv:0711.2593 [astro-ph]} \BibitemShut {NoStop}%
%%CITATION = ARXIV:0711.2593;%%
\bibitem [{\citenamefont {Gogoberidze}\ \emph {et~al.}(2007)\citenamefont
  {Gogoberidze}, \citenamefont {Kahniashvili},\ and\ \citenamefont
  {Kosowsky}}]{Gogoberidze:2007an}%
  \BibitemOpen
  \bibfield  {author} {\bibinfo {author} {\bibfnamefont {G.}~\bibnamefont
  {Gogoberidze}}, \bibinfo {author} {\bibfnamefont {T.}~\bibnamefont
  {Kahniashvili}}, \ and\ \bibinfo {author} {\bibfnamefont {A.}~\bibnamefont
  {Kosowsky}},\ }\href {\doibase 10.1103/PhysRevD.76.083002} {\bibfield
  {journal} {\bibinfo  {journal} {Phys.Rev.}\ }\textbf {\bibinfo {volume}
  {D76}},\ \bibinfo {pages} {083002} (\bibinfo {year} {2007})},\ \Eprint
  {http://arxiv.org/abs/0705.1733} {arXiv:0705.1733 [astro-ph]} \BibitemShut
  {NoStop}%
%%CITATION = ARXIV:0705.1733;%%
\bibitem [{\citenamefont {Caprini}\ \emph
  {et~al.}(2009{\natexlab{a}})\citenamefont {Caprini}, \citenamefont {Durrer},\
  and\ \citenamefont {Servant}}]{Caprini:2009yp}%
  \BibitemOpen
  \bibfield  {author} {\bibinfo {author} {\bibfnamefont {C.}~\bibnamefont
  {Caprini}}, \bibinfo {author} {\bibfnamefont {R.}~\bibnamefont {Durrer}}, \
  and\ \bibinfo {author} {\bibfnamefont {G.}~\bibnamefont {Servant}},\ }\href
  {\doibase 10.1088/1475-7516/2009/12/024} {\bibfield  {journal} {\bibinfo
  {journal} {JCAP}\ }\textbf {\bibinfo {volume} {0912}},\ \bibinfo {pages}
  {024} (\bibinfo {year} {2009}{\natexlab{a}})},\ \Eprint
  {http://arxiv.org/abs/0909.0622} {arXiv:0909.0622 [astro-ph.CO]} \BibitemShut
  {NoStop}%
%%CITATION = ARXIV:0909.0622;%%
\bibitem [{\citenamefont {Caprini}\ \emph
  {et~al.}(2009{\natexlab{b}})\citenamefont {Caprini}, \citenamefont {Durrer},
  \citenamefont {Konstandin},\ and\ \citenamefont {Servant}}]{Caprini:2009fx}%
  \BibitemOpen
  \bibfield  {author} {\bibinfo {author} {\bibfnamefont {C.}~\bibnamefont
  {Caprini}}, \bibinfo {author} {\bibfnamefont {R.}~\bibnamefont {Durrer}},
  \bibinfo {author} {\bibfnamefont {T.}~\bibnamefont {Konstandin}}, \ and\
  \bibinfo {author} {\bibfnamefont {G.}~\bibnamefont {Servant}},\ }\href
  {\doibase 10.1103/PhysRevD.79.083519} {\bibfield  {journal} {\bibinfo
  {journal} {Phys.Rev.}\ }\textbf {\bibinfo {volume} {D79}},\ \bibinfo {pages}
  {083519} (\bibinfo {year} {2009}{\natexlab{b}})},\ \Eprint
  {http://arxiv.org/abs/0901.1661} {arXiv:0901.1661 [astro-ph.CO]} \BibitemShut
  {NoStop}%
%%CITATION = ARXIV:0901.1661;%%
\bibitem [{\citenamefont {Niksa}\ \emph {et~al.}(2018)\citenamefont {Niksa},
  \citenamefont {Schlederer},\ and\ \citenamefont {Sigl}}]{Niksa:2018ofa}%
  \BibitemOpen
  \bibfield  {author} {\bibinfo {author} {\bibfnamefont {P.}~\bibnamefont
  {Niksa}}, \bibinfo {author} {\bibfnamefont {M.}~\bibnamefont {Schlederer}}, \
  and\ \bibinfo {author} {\bibfnamefont {G.}~\bibnamefont {Sigl}},\ }\href
  {\doibase 10.1088/1361-6382/aac89c} {\bibfield  {journal} {\bibinfo
  {journal} {Class. Quant. Grav.}\ }\textbf {\bibinfo {volume} {35}},\ \bibinfo
  {pages} {144001} (\bibinfo {year} {2018})},\ \Eprint
  {http://arxiv.org/abs/1803.02271} {arXiv:1803.02271 [astro-ph.CO]}
  \BibitemShut {NoStop}%
%%CITATION = ARXIV:1803.02271;%%
\bibitem [{\citenamefont {Pol}\ \emph {et~al.}(2019)\citenamefont {Pol},
  \citenamefont {Mandal}, \citenamefont {Brandenburg}, \citenamefont
  {Kahniashvili},\ and\ \citenamefont {Kosowsky}}]{Pol:2019yex}%
  \BibitemOpen
  \bibfield  {author} {\bibinfo {author} {\bibfnamefont {A.~R.}\ \bibnamefont
  {Pol}}, \bibinfo {author} {\bibfnamefont {S.}~\bibnamefont {Mandal}},
  \bibinfo {author} {\bibfnamefont {A.}~\bibnamefont {Brandenburg}}, \bibinfo
  {author} {\bibfnamefont {T.}~\bibnamefont {Kahniashvili}}, \ and\ \bibinfo
  {author} {\bibfnamefont {A.}~\bibnamefont {Kosowsky}},\ }\href@noop {} {\
  (\bibinfo {year} {2019})},\ \Eprint {http://arxiv.org/abs/1903.08585}
  {arXiv:1903.08585 [astro-ph.CO]} \BibitemShut {NoStop}%
%%CITATION = ARXIV:1903.08585;%%
\bibitem [{\citenamefont {Child}\ and\ \citenamefont
  {Giblin}(2012)}]{Child:2012qg}%
  \BibitemOpen
  \bibfield  {author} {\bibinfo {author} {\bibfnamefont {H.~L.}\ \bibnamefont
  {Child}}\ and\ \bibinfo {author} {\bibfnamefont {J.}~\bibnamefont {Giblin},
  \bibfnamefont {John~T.}},\ }\href {\doibase 10.1088/1475-7516/2012/10/001}
  {\bibfield  {journal} {\bibinfo  {journal} {JCAP}\ }\textbf {\bibinfo
  {volume} {1210}},\ \bibinfo {pages} {001} (\bibinfo {year} {2012})},\ \Eprint
  {http://arxiv.org/abs/1207.6408} {arXiv:1207.6408 [astro-ph.CO]} \BibitemShut
  {NoStop}%
%%CITATION = ARXIV:1207.6408;%%
\bibitem [{\citenamefont {Cutting}\ \emph {et~al.}(2018)\citenamefont
  {Cutting}, \citenamefont {Hindmarsh},\ and\ \citenamefont
  {Weir}}]{Cutting:2018tjt}%
  \BibitemOpen
  \bibfield  {author} {\bibinfo {author} {\bibfnamefont {D.}~\bibnamefont
  {Cutting}}, \bibinfo {author} {\bibfnamefont {M.}~\bibnamefont {Hindmarsh}},
  \ and\ \bibinfo {author} {\bibfnamefont {D.~J.}\ \bibnamefont {Weir}},\
  }\href {\doibase 10.1103/PhysRevD.97.123513} {\bibfield  {journal} {\bibinfo
  {journal} {Phys. Rev.}\ }\textbf {\bibinfo {volume} {D97}},\ \bibinfo {pages}
  {123513} (\bibinfo {year} {2018})},\ \Eprint
  {http://arxiv.org/abs/1802.05712} {arXiv:1802.05712 [astro-ph.CO]}
  \BibitemShut {NoStop}%
%%CITATION = ARXIV:1802.05712;%%
\bibitem [{\citenamefont {Hawking}\ \emph {et~al.}(1982)\citenamefont
  {Hawking}, \citenamefont {Moss},\ and\ \citenamefont
  {Stewart}}]{Hawking:1982ga}%
  \BibitemOpen
  \bibfield  {author} {\bibinfo {author} {\bibfnamefont {S.}~\bibnamefont
  {Hawking}}, \bibinfo {author} {\bibfnamefont {I.}~\bibnamefont {Moss}}, \
  and\ \bibinfo {author} {\bibfnamefont {J.}~\bibnamefont {Stewart}},\ }\href
  {\doibase 10.1103/PhysRevD.26.2681} {\bibfield  {journal} {\bibinfo
  {journal} {Phys.Rev.}\ }\textbf {\bibinfo {volume} {D26}},\ \bibinfo {pages}
  {2681} (\bibinfo {year} {1982})}\BibitemShut {NoStop}%
%%CITATION = PHRVA,D26,2681;%%
\bibitem [{\citenamefont {Kosowsky}\ \emph {et~al.}(1992)\citenamefont
  {Kosowsky}, \citenamefont {Turner},\ and\ \citenamefont
  {Watkins}}]{Kosowsky:1991ua}%
  \BibitemOpen
  \bibfield  {author} {\bibinfo {author} {\bibfnamefont {A.}~\bibnamefont
  {Kosowsky}}, \bibinfo {author} {\bibfnamefont {M.~S.}\ \bibnamefont
  {Turner}}, \ and\ \bibinfo {author} {\bibfnamefont {R.}~\bibnamefont
  {Watkins}},\ }\href {\doibase 10.1103/PhysRevD.45.4514} {\bibfield  {journal}
  {\bibinfo  {journal} {Phys. Rev.}\ }\textbf {\bibinfo {volume} {D45}},\
  \bibinfo {pages} {4514} (\bibinfo {year} {1992})}\BibitemShut {NoStop}%
%%CITATION = PHRVA,D45,4514;%%
\bibitem [{\citenamefont {Watkins}\ and\ \citenamefont
  {Widrow}(1992)}]{Watkins:1991zt}%
  \BibitemOpen
  \bibfield  {author} {\bibinfo {author} {\bibfnamefont {R.}~\bibnamefont
  {Watkins}}\ and\ \bibinfo {author} {\bibfnamefont {L.~M.}\ \bibnamefont
  {Widrow}},\ }\href {\doibase 10.1016/0550-3213(92)90362-F} {\bibfield
  {journal} {\bibinfo  {journal} {Nucl. Phys. B}\ }\textbf {\bibinfo {volume}
  {374}},\ \bibinfo {pages} {446} (\bibinfo {year} {1992})}\BibitemShut
  {NoStop}%
\bibitem [{\citenamefont {Braden}\ \emph {et~al.}(2015)\citenamefont {Braden},
  \citenamefont {Bond},\ and\ \citenamefont
  {Mersini-Houghton}}]{Braden:2014cra}%
  \BibitemOpen
  \bibfield  {author} {\bibinfo {author} {\bibfnamefont {J.}~\bibnamefont
  {Braden}}, \bibinfo {author} {\bibfnamefont {J.~R.}\ \bibnamefont {Bond}}, \
  and\ \bibinfo {author} {\bibfnamefont {L.}~\bibnamefont {Mersini-Houghton}},\
  }\href {\doibase 10.1088/1475-7516/2015/03/007} {\bibfield  {journal}
  {\bibinfo  {journal} {JCAP}\ }\textbf {\bibinfo {volume} {1503}},\ \bibinfo
  {pages} {007} (\bibinfo {year} {2015})},\ \Eprint
  {http://arxiv.org/abs/1412.5591} {arXiv:1412.5591 [hep-th]} \BibitemShut
  {NoStop}%
%%CITATION = ARXIV:1412.5591;%%
\bibitem [{\citenamefont {Jinno}\ \emph
  {et~al.}(2019{\natexlab{b}})\citenamefont {Jinno}, \citenamefont
  {Konstandin},\ and\ \citenamefont {Takimoto}}]{Jinno:2019bxw}%
  \BibitemOpen
  \bibfield  {author} {\bibinfo {author} {\bibfnamefont {R.}~\bibnamefont
  {Jinno}}, \bibinfo {author} {\bibfnamefont {T.}~\bibnamefont {Konstandin}}, \
  and\ \bibinfo {author} {\bibfnamefont {M.}~\bibnamefont {Takimoto}},\ }\href
  {\doibase 10.1088/1475-7516/2019/09/035} {\bibfield  {journal} {\bibinfo
  {journal} {JCAP}\ }\textbf {\bibinfo {volume} {09}},\ \bibinfo {pages} {035}
  (\bibinfo {year} {2019}{\natexlab{b}})},\ \Eprint
  {http://arxiv.org/abs/1906.02588} {arXiv:1906.02588 [hep-ph]} \BibitemShut
  {NoStop}%
\bibitem [{\citenamefont {Lewicki}\ and\ \citenamefont
  {Vaskonen}(2019)}]{Lewicki:2019gmv}%
  \BibitemOpen
  \bibfield  {author} {\bibinfo {author} {\bibfnamefont {M.}~\bibnamefont
  {Lewicki}}\ and\ \bibinfo {author} {\bibfnamefont {V.}~\bibnamefont
  {Vaskonen}},\ }\href@noop {} {\  (\bibinfo {year} {2019})},\ \Eprint
  {http://arxiv.org/abs/1912.00997} {arXiv:1912.00997 [astro-ph.CO]}
  \BibitemShut {NoStop}%
\bibitem [{\citenamefont {Enqvist}\ \emph {et~al.}(1992)\citenamefont
  {Enqvist}, \citenamefont {Ignatius}, \citenamefont {Kajantie},\ and\
  \citenamefont {Rummukainen}}]{Enqvist:1991xw}%
  \BibitemOpen
  \bibfield  {author} {\bibinfo {author} {\bibfnamefont {K.}~\bibnamefont
  {Enqvist}}, \bibinfo {author} {\bibfnamefont {J.}~\bibnamefont {Ignatius}},
  \bibinfo {author} {\bibfnamefont {K.}~\bibnamefont {Kajantie}}, \ and\
  \bibinfo {author} {\bibfnamefont {K.}~\bibnamefont {Rummukainen}},\ }\href
  {\doibase 10.1103/PhysRevD.45.3415} {\bibfield  {journal} {\bibinfo
  {journal} {Phys.Rev.}\ }\textbf {\bibinfo {volume} {D45}},\ \bibinfo {pages}
  {3415} (\bibinfo {year} {1992})}\BibitemShut {NoStop}%
%%CITATION = PHRVA,D45,3415;%%
\bibitem [{\citenamefont {Jinno}\ \emph {et~al.}(2017)\citenamefont {Jinno},
  \citenamefont {Lee}, \citenamefont {Seong},\ and\ \citenamefont
  {Takimoto}}]{Jinno:2017ixd}%
  \BibitemOpen
  \bibfield  {author} {\bibinfo {author} {\bibfnamefont {R.}~\bibnamefont
  {Jinno}}, \bibinfo {author} {\bibfnamefont {S.}~\bibnamefont {Lee}}, \bibinfo
  {author} {\bibfnamefont {H.}~\bibnamefont {Seong}}, \ and\ \bibinfo {author}
  {\bibfnamefont {M.}~\bibnamefont {Takimoto}},\ }\href {\doibase
  10.1088/1475-7516/2017/11/050} {\bibfield  {journal} {\bibinfo  {journal}
  {JCAP}\ }\textbf {\bibinfo {volume} {1711}},\ \bibinfo {pages} {050}
  (\bibinfo {year} {2017})},\ \Eprint {http://arxiv.org/abs/1708.01253}
  {arXiv:1708.01253 [hep-ph]} \BibitemShut {NoStop}%
%%CITATION = ARXIV:1708.01253;%%
\bibitem [{\citenamefont {Garcia~Garcia}\ \emph {et~al.}(2016)\citenamefont
  {Garcia~Garcia}, \citenamefont {Krippendorf},\ and\ \citenamefont
  {March-Russell}}]{GarciaGarcia:2016xgv}%
  \BibitemOpen
  \bibfield  {author} {\bibinfo {author} {\bibfnamefont {I.}~\bibnamefont
  {Garcia~Garcia}}, \bibinfo {author} {\bibfnamefont {S.}~\bibnamefont
  {Krippendorf}}, \ and\ \bibinfo {author} {\bibfnamefont {J.}~\bibnamefont
  {March-Russell}},\ }\href@noop {} {\  (\bibinfo {year} {2016})},\ \Eprint
  {http://arxiv.org/abs/1607.06813} {arXiv:1607.06813 [hep-ph]} \BibitemShut
  {NoStop}%
%%CITATION = ARXIV:1607.06813;%%
\bibitem [{\citenamefont {Aarts}\ \emph {et~al.}(2000)\citenamefont {Aarts},
  \citenamefont {Bonini},\ and\ \citenamefont {Wetterich}}]{Aarts:2000mg}%
  \BibitemOpen
  \bibfield  {author} {\bibinfo {author} {\bibfnamefont {G.}~\bibnamefont
  {Aarts}}, \bibinfo {author} {\bibfnamefont {G.~F.}\ \bibnamefont {Bonini}}, \
  and\ \bibinfo {author} {\bibfnamefont {C.}~\bibnamefont {Wetterich}},\ }\href
  {\doibase 10.1016/S0550-3213(00)00447-8} {\bibfield  {journal} {\bibinfo
  {journal} {Nucl. Phys.}\ }\textbf {\bibinfo {volume} {B587}},\ \bibinfo
  {pages} {403} (\bibinfo {year} {2000})},\ \Eprint
  {http://arxiv.org/abs/hep-ph/0003262} {arXiv:hep-ph/0003262 [hep-ph]}
  \BibitemShut {NoStop}%
%%CITATION = HEP-PH/0003262;%%
\bibitem [{\citenamefont {Micha}\ and\ \citenamefont
  {Tkachev}(2003)}]{Micha:2002ey}%
  \BibitemOpen
  \bibfield  {author} {\bibinfo {author} {\bibfnamefont {R.}~\bibnamefont
  {Micha}}\ and\ \bibinfo {author} {\bibfnamefont {I.~I.}\ \bibnamefont
  {Tkachev}},\ }\href {\doibase 10.1103/PhysRevLett.90.121301} {\bibfield
  {journal} {\bibinfo  {journal} {Phys. Rev. Lett.}\ }\textbf {\bibinfo
  {volume} {90}},\ \bibinfo {pages} {121301} (\bibinfo {year} {2003})},\
  \Eprint {http://arxiv.org/abs/hep-ph/0210202} {arXiv:hep-ph/0210202 [hep-ph]}
  \BibitemShut {NoStop}%
%%CITATION = HEP-PH/0210202;%%
\bibitem [{\citenamefont {Arrizabalaga}\ \emph {et~al.}(2005)\citenamefont
  {Arrizabalaga}, \citenamefont {Smit},\ and\ \citenamefont
  {Tranberg}}]{Arrizabalaga:2005tf}%
  \BibitemOpen
  \bibfield  {author} {\bibinfo {author} {\bibfnamefont {A.}~\bibnamefont
  {Arrizabalaga}}, \bibinfo {author} {\bibfnamefont {J.}~\bibnamefont {Smit}},
  \ and\ \bibinfo {author} {\bibfnamefont {A.}~\bibnamefont {Tranberg}},\
  }\href {\doibase 10.1103/PhysRevD.72.025014} {\bibfield  {journal} {\bibinfo
  {journal} {Phys. Rev.}\ }\textbf {\bibinfo {volume} {D72}},\ \bibinfo {pages}
  {025014} (\bibinfo {year} {2005})},\ \Eprint
  {http://arxiv.org/abs/hep-ph/0503287} {arXiv:hep-ph/0503287 [hep-ph]}
  \BibitemShut {NoStop}%
%%CITATION = HEP-PH/0503287;%%
\bibitem [{\citenamefont {Garcia-Bellido}\ \emph {et~al.}(2008)\citenamefont
  {Garcia-Bellido}, \citenamefont {Figueroa},\ and\ \citenamefont
  {Sastre}}]{GarciaBellido:2007af}%
  \BibitemOpen
  \bibfield  {author} {\bibinfo {author} {\bibfnamefont {J.}~\bibnamefont
  {Garcia-Bellido}}, \bibinfo {author} {\bibfnamefont {D.~G.}\ \bibnamefont
  {Figueroa}}, \ and\ \bibinfo {author} {\bibfnamefont {A.}~\bibnamefont
  {Sastre}},\ }\href {\doibase 10.1103/PhysRevD.77.043517} {\bibfield
  {journal} {\bibinfo  {journal} {Phys.Rev.}\ }\textbf {\bibinfo {volume}
  {D77}},\ \bibinfo {pages} {043517} (\bibinfo {year} {2008})},\ \Eprint
  {http://arxiv.org/abs/0707.0839} {arXiv:0707.0839 [hep-ph]} \BibitemShut
  {NoStop}%
%%CITATION = ARXIV:0707.0839;%%
\bibitem [{\citenamefont {Daverio}\ \emph {et~al.}(2015)\citenamefont
  {Daverio}, \citenamefont {Hindmarsh},\ and\ \citenamefont
  {Bevis}}]{David:2015eya}%
  \BibitemOpen
  \bibfield  {author} {\bibinfo {author} {\bibfnamefont {D.}~\bibnamefont
  {Daverio}}, \bibinfo {author} {\bibfnamefont {M.}~\bibnamefont {Hindmarsh}},
  \ and\ \bibinfo {author} {\bibfnamefont {N.}~\bibnamefont {Bevis}},\
  }\href@noop {} {\  (\bibinfo {year} {2015})},\ \Eprint
  {http://arxiv.org/abs/1508.05610} {arXiv:1508.05610 [physics.comp-ph]}
  \BibitemShut {NoStop}%
%%CITATION = ARXIV:1508.05610;%%
\end{thebibliography}%

\newpage

\appendix

\section{Convergence tests}
\label{app:convergence}

\subsection{Gravitational waves}
Our convergence tests for the gravitational-wave power spectrum
consist of performing a series of simulations in which the bubbles are
nucleated in the same position, but the lattice spacing $\Delta x$ is
varied while keeping the timestep $\Delta t = \Delta x / 5$. We refer
to the value of $\Delta x$ used in the main paper as
$\Delta x_\mathrm{ref}$.  Tables~\ref{table:simGW} and
\ref{table:late-sim} contain the values of this and other important
simulation parameters.  The captions to these tables are also useful
as reminders of the symbols used in the following discussion.

\begin{figure}[htb]
  \centering
  \includegraphics[width=0.485\textwidth,clip]{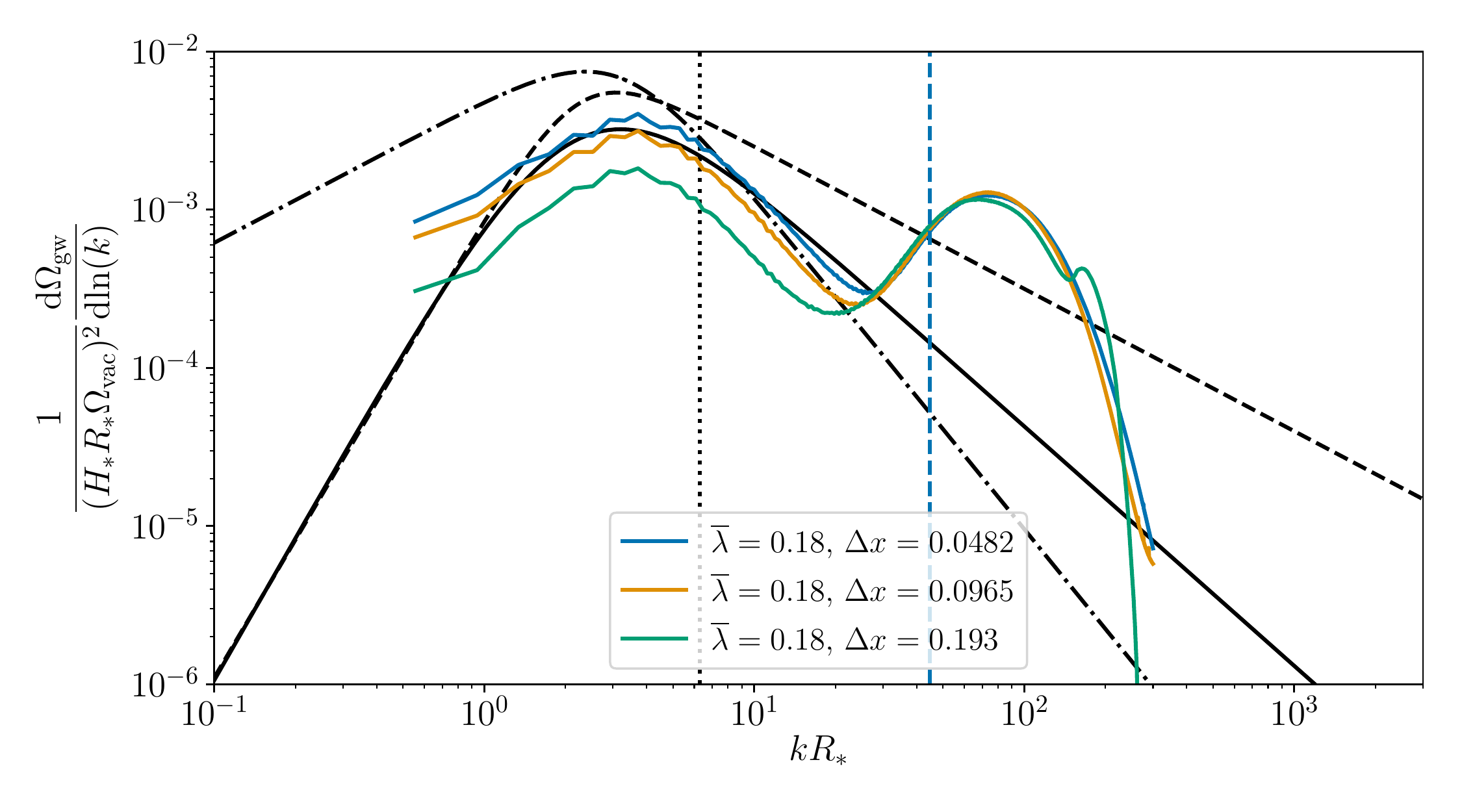}
  \caption{ Variation of the gravitational-wave power spectrum with
    lattice spacing at $t/\Rstar = 8.0$ for $\lambar=0.18$ with
    $\Nb=4096$. The black dashed line gives the result from the
    envelope approximation~\cite{Konstandin:2017sat}, the black
    dash-dot line gives the prediction from the bulk flow
    model~\cite{Konstandin:2017sat}, and the solid black line
    indicates the previous fit provided in
    Ref.~\cite{Cutting:2018tjt}. The vertical dotted line gives the
    location of $k = 2\pi / \Rstar$, whereas the red dashed line
    indicates where $k=\Mb$. At high wavenumbers the signal is
  overwhelmed by noise arising from single-precision floating point
  numerical errors. This noise is identified by comparing a smaller
  single-precision and double-precision run. We therefore apply a cut off
  in the UV at $k=\pi/2\Delta x$.}
\label{fig:gwSpeclat}
\end{figure}

In Fig.~\ref{fig:gwSpeclat} we plot the gravitational-wave power
spectrum at $t/\Rstar = 8.0$ for $\lambar = 0.18$ with $\Nb=4096$. We
do this for $\Delta x/\Delta x_\mathrm{ref}$ equal to $1$, $2$, and
$4$. From this plot, we can see that the gravitational-wave power
generated by oscillations around the mass scale is well behaved at
these lattice spacings. The spectrum due to bubble collisions varies
more substantially. The peak location remains fairly fixed, and the IR
and UV power laws seem consistent across lattice spacings. The
amplitude of the spectrum increases as $\Delta x$ is decreased. From
this, we can clearly see that it is the total gravitational-wave power
rather than the peak location or power law exponents that is most
sensitive to the lattice spacing.

\begin{figure*}
\subfigure[]{\includegraphics[width=0.45\textwidth,clip]{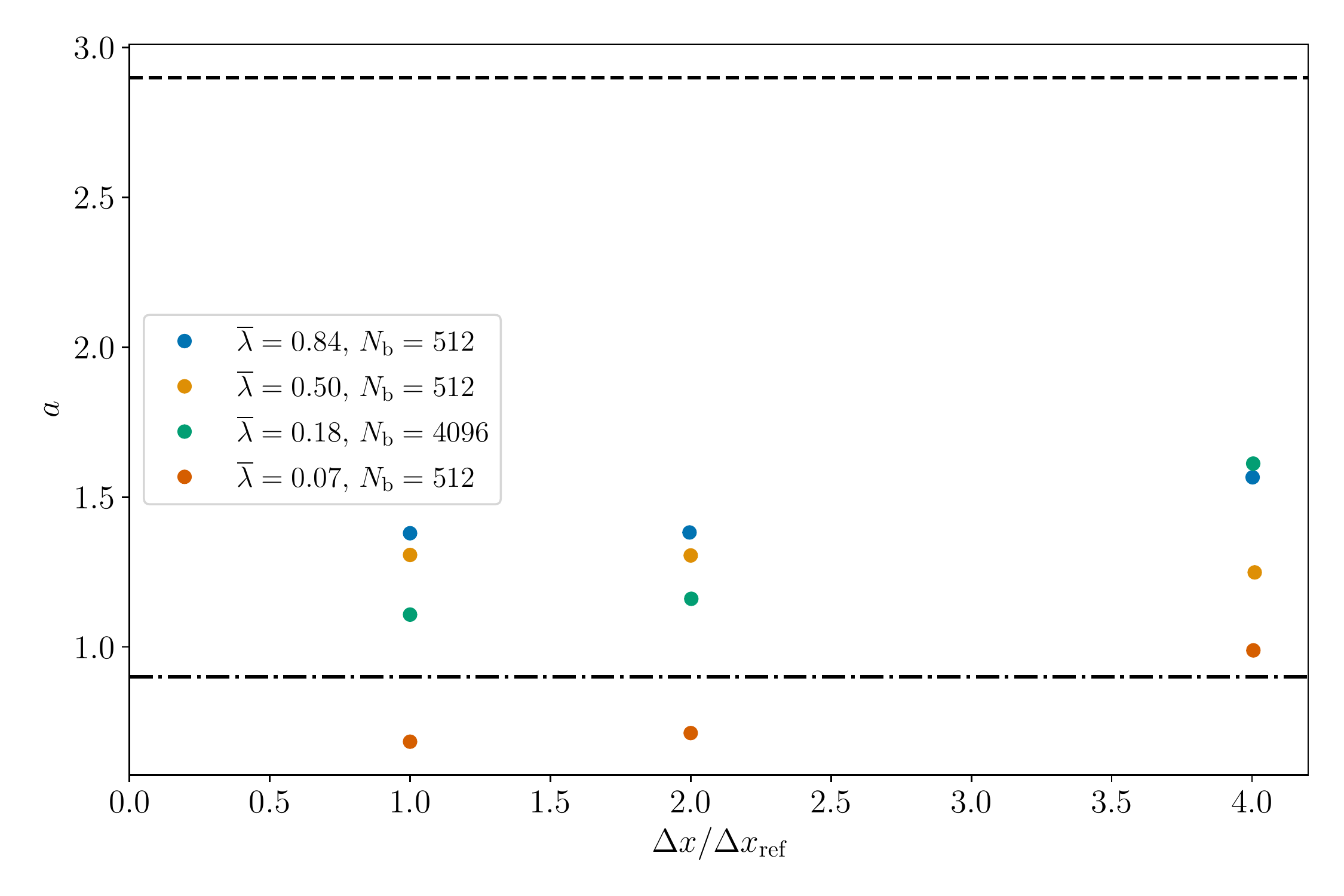}}
\hfill
\subfigure[]{\includegraphics[width=0.45\textwidth,clip]{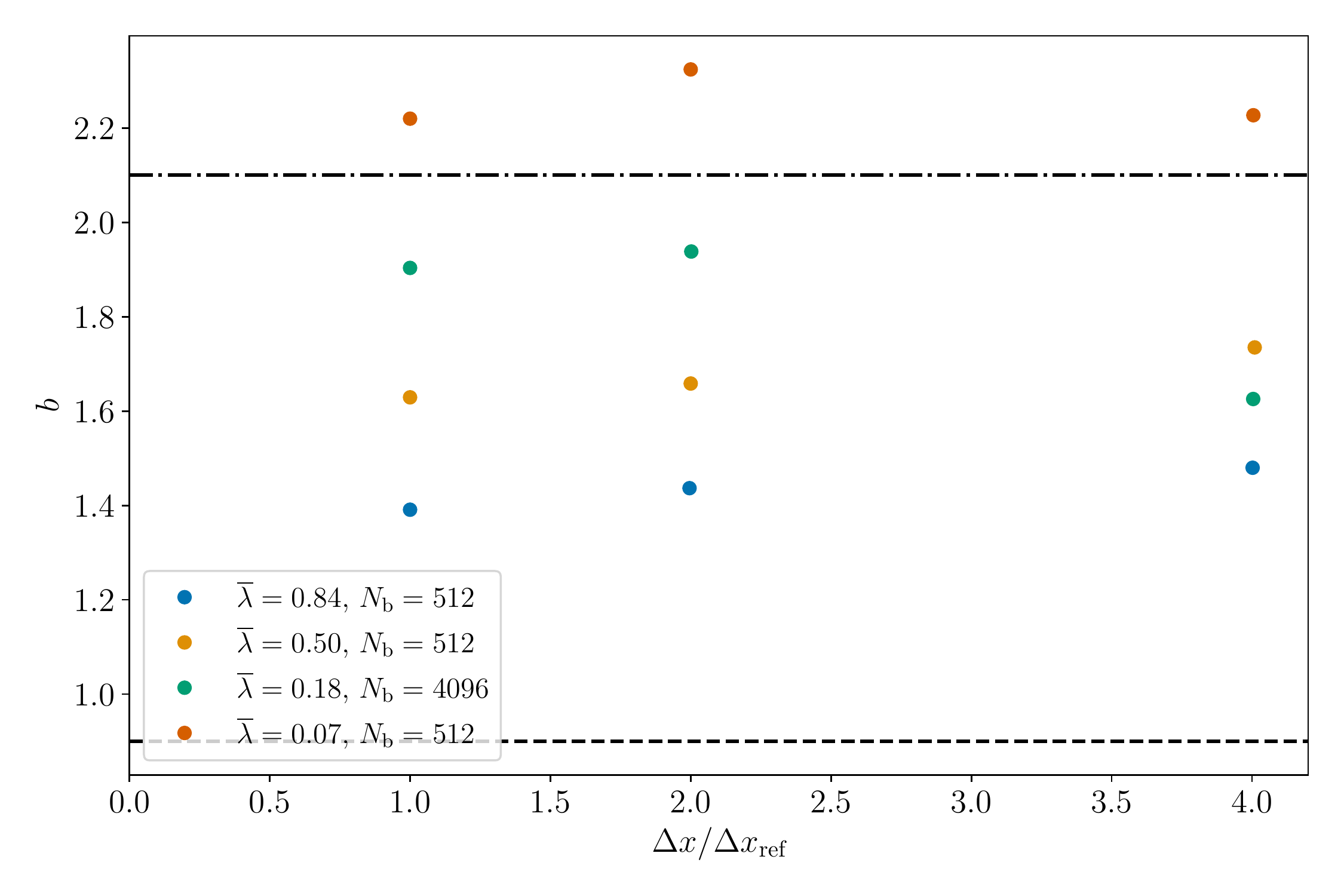}}
\\
\subfigure[]{\includegraphics[width=0.45\textwidth,clip]{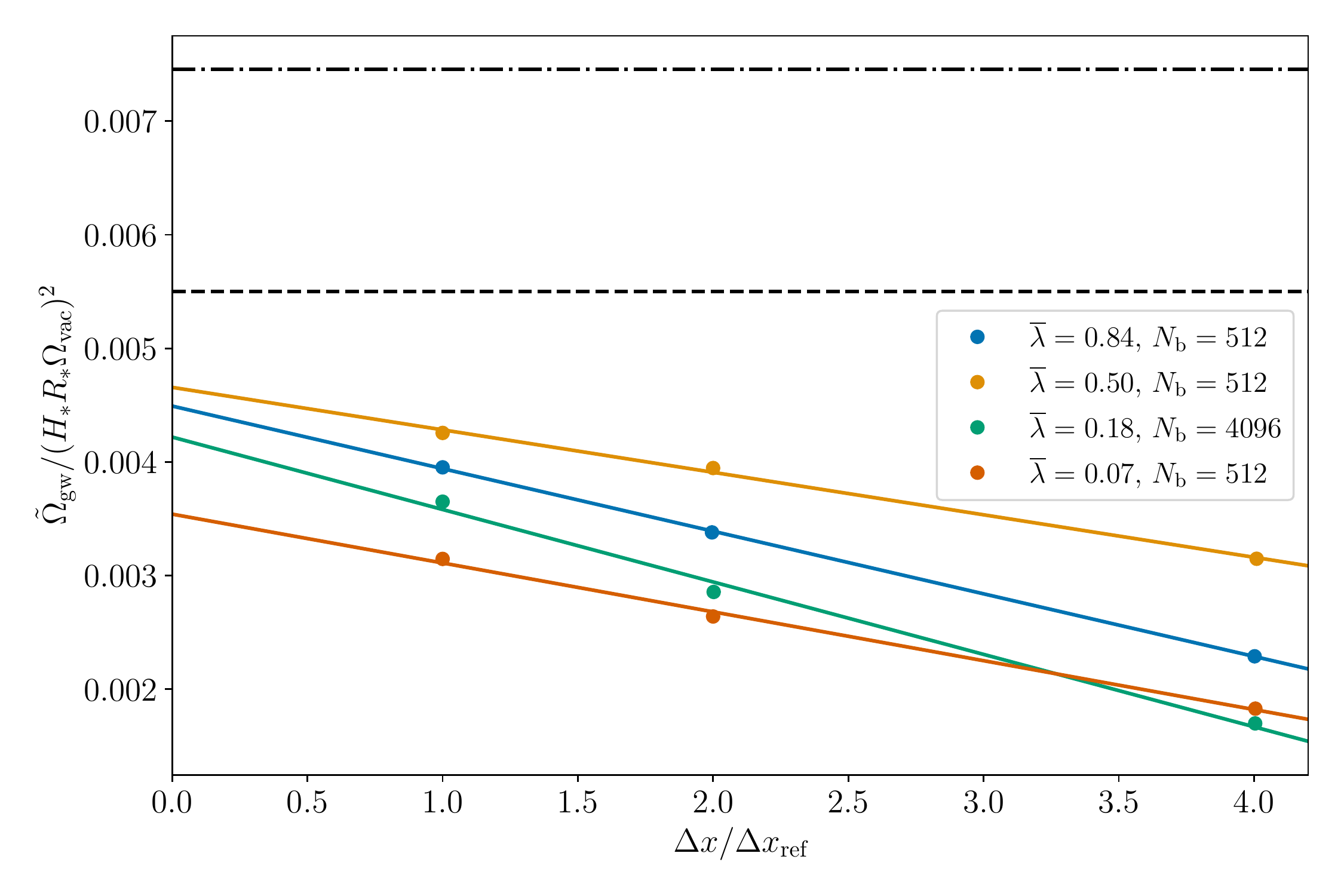}}
\hfill
\subfigure[]{\includegraphics[width=0.45\textwidth,clip]{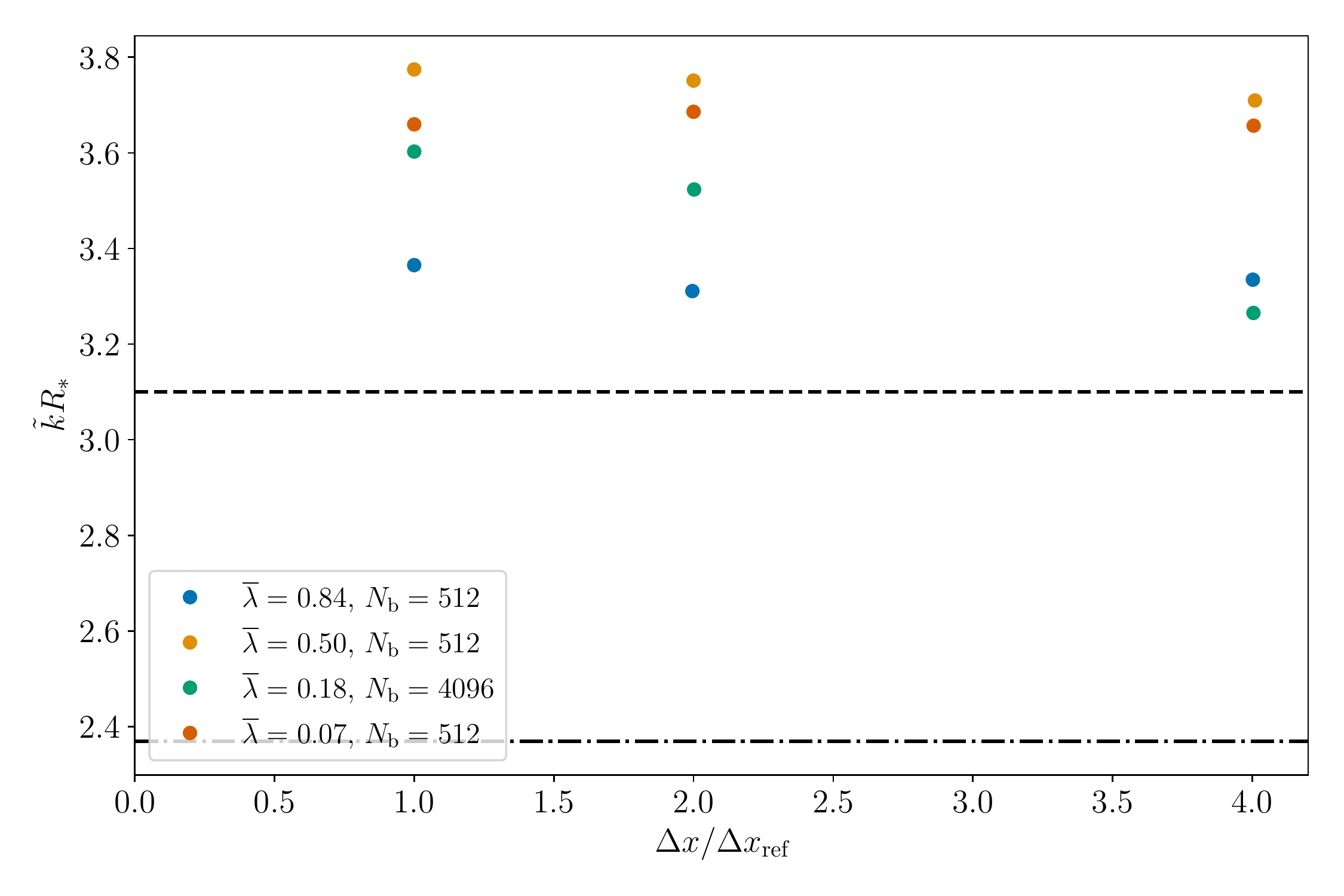}}
\caption{Convergence of the fitting parameters in Eq.~\ref{eq:thisfit}
  calculated at the end of each simulation. We plot how the fitting
  parameters vary with $\Delta x/\Delta x_\mathrm{ref}$, where
  $\Delta x_\mathrm{ref}$ corresponds to the value of $\Delta x$ used
  in Table~\ref{table:simGW}. In (a) we show the IR power law $a$, in
  (b) we show the UV power law $b$, in (c) we plot the peak amplitude
  $\tilde{\Omega}_\mathrm{gw}$, and in (d) we plot the peak frequency
  $\tilde{k}$. For the peak amplitude we also plot a linear fit to the
  continuum value. In each plot, we highlight the prediction for each
  parameter by the envelope approximation by a horizontal dashed black
  line, and for the bulk flow model by a dash-dot black line. }
\label{fig:FittingLat}
\end{figure*}

From our convergence tests, we can estimate the lattice errors on the
fitting parameters reported in Table~\ref{table:fitting}. To
do this, we must vary the lattice spacing of the simulations with the
most bubbles. This corresponds to $\Nb=4096$ for $\lambar=0.18$, and
$\Nb=512$ for all other $\lambar$. We perform additional simulations
with $\Delta x/\Delta x_\mathrm{ref} = 2$ and $4$. We find the fitting
parameters in Eq.~\ref{eq:thisfit} at the end of each simulation. We
then plot how the fitting parameters vary with $\Delta x$ in
Fig.~\ref{fig:FittingLat}. Differences between the parameter values at
$\Delta x_\mathrm{ref}$ and the values quoted in
Table~\ref{table:fitting} arise as we use a uniform uncertainty across
all bins\footnote{We use the SciPy library function
  $\texttt{optimize.curve\_fit}$ with arguments $\texttt{sigma=None}$
  and $\texttt{absolute\_sigma=False}$. This weights each bin power
  spectrum bin used in the fit equally with a uniform
  uncertainty.}. This differs to the results listed in the main body
of the paper where the difference in power at each bin between high
and mid resolutions runs was used as the uncertainty.

We see that the change in $a$, $b$ and $\tilde{k}$ between
$\Delta x/\Delta x_\mathrm{ref}=2$ and
$\Delta x/\Delta x_\mathrm{ref}=1$ is at the $\sim 1\%$ level, whereas
it is at the $\sim 10\%$ level for $\tilde{\Om}_\mathrm{gw}$.
Extrapolating a linear fit on $\tilde{\Om}_\mathrm{gw}$ as a function
of $\Delta x$ to the continuum shows us that the error on
$\tilde{\Om}_\mathrm{gw}$ at $\Delta x_\mathrm{ref}$ is on the order
of $10\%$. Even in the continuum limit $\tilde{\Om}_\mathrm{gw}$ is
smaller than the envelope prediction.

\begin{figure*}
\subfigure[]{\includegraphics[width=0.45\textwidth,clip]{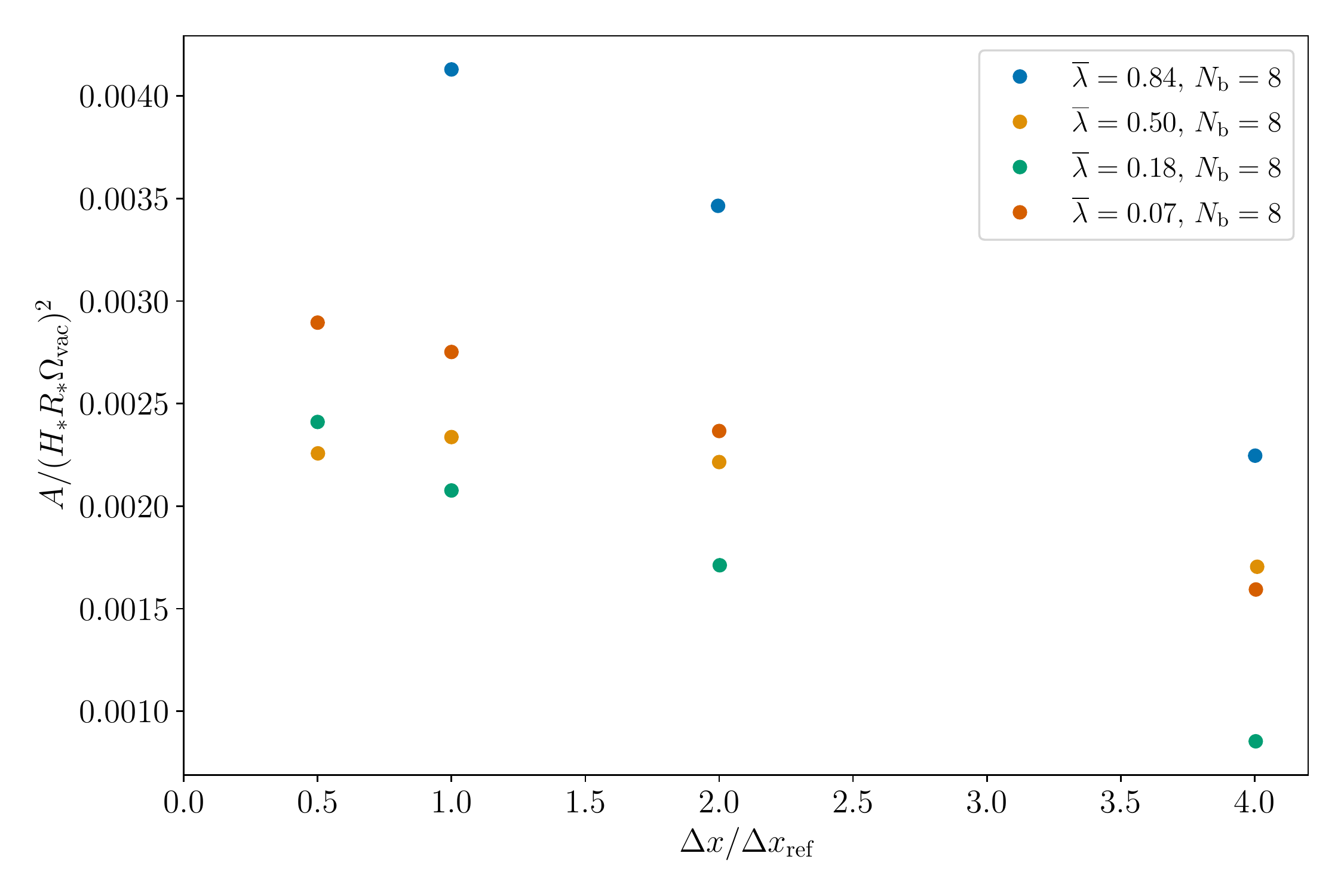}}
\hfill
\subfigure[]{\includegraphics[width=0.45\textwidth,clip]{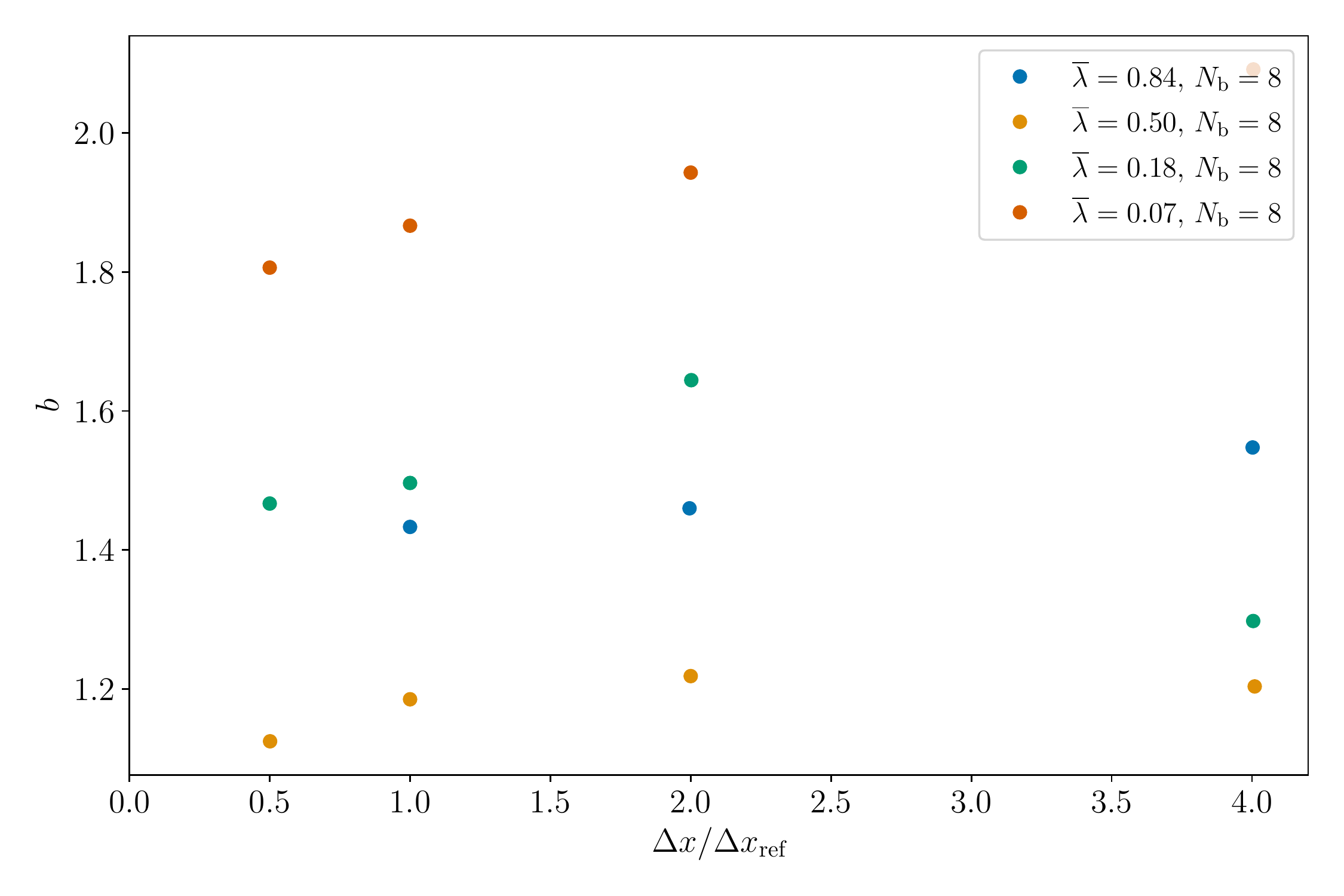}}
\caption{Convergence of the fitting parameters in
  Eq.~\ref{eq:uvplawfit} calculated at the end of each simulation. We
  plot how the fitting parameters vary with
  $\Delta x/\Delta x_\mathrm{ref}$, where $\Delta x_\mathrm{ref}$
  corresponds to the value of $\Delta x$ used in
  Table~\ref{table:simGW}. In (a) we plot the amplitude of the power
  spectrum at $k=2\pi/\Rstar$, $A$ and in (b) we show the UV power law
  $b$.  }
\label{fig:FittingLatUVplaw}
\end{figure*}

To check the behaviour of the gravitational-wave power
spectrum for $\Delta x/\Delta x_\mathrm{ref}<1$, we must reduce the
size of the simulations and number of bubbles. We perform a series of
simulations with $\Nb=8$ for each $\lambar$. In this case we can no
longer fit the power spectrum according to Eq.~\ref{eq:thisfit},
as the peak of the spectrum is not resolved. Instead we fit the UV
power law according to the following equation,
\begin{equation}
  \label{eq:uvplawfit}
  \dfrac{d \OmGW}{d \mathrm{ln}(k)} = A \left(\dfrac{R_* }{2\pi}k\right)^{-b}\text,
\end{equation}
where $b$ is the UV power law exponent and $A$ corresponds to the amplitude of
the spectrum at $k=2\pi/\Rstar$. We provide the resulting evaluation of
$A$ and $b$ at $t/\Rstar = 8.0$ in Fig.~\ref{fig:FittingLatUVplaw}. We
do not see any indication of a change in behaviour at smaller lattice
spacing than $\Delta x_\mathrm{ref}$.

\subsection{Scalar field}

\begin{figure*}
\subfigure[\,$\lambar = 0.84$]{\includegraphics[width=0.485\textwidth,clip]{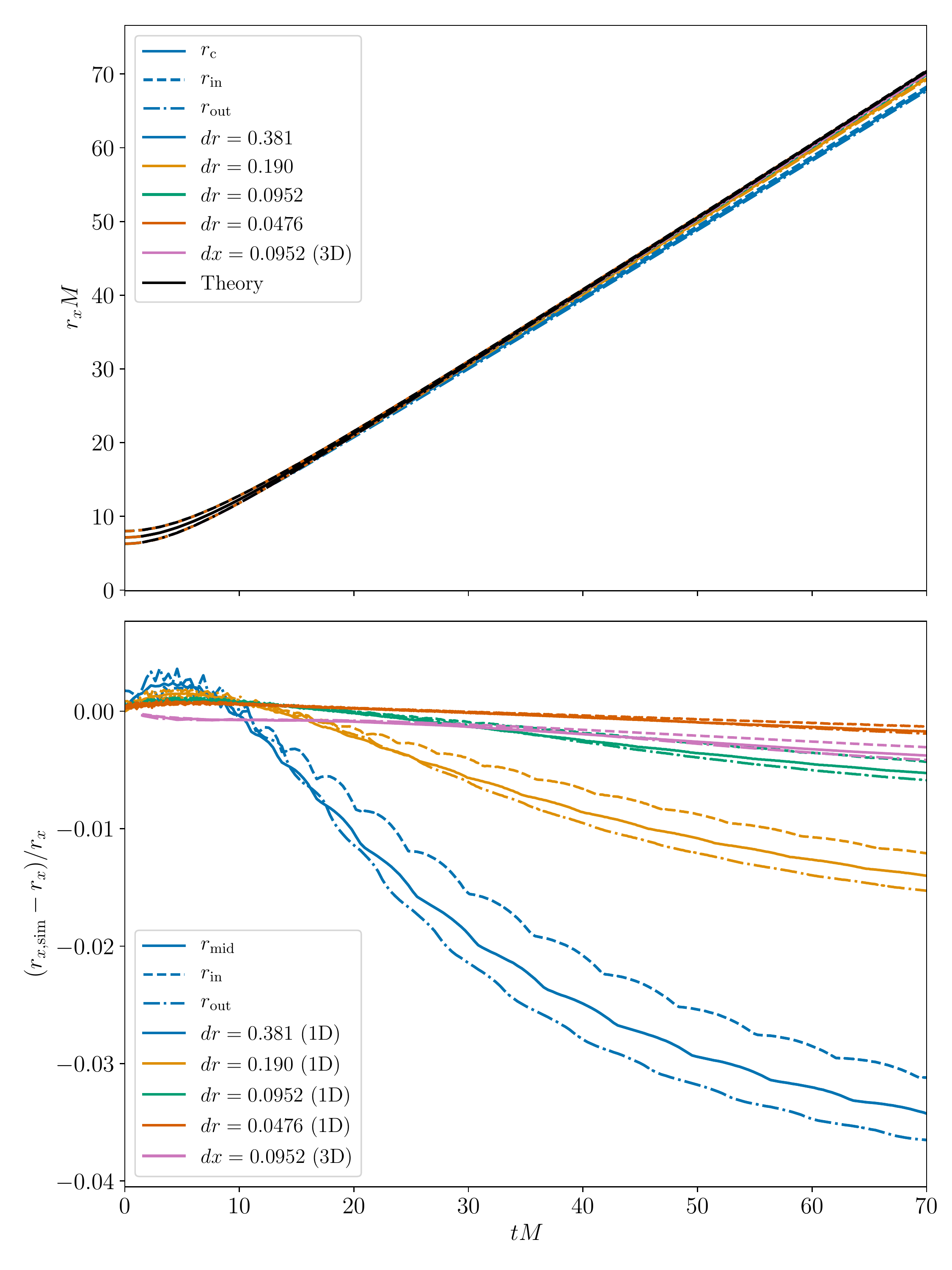}}
\hfill
\subfigure[\,$\lambar = 0.07$]{\includegraphics[width=0.485\textwidth,clip]{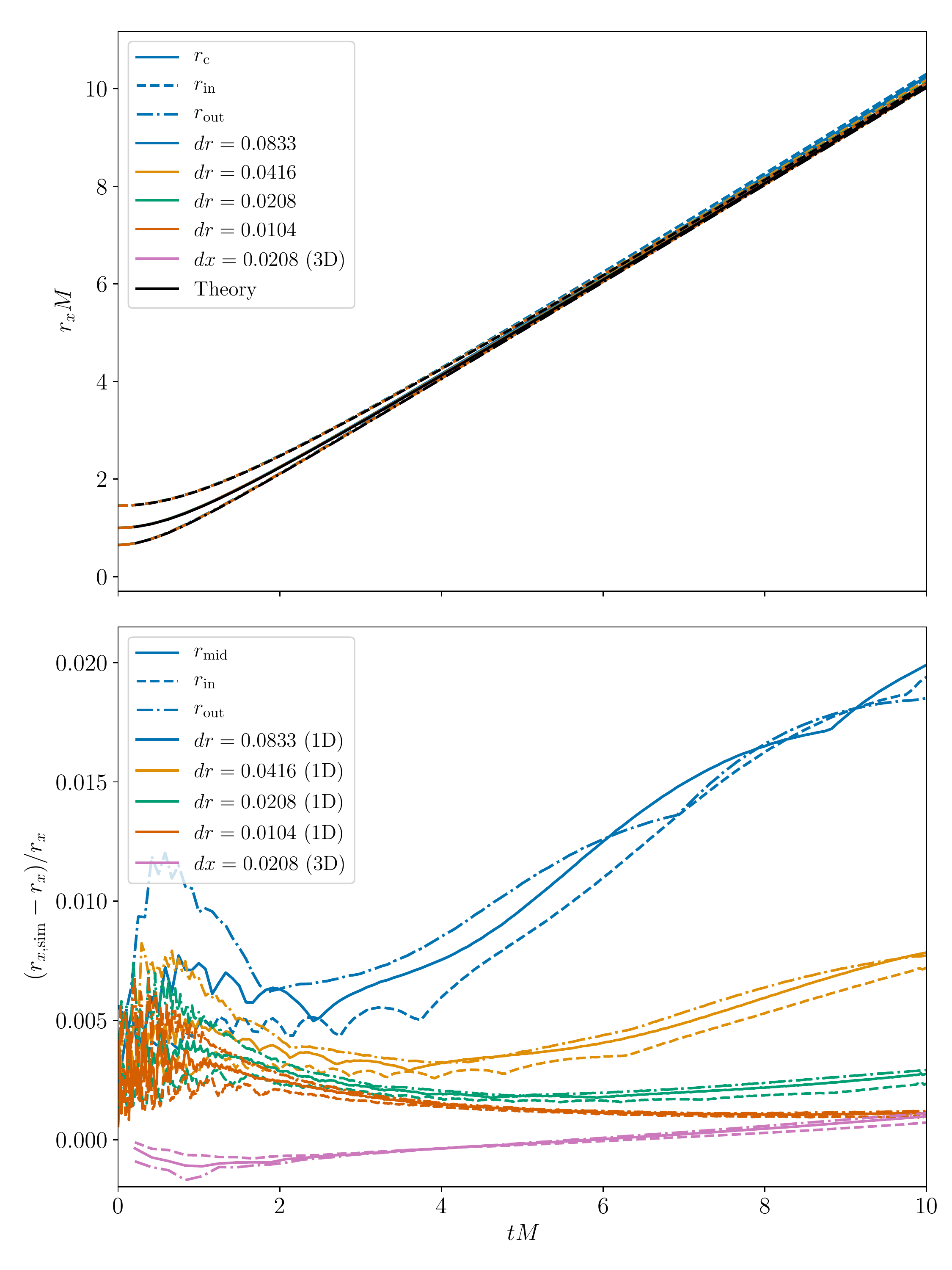}}
\caption{ In the top plots we show the evolution of the bubble radius
  parameters $\rMid$, $\rIn$ and $\rOut$ (defined in subsection
  \ref{ss:CriPro}) for an isolated bubble.  These are given for 1D
  simulations with various lattice spacings as well as the theoretical
  behaviour. The bottom panels give the fractional deviation from the
  theoretical value for each lattice spacing. We also include the
  result of an isolated bubble left to expand in a 3D simulation.}
\label{fig:Radii}
\end{figure*}

\begin{figure*}
\subfigure[\,$\lambar = 0.84$]{\includegraphics[width=0.485\textwidth,clip]{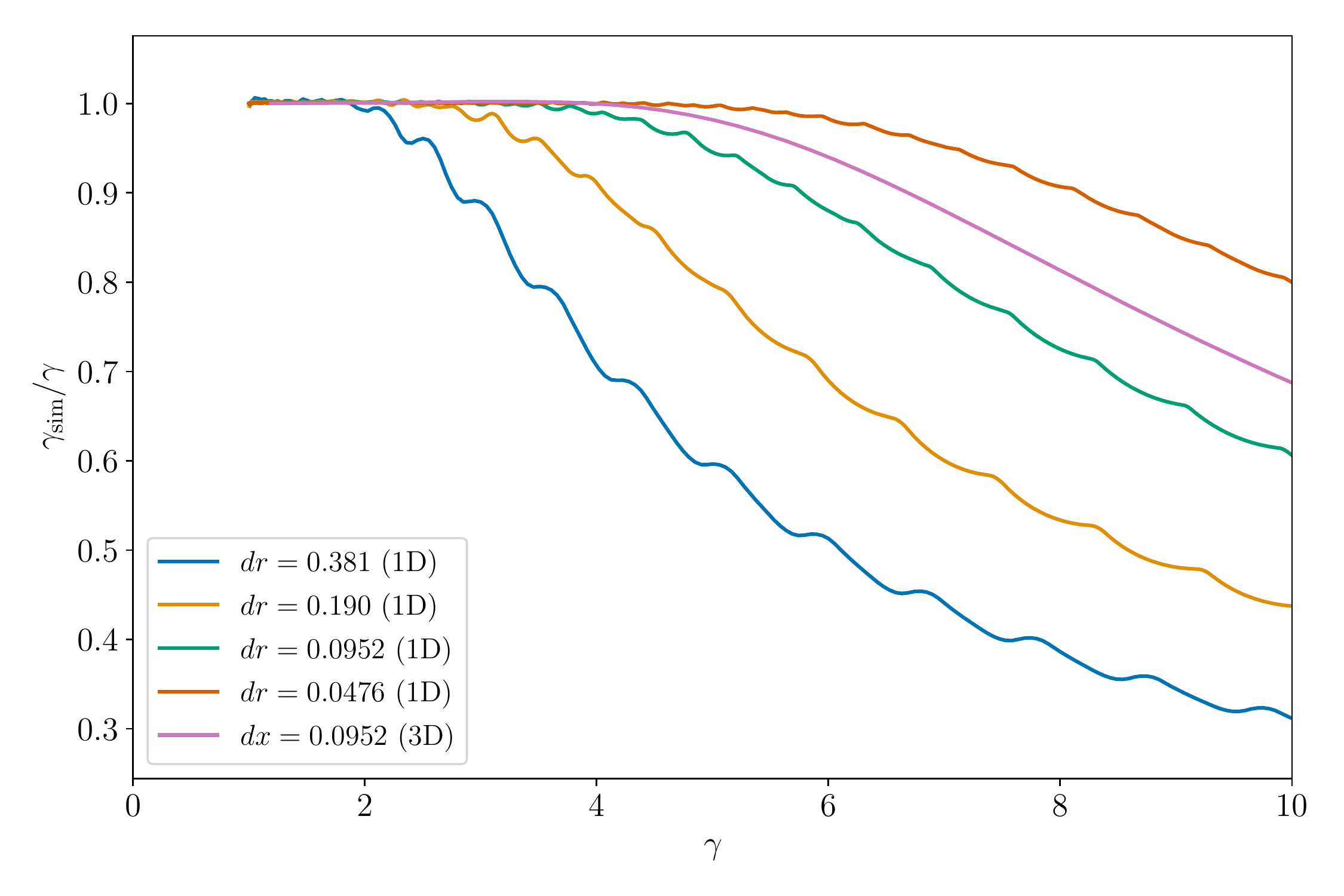}}
\hfill
\subfigure[\,$\lambar = 0.07$]{\includegraphics[width=0.485\textwidth,clip]{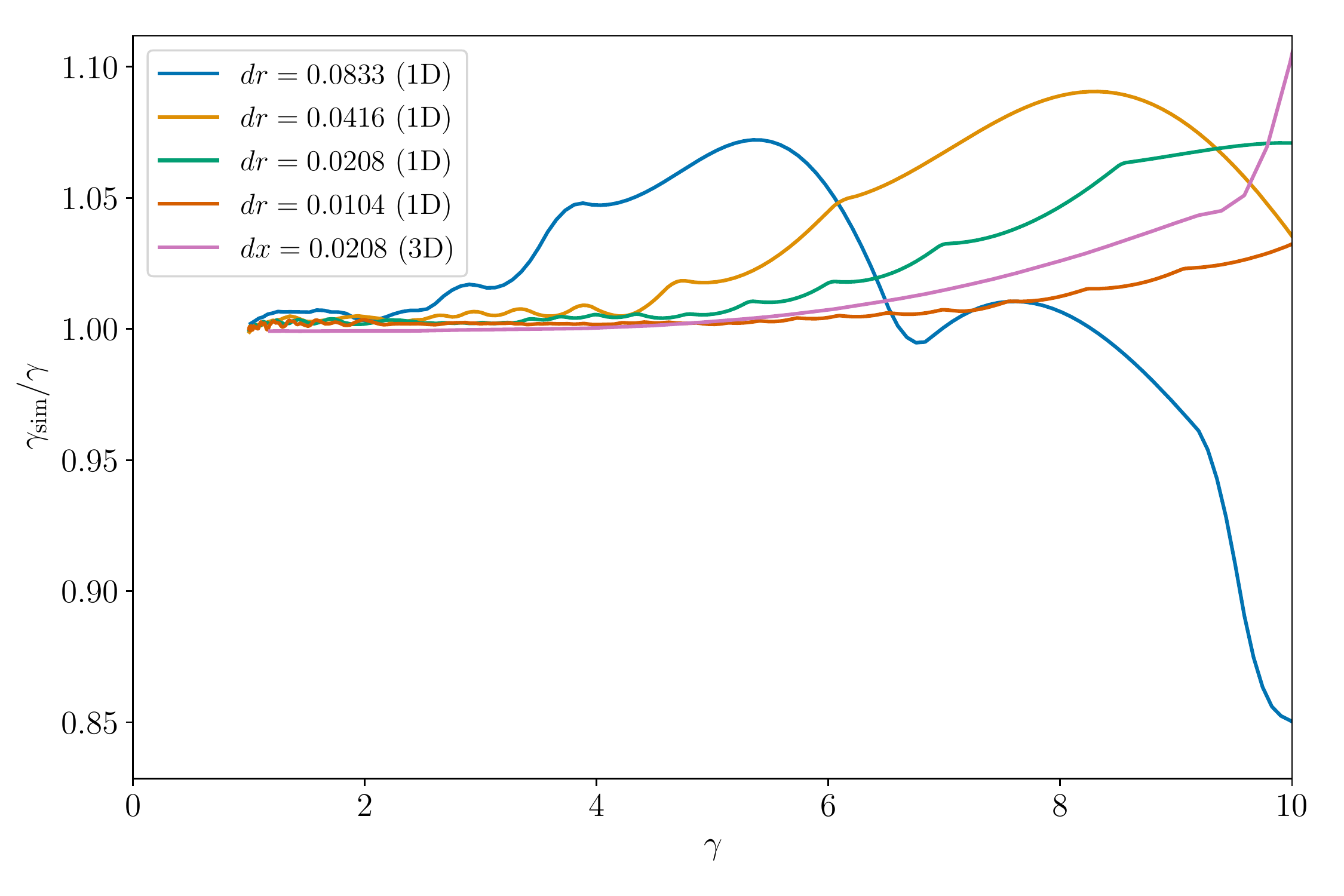}}
\caption{ Deviation of the bubble wall Lorentz factor
  $\gamma$ from its theoretical value in 1D simulations of isolated bubbles for
  a variety of lattice spacings. We also include the result of an
  isolated bubble left to expand in a 3D simulation.}
\label{fig:Gamma}
\end{figure*}

To study the effect of the lattice spacing on the scalar
field, we perform a series of simulations of isolated bubbles. We do
this both in a simplified 1D code with spherical symmetry and compare
the results to an isolated bubble expanding in our 3D code. From this,
we are able to measure the deviation of the scalar field profile from
its expected behaviour outlined in Section \ref{sec:scalarExp}. The
deviation then provides some measure of the lattice effects. We show
the deviation of bubble radius parameters
$\rIn$, $\rOut$ and $\rMid$ for a series of lattice
spacings and two $\lambar$ in Fig.~\ref{fig:Radii}. The resulting
effect on $\gamma$ estimated from the wall thickness is shown in
Fig.~\ref{fig:Gamma}.

We see that even a small deviation in $\rIn$, $\rOut$ and $\rMid$ can
result in a large change in the measured value of
$\gamma_\mathrm{sim}$. The finer the lattice spacing the larger
$\gamma$ can grow with $\gamma_\mathrm{sim}$ remaining close to the
theoretical value. We also see that for the same lattice spacing, the
3D runs show smaller lattice effects during expansion. For large
$\lambar$, as $\gamma$ increases $\gamma_\mathrm{sim}/\gamma$ will
decrease, whereas for small $\lambar$ we see that first, the lattice
effects cause the ratio $\gamma_\mathrm{sim}/\gamma$ to grow before eventually
it also decreases below unity.

\section{Slices}
\label{app:slices}

\begin{figure}
  \centering
  \hfill
  \subfigure
  {\includegraphics[width=0.32\textwidth,clip]{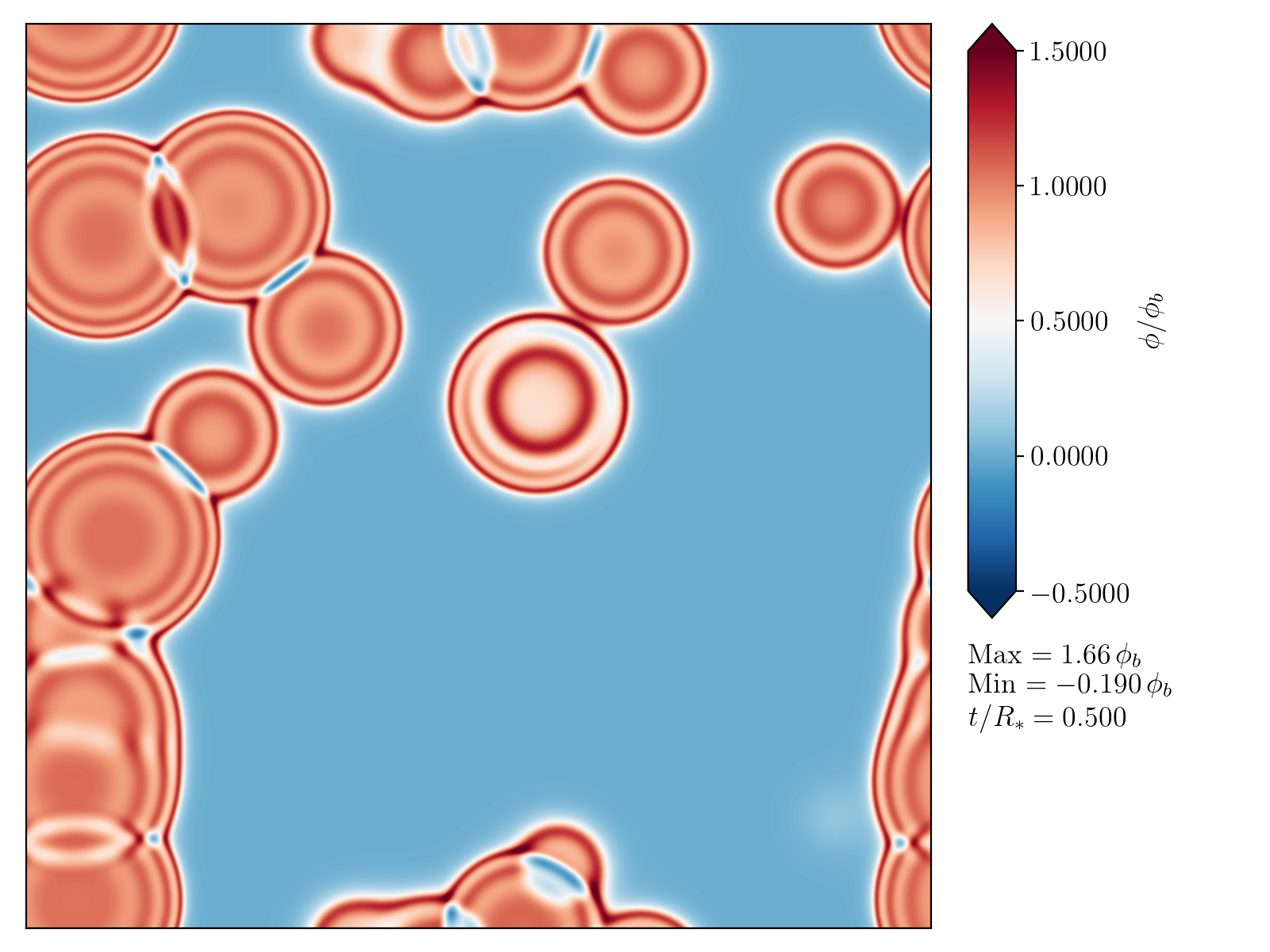}}
  \hfill
  \subfigure
  {\includegraphics[width=0.32\textwidth,clip]{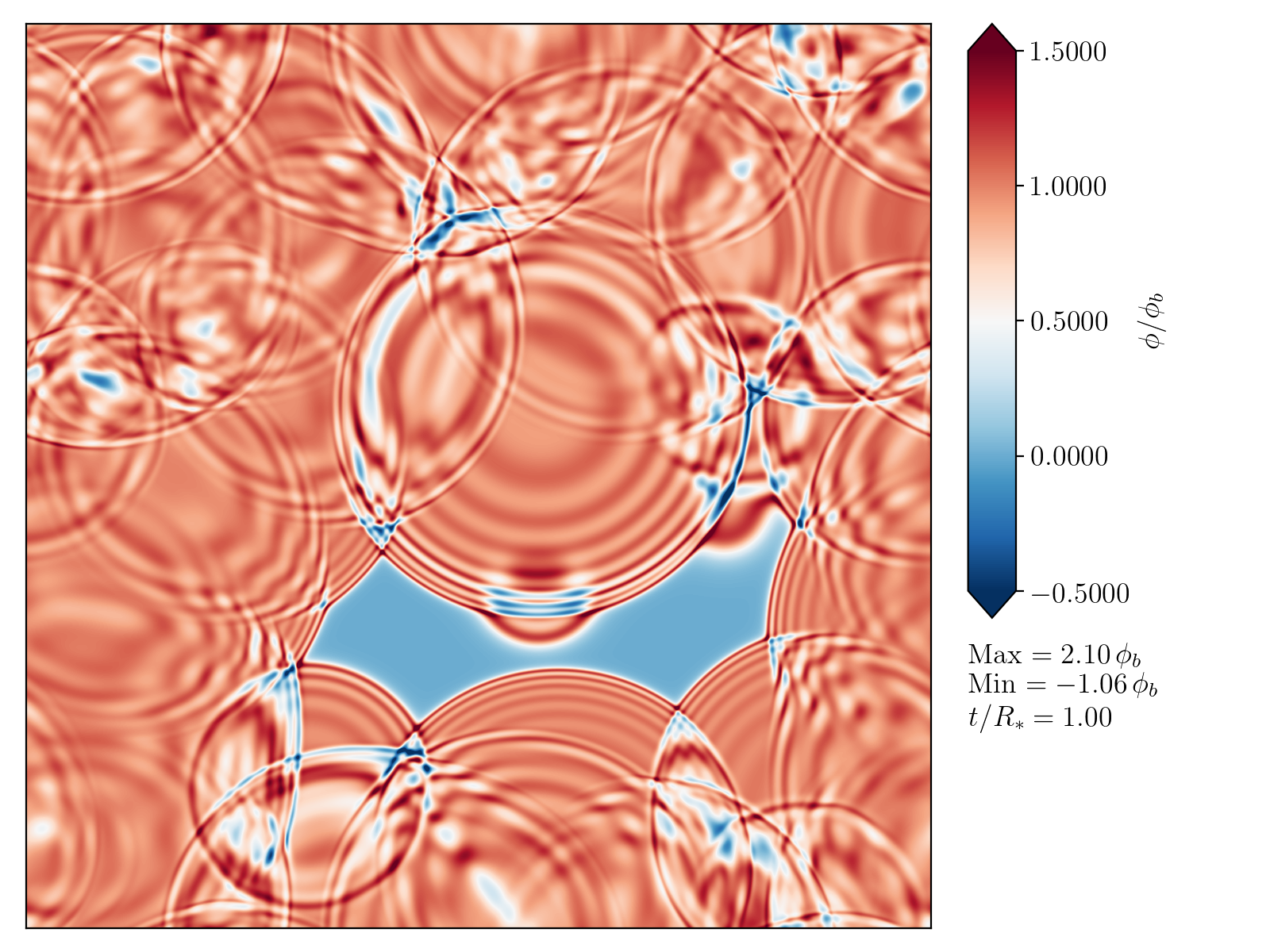}}
  \hfill
  \subfigure
  {\includegraphics[width=0.32\textwidth,clip]{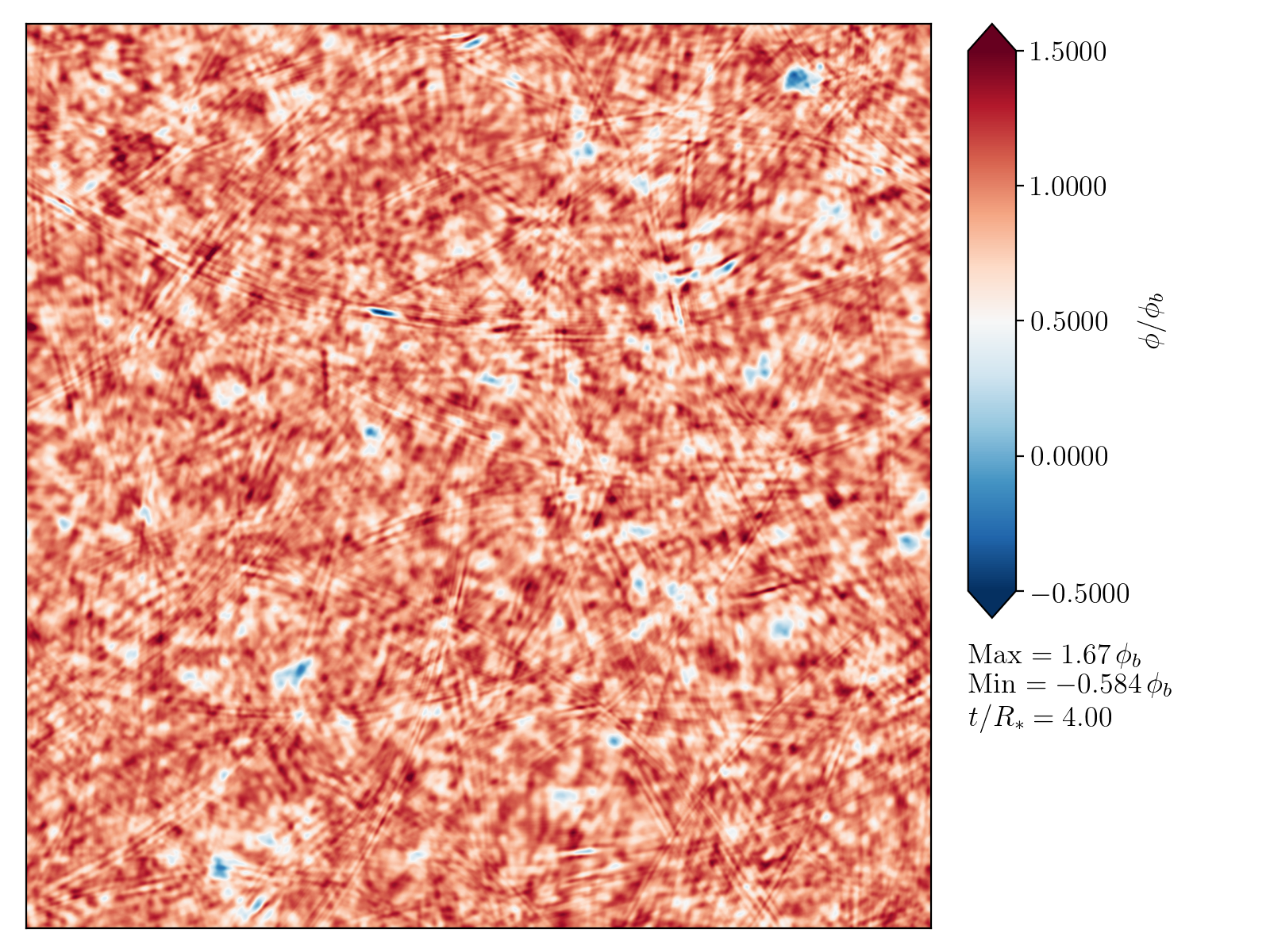}}
  \hfill
  \\
  \hfill
  \subfigure
  {\includegraphics[width=0.32\textwidth,clip]{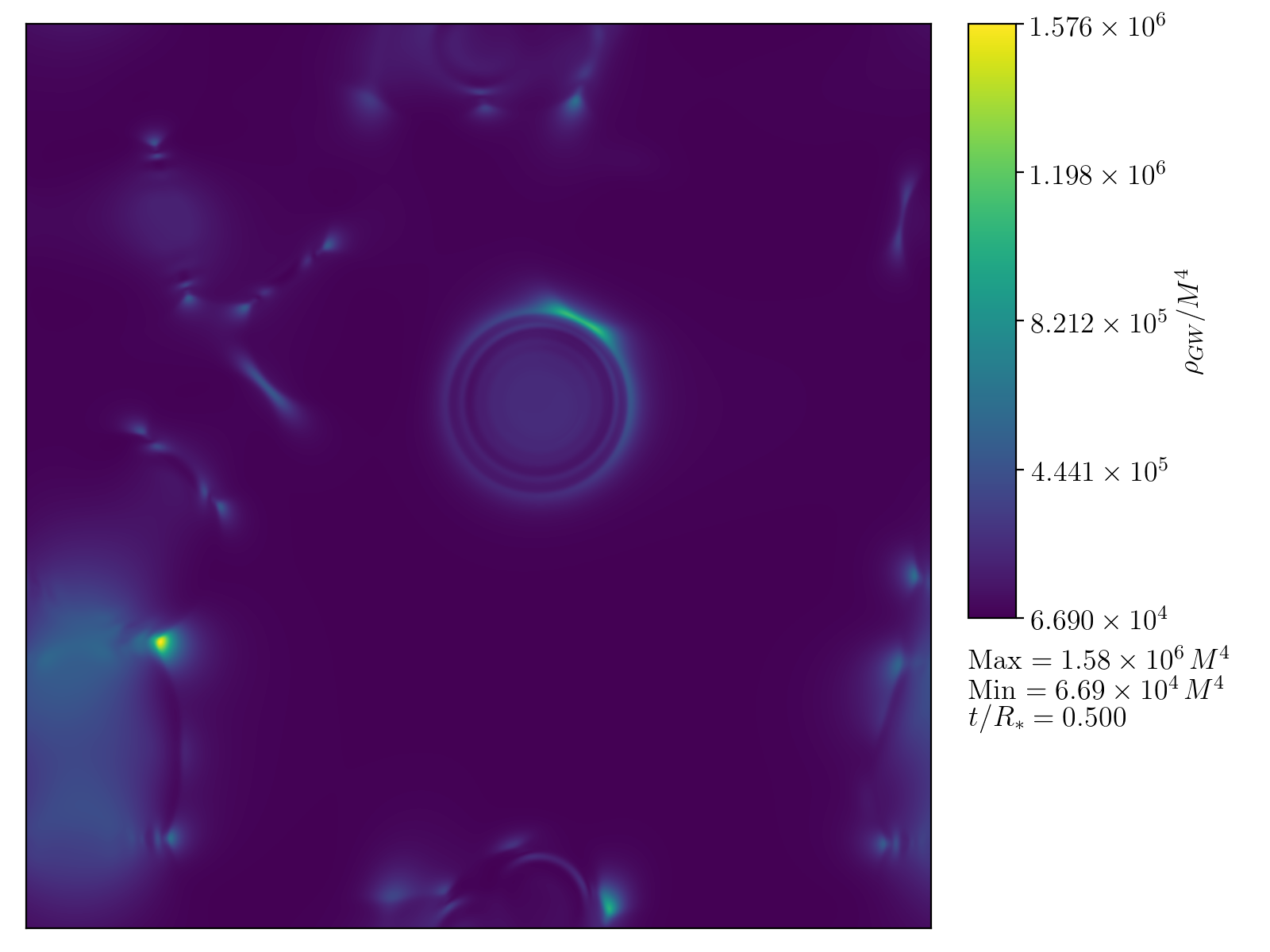}}
  \hfill
  \subfigure
  {\includegraphics[width=0.32\textwidth,clip]{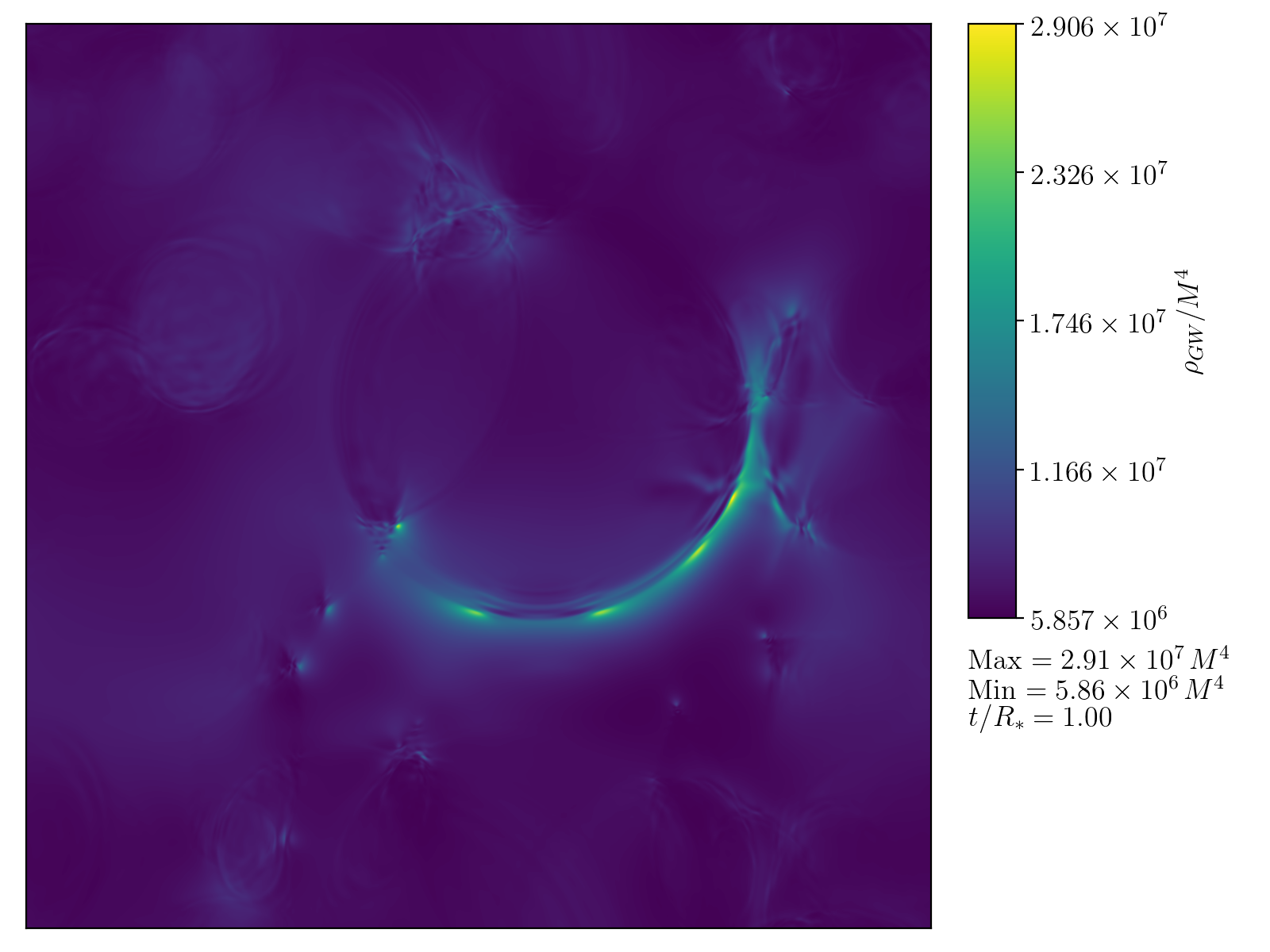}}
  \hfill
  \subfigure
  {\includegraphics[width=0.32\textwidth,clip]{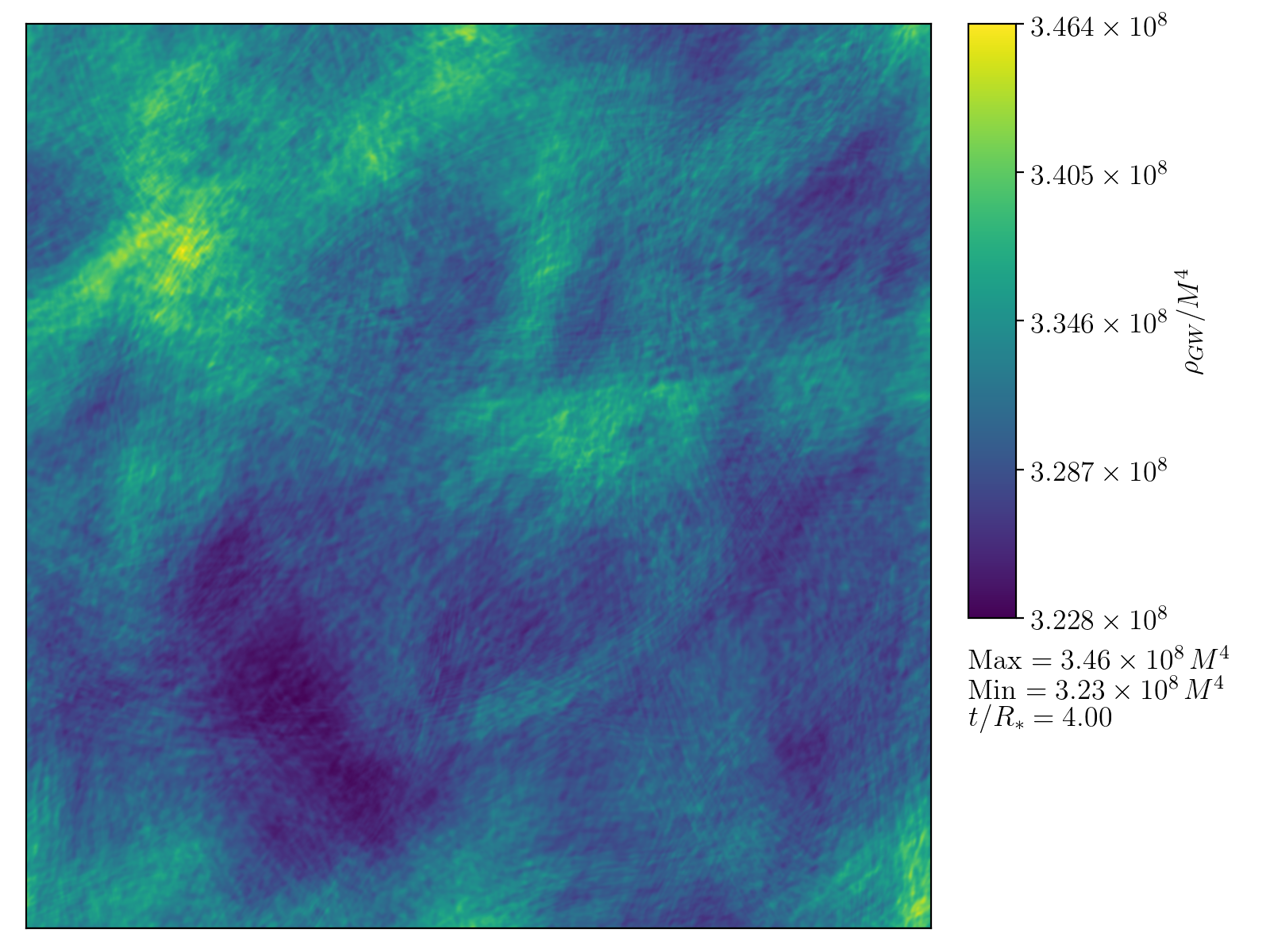}}
  \hfill
  \\
  \hfill
  \subfigure
  {\includegraphics[width=0.32\textwidth,clip]{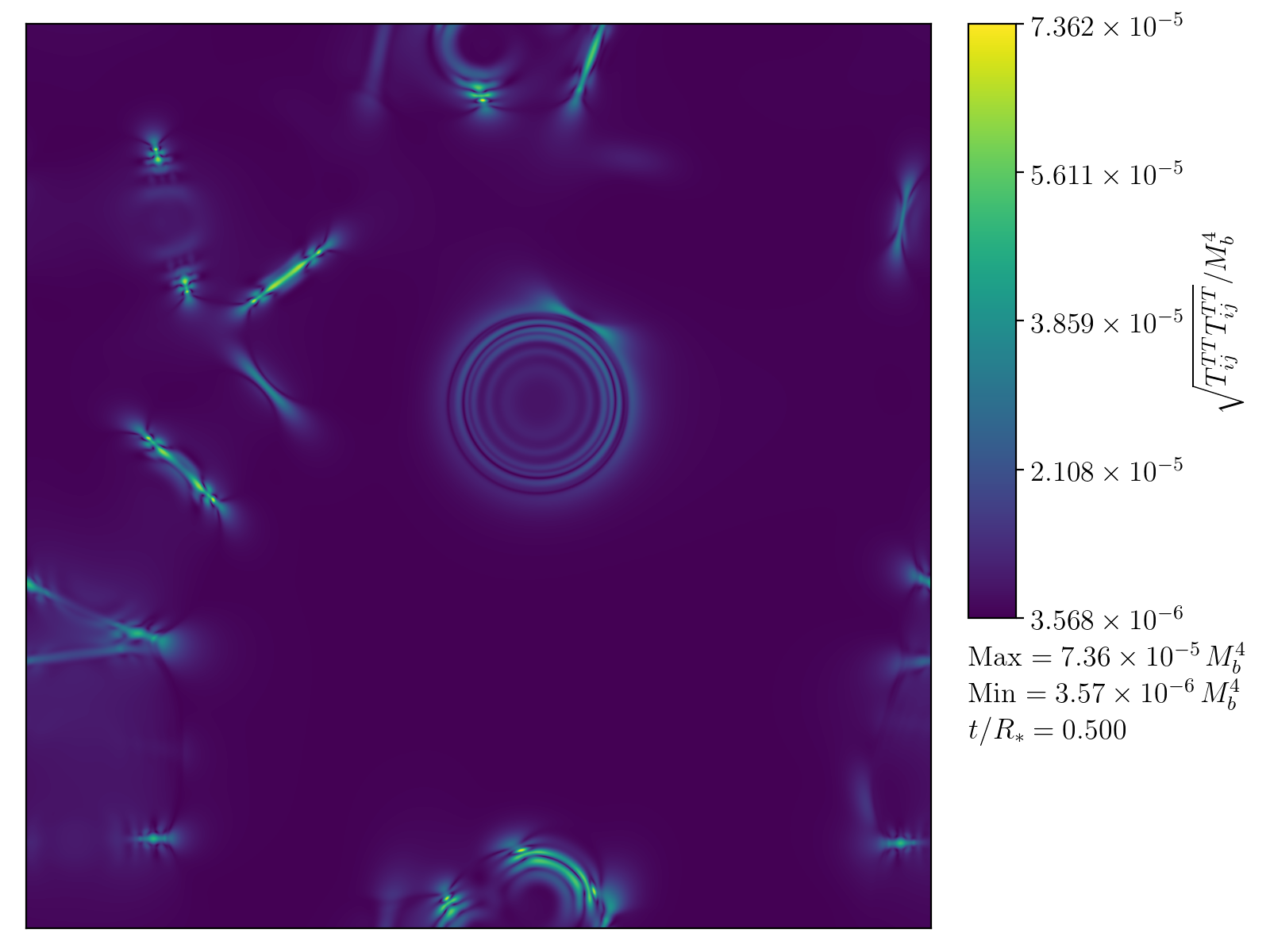}}
  \hfill
  \subfigure
  {\includegraphics[width=0.32\textwidth,clip]{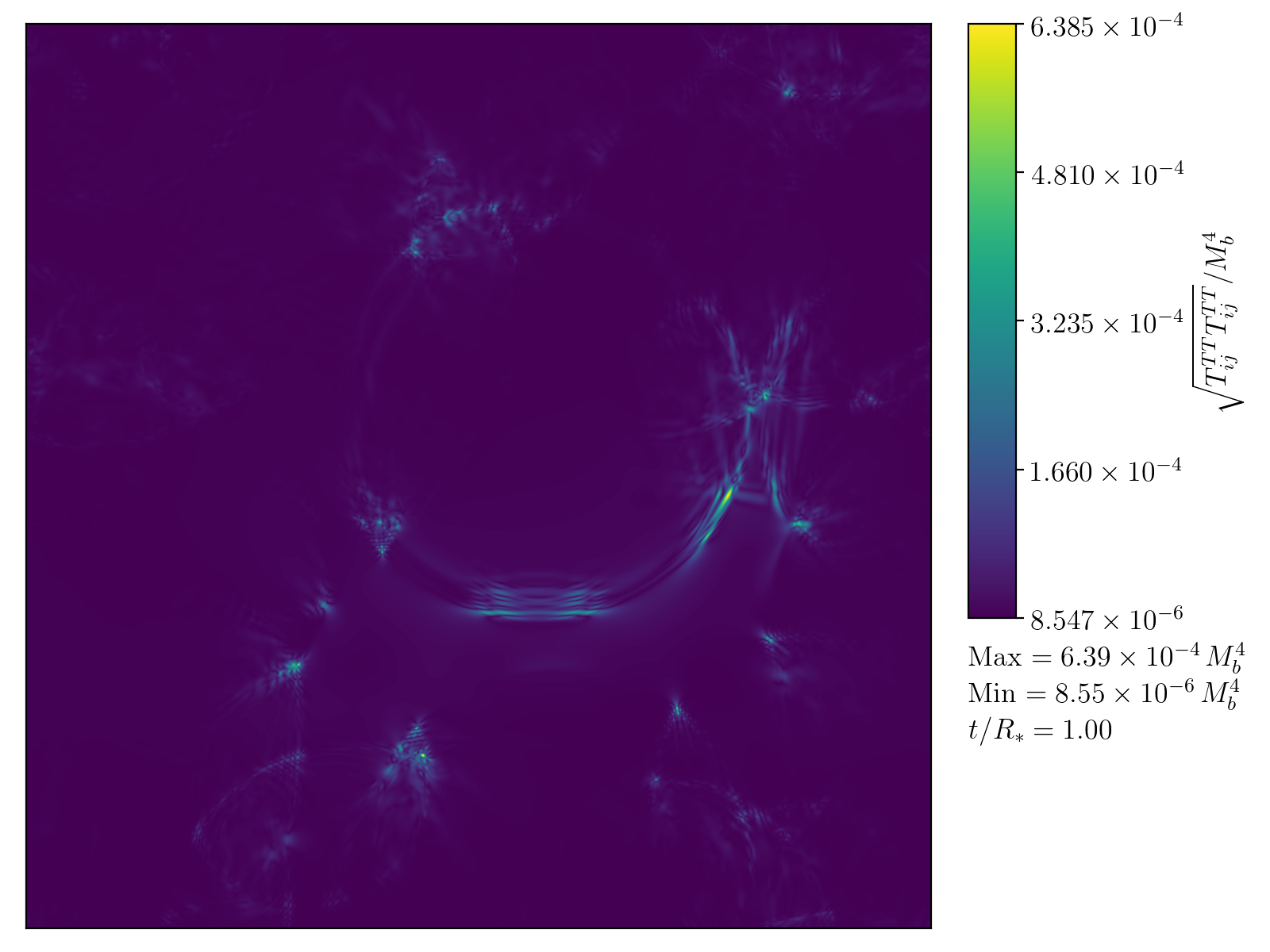}}
  \hfill
  \subfigure
  {\includegraphics[width=0.32\textwidth,clip]{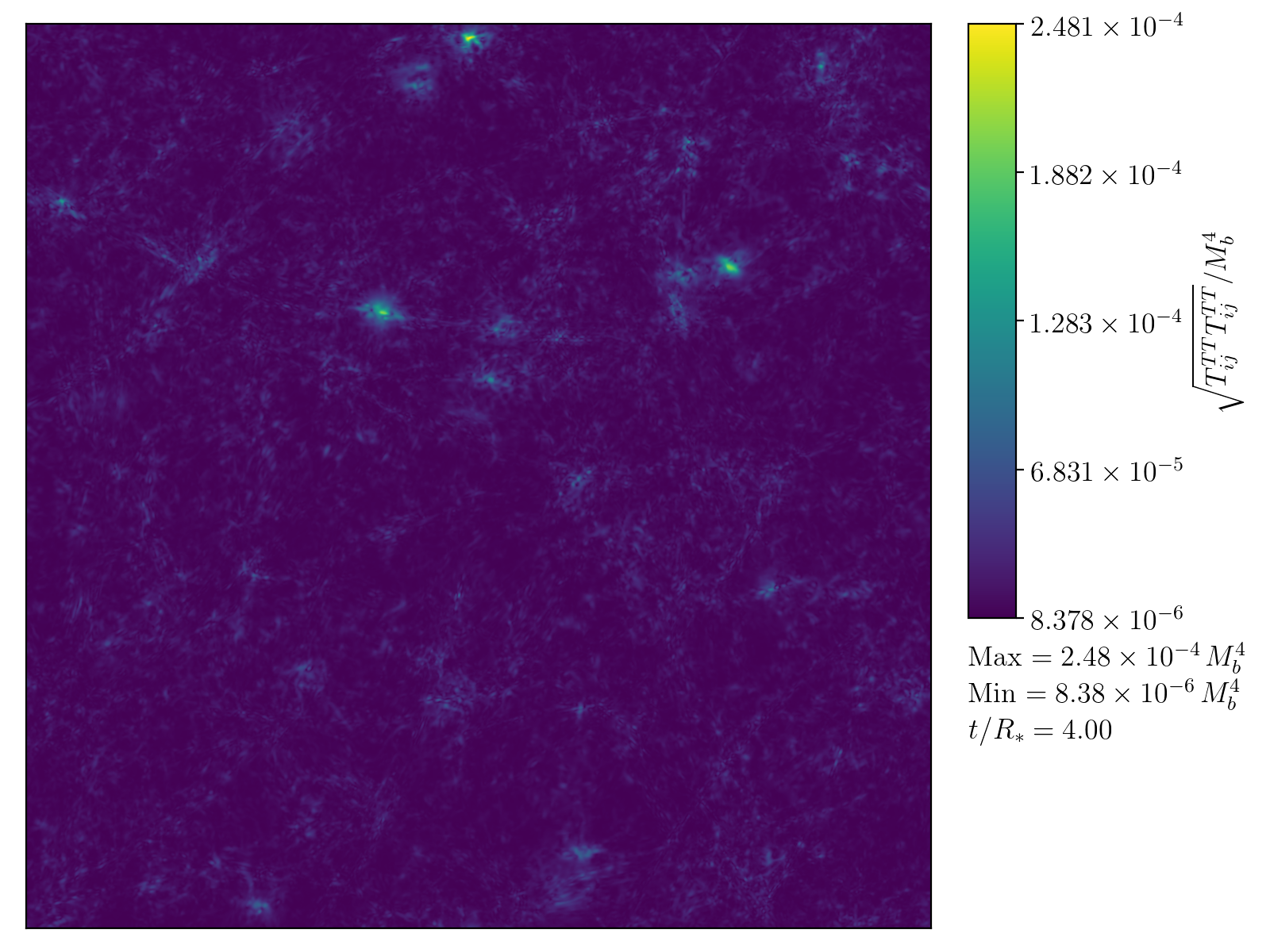}}
  \hfill
  \caption{Slices  $(0,y,z)$ for a simulation with $\lambar=0.07$ and $\Nb=64$.  In the top row we
    plot the scalar field normalised by the broken phase value. The middle row shows the energy density in gravitational waves
    $\rGW$. The bottom row shows the modulus of the transverse traceless shear-stress.}
  \label{fig:slices-0.18}
\end{figure}

\begin{figure}
  \centering
  \hfill
  \subfigure
  {\includegraphics[width=0.32\textwidth,clip]{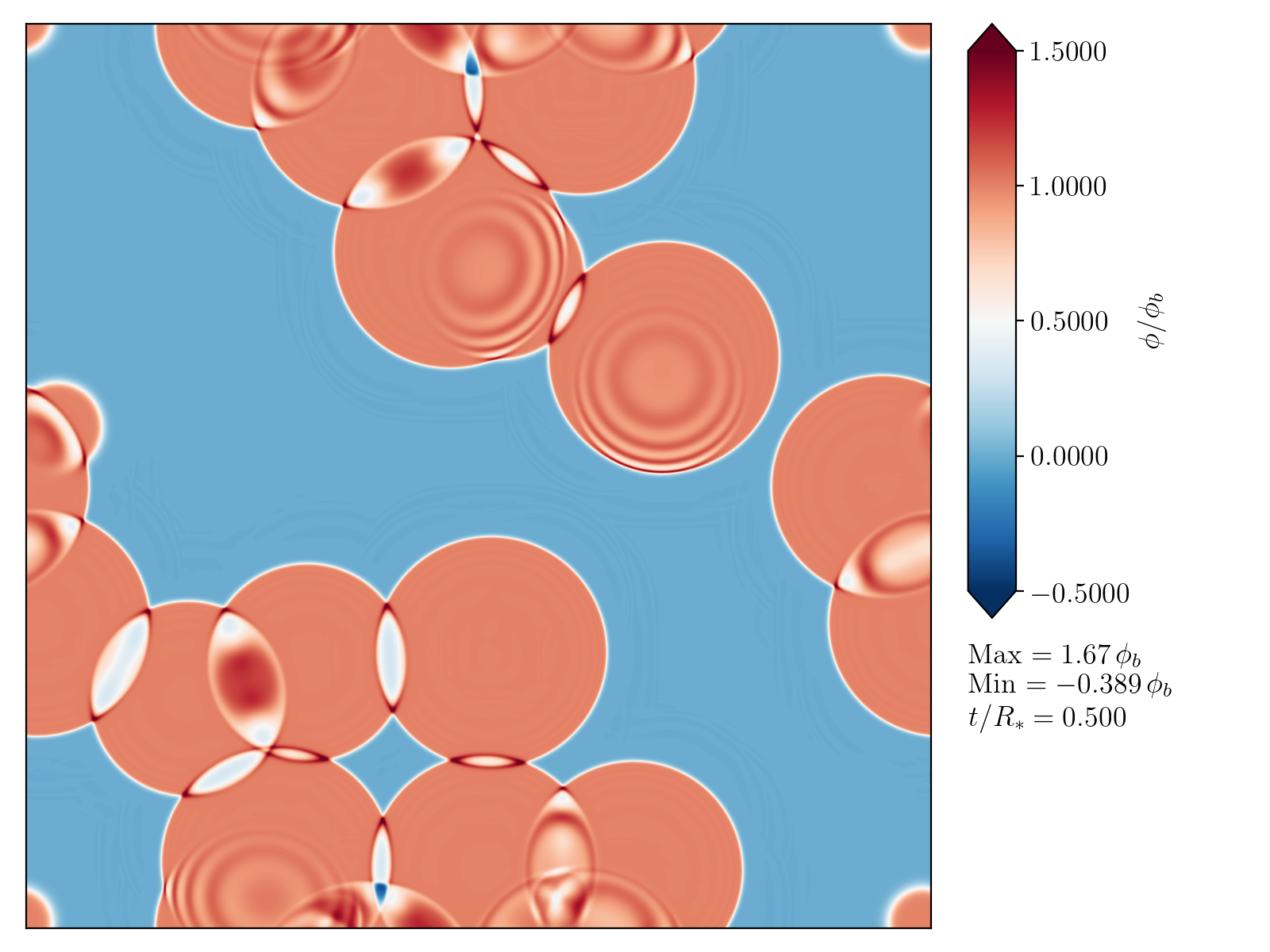}}
  \hfill
  \subfigure
  {\includegraphics[width=0.32\textwidth,clip]{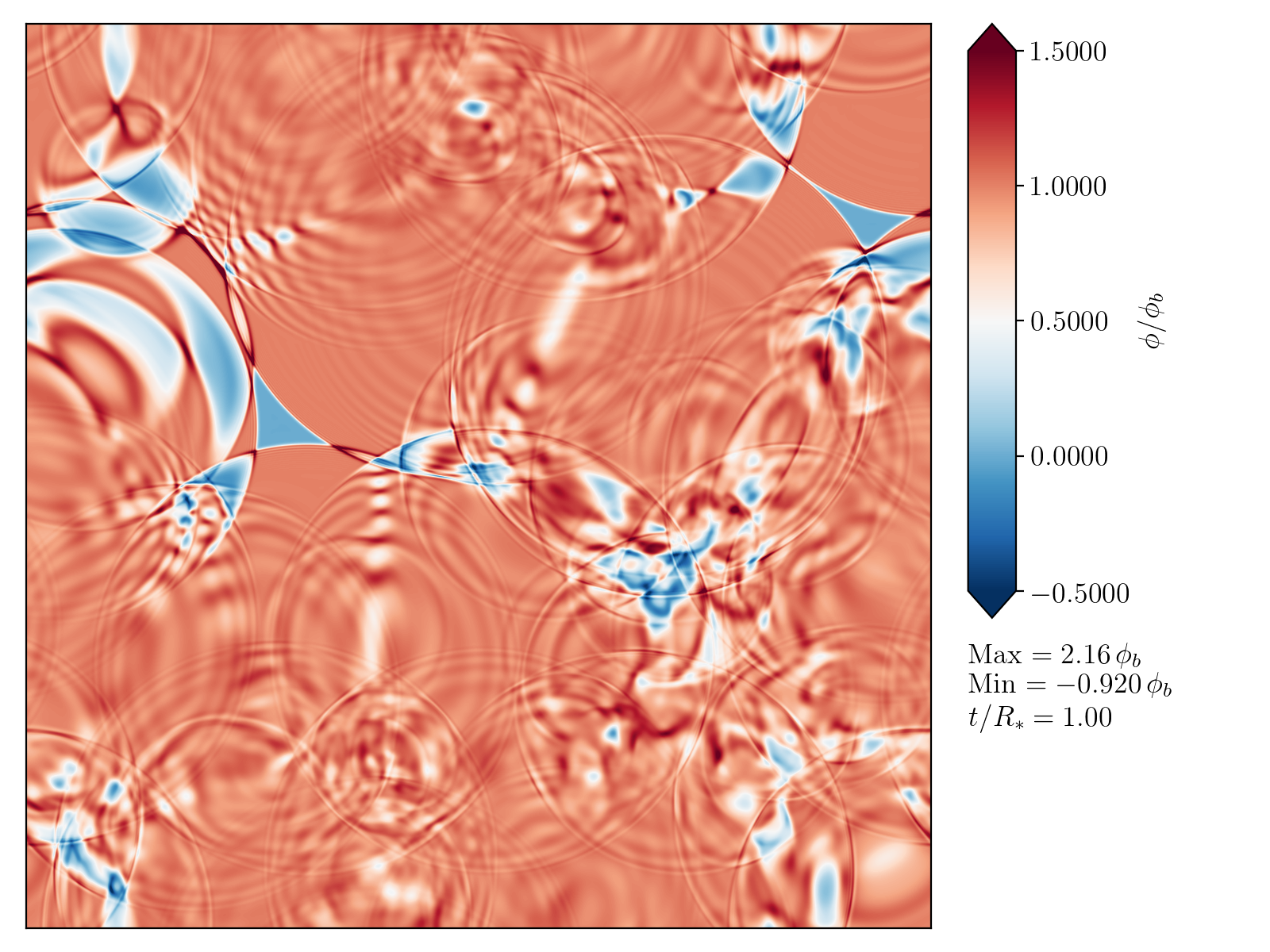}}
  \hfill
  \subfigure
  {\includegraphics[width=0.32\textwidth,clip]{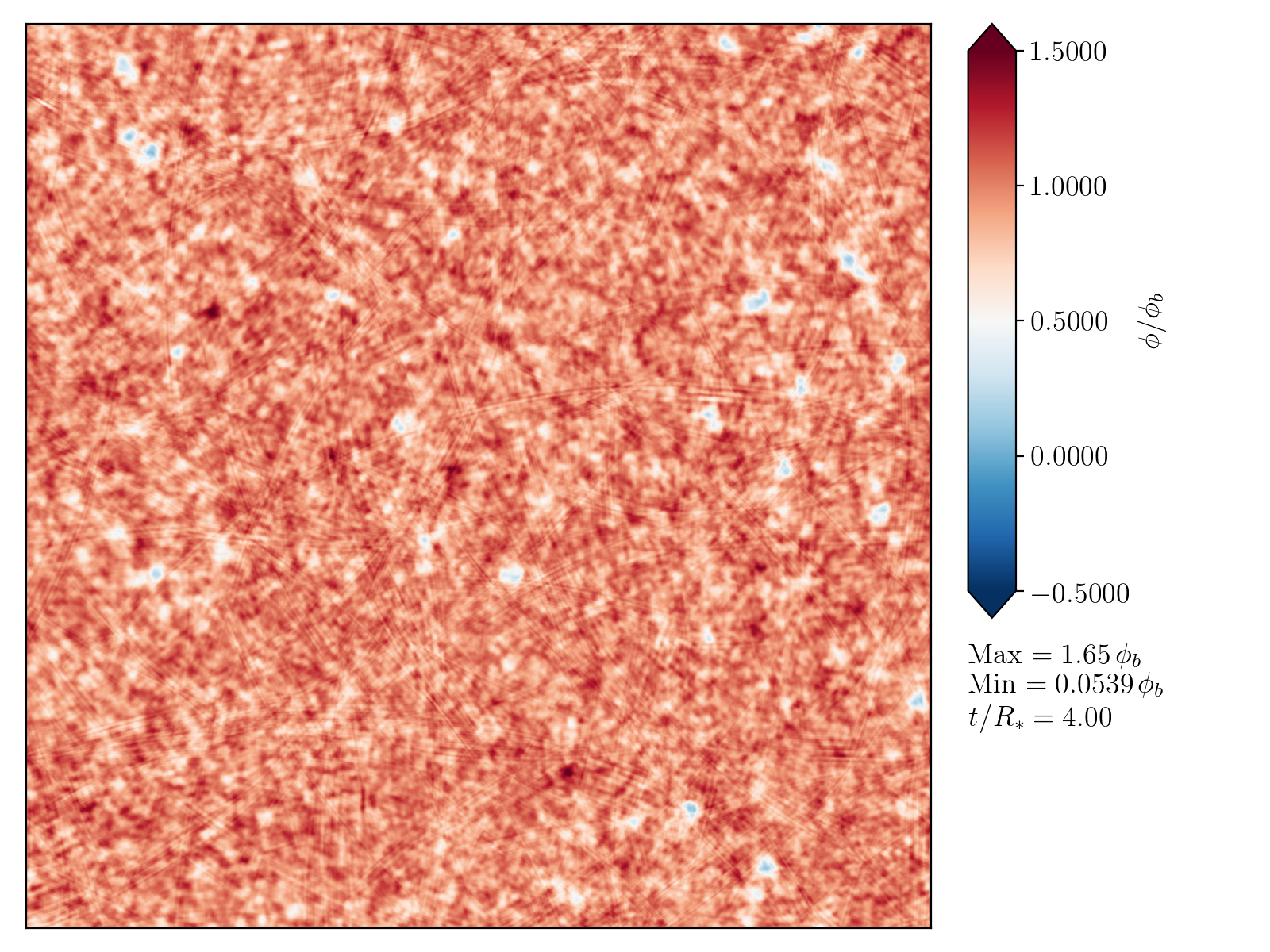}}
  \hfill
  \\
  \hfill
  \subfigure
  {\includegraphics[width=0.32\textwidth,clip]{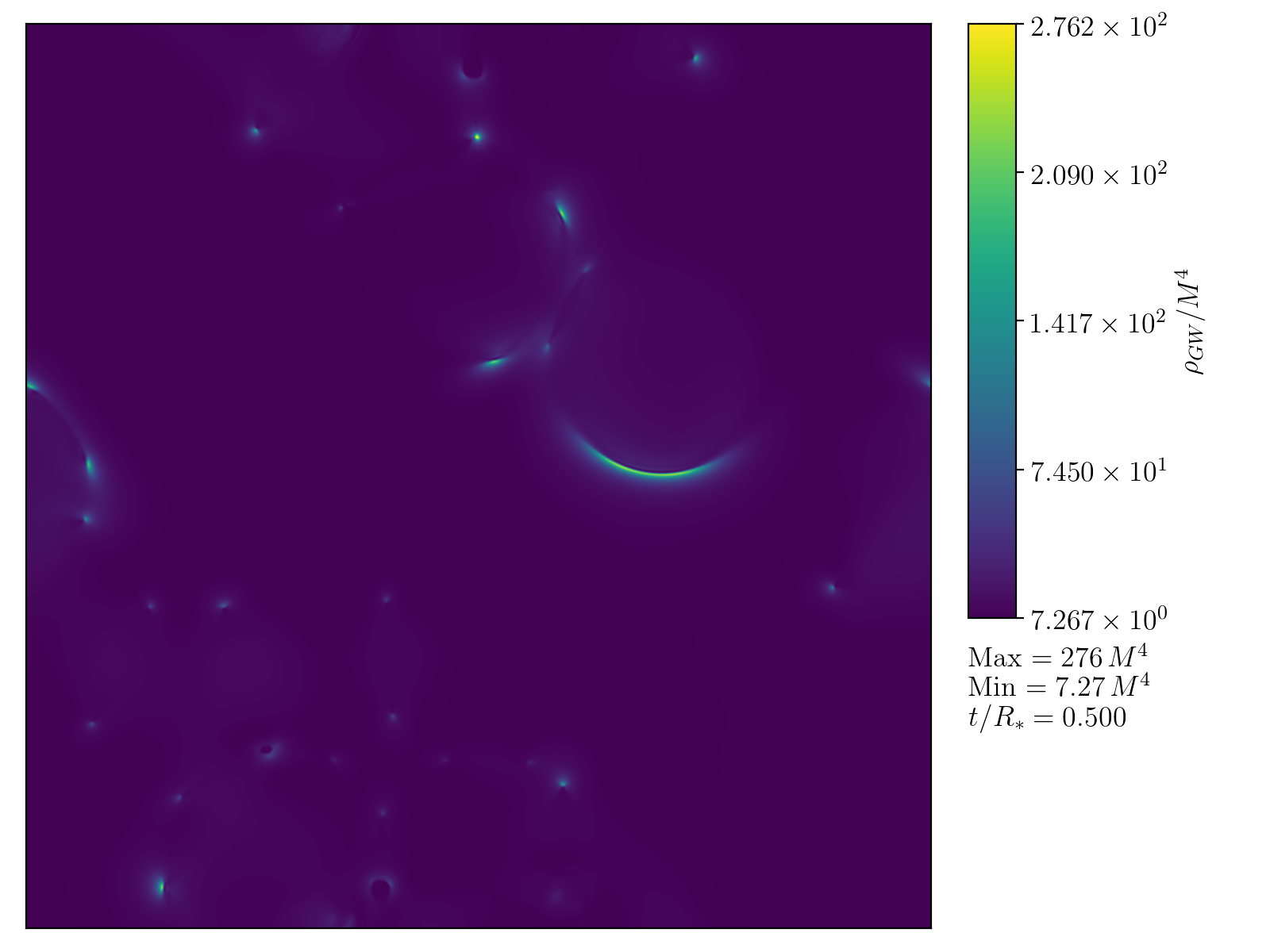}}
  \hfill
  \subfigure
  {\includegraphics[width=0.32\textwidth,clip]{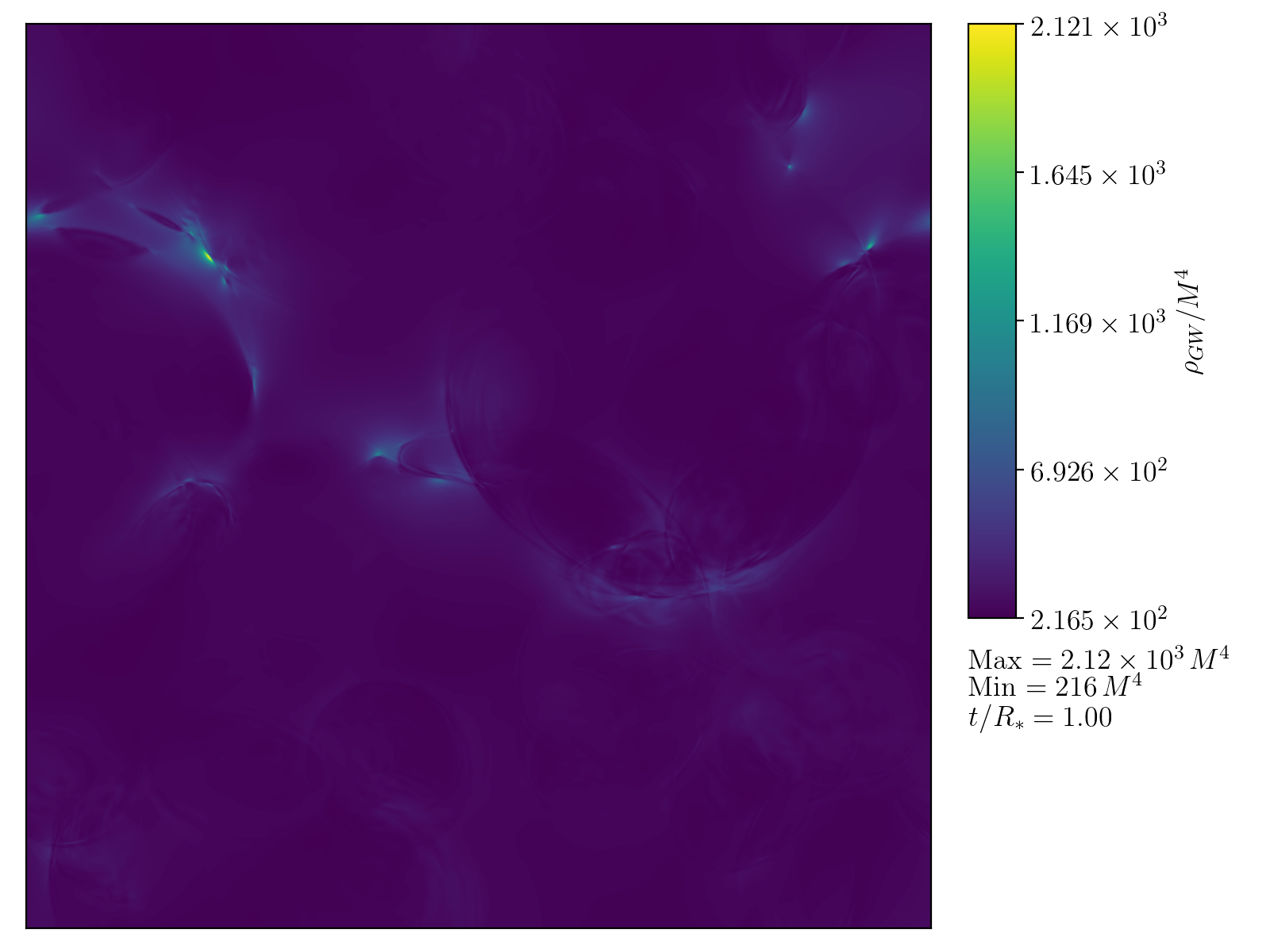}}
  \hfill
  \subfigure
  {\includegraphics[width=0.32\textwidth,clip]{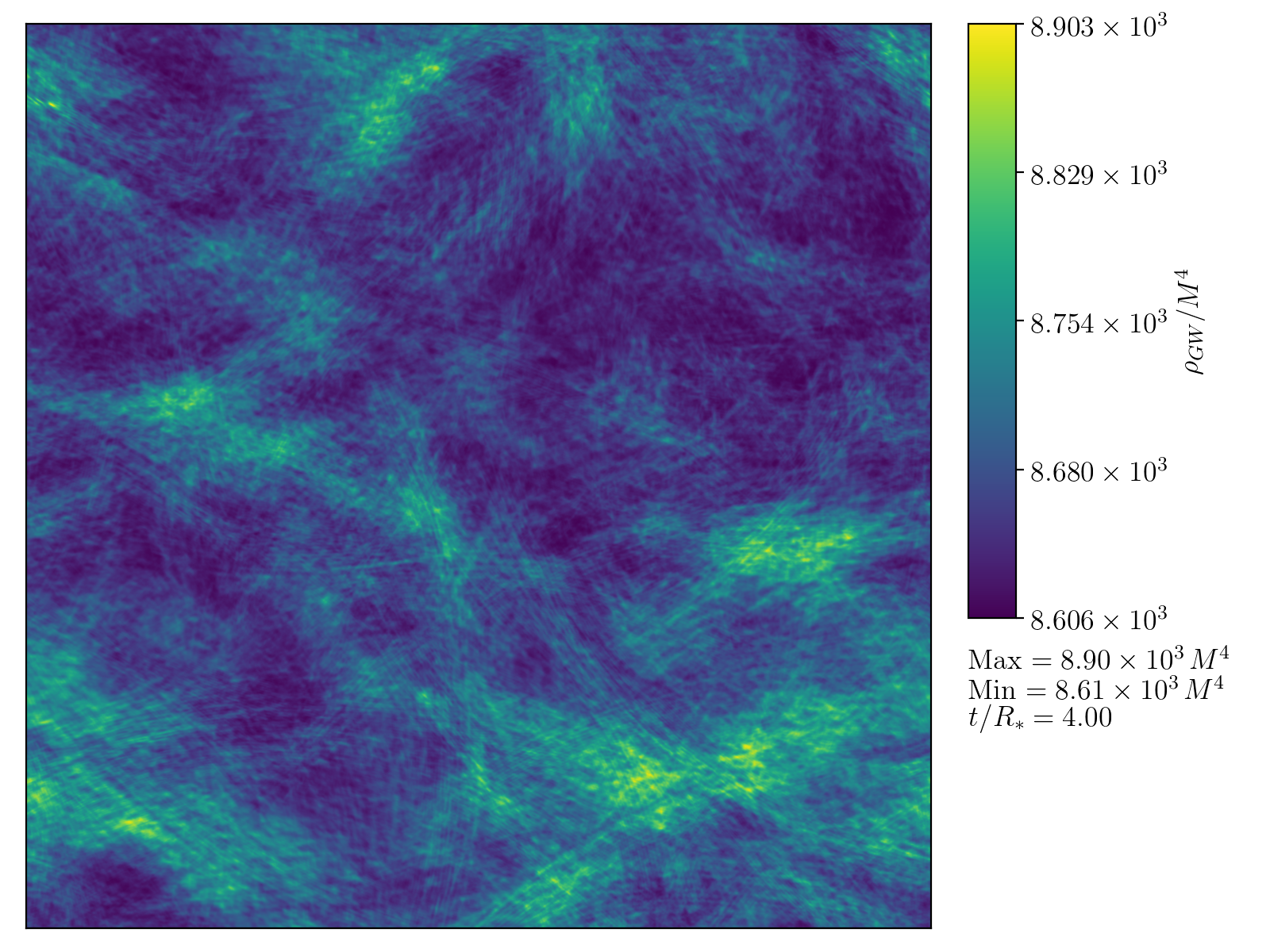}}
  \hfill
  \\
  \hfill
  \subfigure
  {\includegraphics[width=0.32\textwidth,clip]{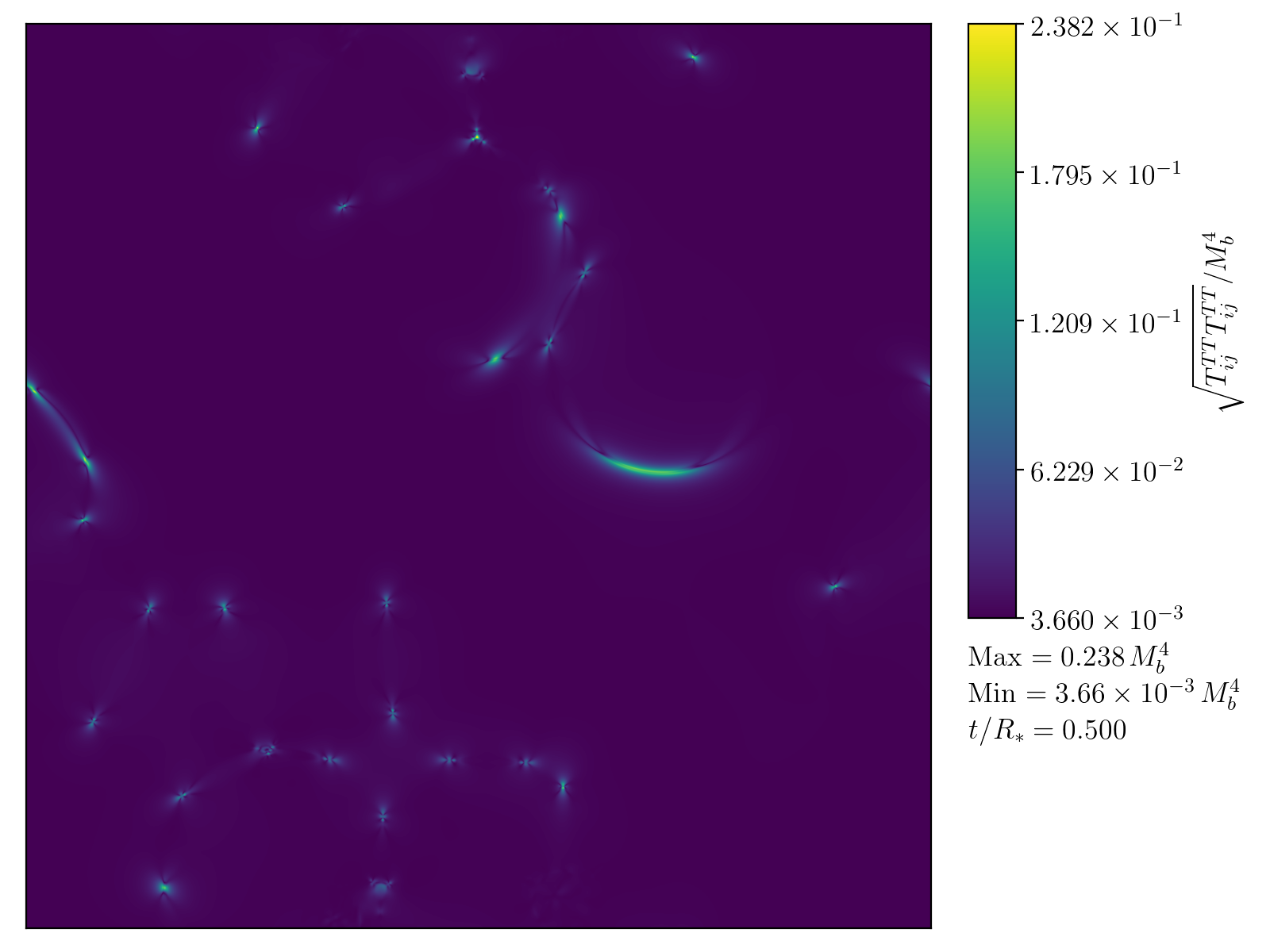}}
  \hfill
  \subfigure
  {\includegraphics[width=0.32\textwidth,clip]{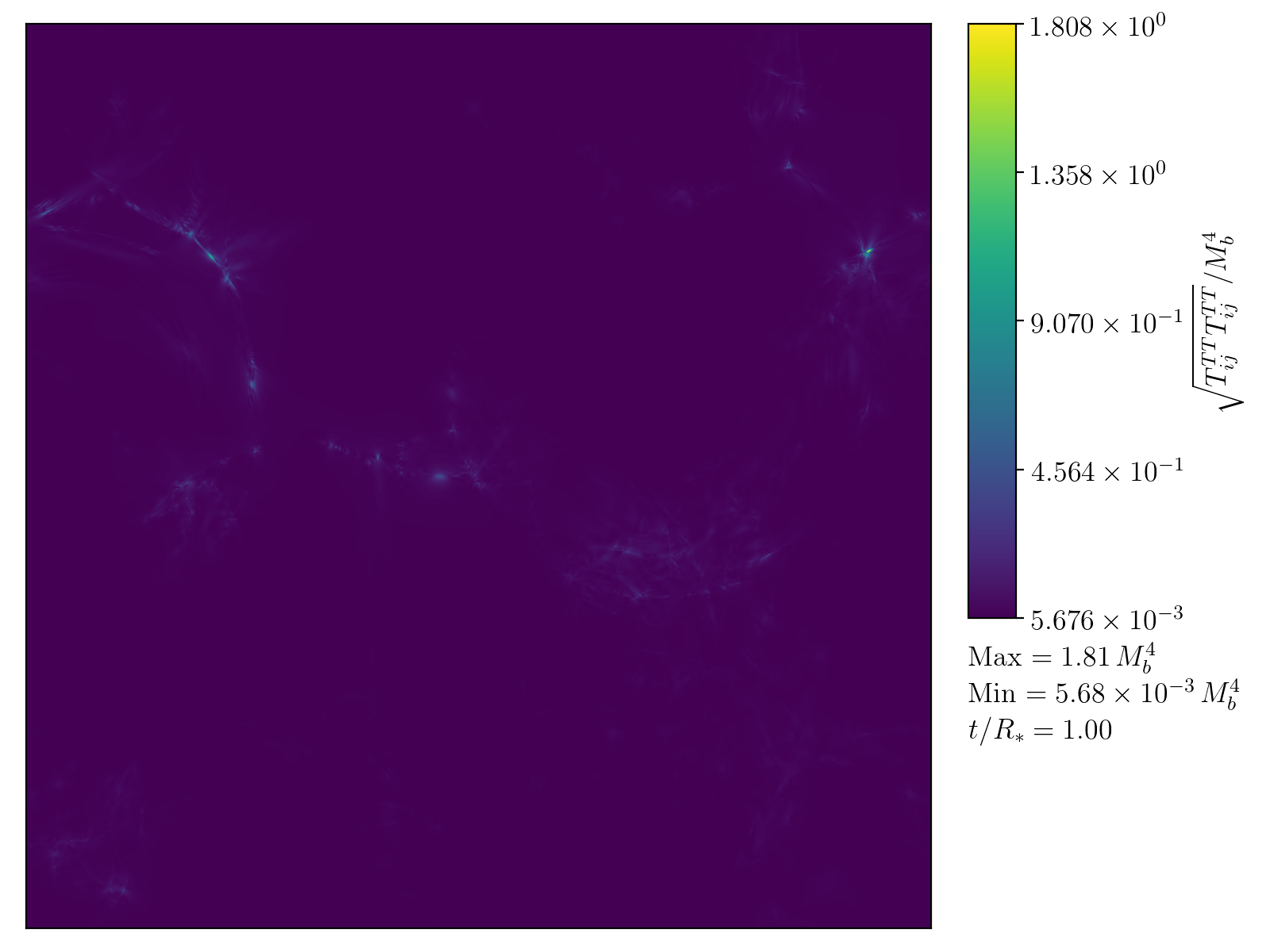}}
  \hfill
  \subfigure
  {\includegraphics[width=0.32\textwidth,clip]{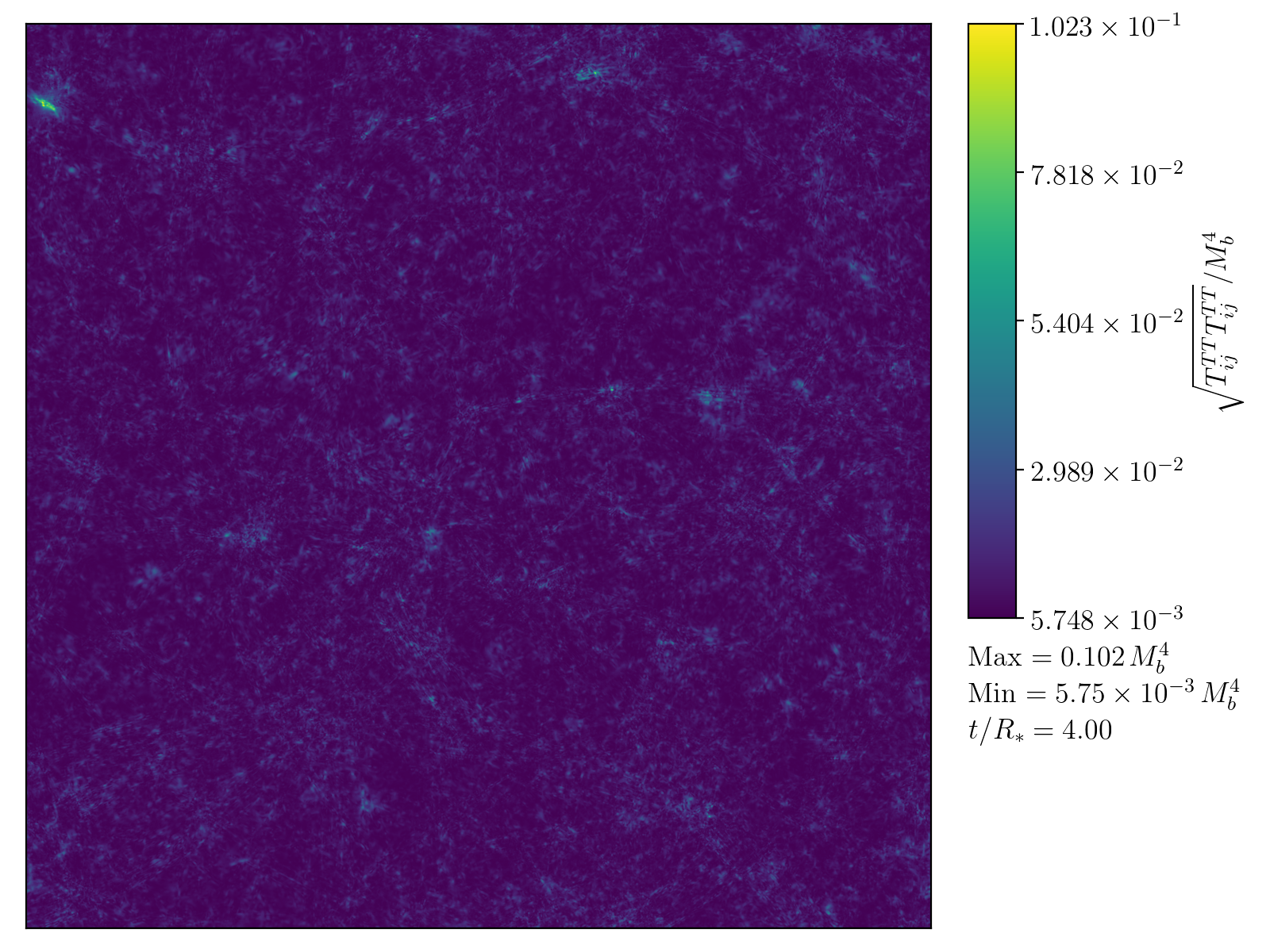}}
  \hfill
  \caption{Slices  $(0,y,z)$ for a simulation with $\lambar=0.84$ and $\Nb=64$.  In the top row we
    plot the scalar field normalised by the broken phase value. The middle row shows the energy density in gravitational waves
    $\rGW$. The bottom row shows the modulus of the transverse traceless shear-stress.}
  \label{fig:slices-0.84}
\end{figure}

In Fig.~\ref{fig:slices-0.18} and Fig.~\ref{fig:slices-0.84} we show
slices through simulations with $\lambar =0.07$ and $\lambar=0.84$
respectively. Both simulations have $\Nb=64$, and $\gmStar = 4$. The
slices show $\phi$, $\rGW$ and $\sqrt{T^{TT}_{ij}T^{TT}_{ij}}$ at
$t/\Rstar$ equal to $0.5$, $1.0$ and $4.0$.

\onecolumngrid

\twocolumngrid
\end{document}